\documentclass[a4paper]{article}

\usepackage{arxiv}

\usepackage[utf8]{inputenc} 
\usepackage[T1]{fontenc} 
\usepackage{hyperref} 
\usepackage{url} 
\usepackage{booktabs} 
\usepackage{amsfonts} 
\usepackage{nicefrac} 
\usepackage{microtype} 
\usepackage{lipsum}		
\usepackage{graphicx}
\usepackage{doi}
\usepackage{dsfont}
\usepackage{tabularray}
\usepackage[acronym]{glossaries}
\usepackage{rotating}

\usepackage{soul, color}
\usepackage{todonotes}
\usepackage{amsmath}
\usepackage{listings}
\usepackage{xcolor}
\usepackage{amsthm} 
\usepackage{algpseudocode}
\usepackage{algorithm}
\usepackage{longtable}
\graphicspath{{figures/New_simulation_figures}}
\usepackage{setspace}
\usepackage{bm}
\usepackage{float}
\usepackage{endfloat}
\DeclareDelayedFloatFlavor{sidewaystable}{table} \DeclareDelayedFloatFlavor{sidewaysfigure}{figure}
\newcommand{\nc}{\newcommand}
\nc{\ntitle}[1]{
 \begin{center}
   \fbox{\textbf{\Large #1}}
  \end{center}         }

\nc{\tred}[1]{\textcolor[rgb]{0.00,0.00,0.00}{#1}}
\nc{\tb}[1]{\textcolor[rgb]{0.00,0.00,1.00}{#1}}

\nc{\slideline}{\smallskip \hrule\hrule \smallskip}
\nc{\stitle}[1]{
\textbf{\large #1}
\slideline
\slideline
          }

\nc{\nn}{\nonumber}
\nc{\fns}{\footnotesize}

\nc{\revisionline}{\vspace{.1in} \today \vspace{.1in} \hrule\hrule\hrule\vspace{.1in}}
\nc{\newpp}{\vspace{.1in} \noindent}

\nc{\wh}{\widehat}

\nc{\Ef}{ {\rm E}_{\infty} }
\nc{\Ex}{ {\rm E} }
\nc{\Ec}{ {\rm E}_1 }
\nc{\Pf}{ {\rm P}_{\infty} }
\nc{\Pc}{ {\rm P}_{1} }
\nc{\Prb}{ {\rm P} }
\nc{\sd}{\pm \hat{\sigma} }

\nc{\indep}{{\, \perp \! \! \! \perp  \,} }
\nc{\tsps}{^{ {\rm T} } }

\nc{\pu}{\pi_{\rm U}}
\nc{\pbi}{\pi_{\rm B}}
\nc{\pnb}{\pi_{\rm NB}}
\nc{\prp}{\propto}
\nc{\pr}{ {\rm pr} }

\nc{\al}{\alpha}
\nc{\dl}{\delta}
\nc{\la}{\lambda}
\nc{\om}{\omega}
\nc{\vep}{\varepsilon}

\nc{\snf}{\sum_{n=1}^{\infty}}
\nc{\skf}{\sum_{k=1}^{\infty}}
\nc{\sner}{\sum_{n=1}^{86}}
\nc{\sjn}{\sum_{j=1}^{n}}
\nc{\skn}{\sum_{k=1}^{n}}
\nc{\sumim}{\sum_{i=1}^m}
\nc{\sumjn}{\sum_{j=1}^n}
\nc{\sumlL}{\sum_{l=1}^{L}}
\nc{\sumL}{\sum_{l=1}^{L}}
\nc{\sumkK}{\sum_{k=1}^{K_i}}
\nc{\sumrR}{\sum_{r=1}^R}

\nc{\hivp}{\sum_{ {\rm HIV}^+ } }
\nc{\sumiN}{ \sum_{i=1}^N }
\nc{\summM}{ \sum_{m=1}^M }
\nc{\sumjM}{ \sum_{j=1}^M }

\nc{\lsq}{\left[}
\nc{\rsq}{\right]}
\nc{\lbc}{\left \{ }
\nc{\rbc}{\right \} }
\nc{\lp}{\left(}
\nc{\rp}{\right)}

\nc{\imp}{\Rightarrow}
\nc{\lbf}{\lim_{b \rightarrow \infty}}
\nc{\limNinf}{\lim_{N \rightarrow \infty}}
\nc{\limminf}{\lim_{m \rightarrow \infty}}
\nc{\limninf}{\lim_{n \rightarrow \infty}}
\nc{\convd}{\stackrel{D}{\longrightarrow}}
\nc{\convp}{\stackrel{P}{\longrightarrow}}
\nc{\eqd}{\stackrel{{\EuScript D}}{=}}

\nc{\trans}{^{\text T}}
\nc{\ol}{\overline}
\nc{\logit}{\text{logit}\,}

\nc{\rl}{ {\rm {\bf R} } }
\nc{\zah}{ {\rm {\bf Z} } }

\nc{\lkn}{\Lambda^n_k}
\nc{\stp}{ {\cal C}_b }
\nc{\istp}{ {\cal I}_A }
\nc{\snb}{S_{N_b}}
\nc{\stb}{S_{T_b}}
\nc{\ixlog}{I_{ \{ 0 \leq x \leq \log \al \} } }
\nc{\iulog}{I_{ \{ 0 \leq u  \leq \log \al \} } }
\nc{\rgn}{ \Upsilon_n }
\nc{\var}{{\rm var}}
\nc{\cov}{{\rm cov}}
\nc{\corr}{{\rm corr}}
\nc{\dpl}{\partial}
\nc{\half}{ {\textstyle \frac{1}{2}} }
\nc{\tr}{{\rm trace}}
\nc{\real}{\mathbb{R}}
\nc{\bbC}{\mathbb{C}}

\def\boxit#1{\vbox{\hrule\hbox{\vrule\kern6pt\vbox{\kern6pt#1\kern6pt}\kern6pt\vrule}\hrule}}

\nc{\calb}{ {\cal B} }
\nc{\calc}{ {\cal C} }
\nc{\bcalc}{ \mbox{\boldmath{${\cal C}$}}}
\nc{\cald}{ {\cal D} }
\nc{\cale}{ {\cal E} }
\nc{\cali}{ {\cal I} }
\nc{\call}{ {\cal L} }
\nc{\calm}{ {\cal M} }
\nc{\caln}{ {\cal N} }
\nc{\cals}{ {\cal S} }
\nc{\calo}{ {\cal O} }
\nc{\bcalo}{ \mbox{\boldmath{${\cal O}$}}}
\nc{\calt}{ {\cal T} }
\nc{\calv}{ {\cal V} }
\nc{\bcalu}{ \mbox{\boldmath{${\cal U}$}}}
\nc{\calu}{ {\cal U} }
\nc{\calw}{ {\cal W} }
\nc{\calx}{ {\cal X} }

\nc{\sca}{ {\EuScript A} }
\nc{\scb}{ {\EuScript B} }
\nc{\scc}{ {\EuScript C} }
\nc{\scd}{ {\EuScript D} }
\nc{\sce}{ {\EuScript E} }
\nc{\scf}{ {\EuScript F} }
\nc{\scF}{ {\EuScript f} }
\nc{\scg}{ {\EuScript G} }
\nc{\sch}{ {\EuScript H} }
\nc{\sci}{ {\EuScript I} }
\nc{\scj}{ {\EuScript J} }
\nc{\sck}{ {\EuScript K} }
\nc{\scl}{ {\EuScript L} }
\nc{\sclic}{ \scl_i^{\rm c} }
\nc{\scm}{ {\EuScript M} }
\nc{\scn}{ {\EuScript N} }
\nc{\sco}{ {\EuScript O} }
\nc{\scp}{ {\EuScript P} }
\nc{\scq}{ {\EuScript Q} }
\nc{\scr}{ {\EuScript R} }
\nc{\scs}{ {\EuScript S} }
\nc{\sct}{ {\EuScript T} }
\nc{\scu}{ {\EuScript U} }
\nc{\scv}{ {\EuScript V} }
\nc{\scw}{ {\EuScript W} }
\nc{\scx}{ {\EuScript X} }
\nc{\scy}{ {\EuScript Y} }
\nc{\scz}{ {\EuScript Z} }
\nc{\scxo}{ {\EuScript X}_{\rm obs} }
\nc{\Xobs}{ \pmb{\scx}_{\rm obs} }
\nc{\Xcom}{ \pmb{\scx} }
\nc{\Xmis}{ \pmb{\scx}_{\rm mis} }

\nc{\bsci}{ \mbox{\boldmath{$\sci$}}}
\nc{\bscj}{ \mbox{\boldmath{$\scj$}}}

\nc{\sumlic}{\sum_{l \in sclic}}

\nc{\scyo}{ {\EuScript Y}_{\rm obs} }

\nc{\bga}{\begin{array}{c}}
\nc{\ena}{\end{array}}

\nc{\mhat}{ {\hat{p}}_M }
\nc{\fhat}{ {\hat{p}}_F }
\nc{\ph} { \hat{p} }

\nc{\ta}{ {\tilde{a}} }
\nc{\tc}{ {\tilde{c}} }

\nc{\bal}{\mbox{\boldmath{$\alpha$}}}
\nc{\balpha}{\mbox{\boldmath{$\alpha$}}}
\nc{\bone}{\mbox{\boldmath{$1$}}}
\nc{\bbet}{\mbox{\boldmath{$\beta$}}}
\nc{\bbeta}{\mbox{\boldmath{$\beta$}}}
\nc{\bDel}{\mbox{\boldmath{$\Delta$}}}
\nc{\bDelta}{\mbox{\boldmath{$\Delta$}}}
\nc{\bdel}{\mbox{\boldmath{$\delta$}}}
\nc{\bdelta}{\mbox{\boldmath{$\delta$}}}
\nc{\bet}{\mbox{\boldmath{$\eta$}}}
\nc{\beps}{\mbox{\boldmath{$\epsilon$}}}
\nc{\bvep}{\mbox{\boldmath{$\vep$}}}
\nc{\bgam}{\mbox{\boldmath{$\gamma$}}}
\nc{\bgamma}{\mbox{\boldmath{$\gamma$}}}
\nc{\bGamma}{\mbox{\boldmath{$\Gamma$}}}
\nc{\bLam}{\mbox{\boldmath{$\Lambda$}}}
\nc{\bLambda}{\mbox{\boldmath{$\Lambda$}}}
\nc{\blambda}{\mbox{\boldmath{$\lambda$}}}
\nc{\bmu}{ \mbox{\boldmath{$\mu$}}}
\nc{\bOm}{ \mbox{\boldmath{$\Omega$}}}
\nc{\bOmega}{ \mbox{\boldmath{$\Omega$}}}
\nc{\bom}{ \mbox{\boldmath{$\omega$}}}
\nc{\bomega}{ \mbox{\boldmath{$\omega$}}}
\nc{\bpi}{ \mbox{\boldmath{$\pi$}}}
\nc{\bPi}{ \mbox{\boldmath{$\Pi$}}}
\nc{\bpsi}{ \mbox{\boldmath{$\psi$}}}
\nc{\bPsi}{ \mbox{\boldmath{$\Psi$}}}
\nc{\bphi}{ \mbox{\boldmath{$\phi$}}}
\nc{\bPhi}{ \mbox{\boldmath{$\Phi$}}}
\nc{\bxi}{ \mbox{\boldmath{$\xi$}}}
\nc{\bXi}{ \mbox{\boldmath{$\Xi$}}}
\nc{\bSig}{\mbox{\boldmath{$\Sigma$}}}
\nc{\bSigma}{\mbox{\boldmath{$\Sigma$}}}
\nc{\bsig}{\mbox{\boldmath{$\sigma$}}}
\nc{\btau}{\mbox{\boldmath{$\tau$}}}
\nc{\bThe}{\mbox{\boldmath{$\Theta$}}}
\nc{\bTheta}{\mbox{\boldmath{$\Theta$}}}
\nc{\bthe}{\mbox{\boldmath{$\theta$}}}
\nc{\btheta}{\mbox{\boldmath{$\theta$}}}
\nc{\bzeta}{\mbox{\boldmath{$\zeta$}}}
\nc{\bIm}{\mbox{\boldmath{$\Im$}}}

\nc{\ba}{ { \bf a }}
\nc{\bA}{ { \bf A }}
\nc{\bB}{ { \bf B }}
\nc{\bb}{ { \bf b }}
\nc{\bc}{ { \bf c }}
\nc{\bC}{ { \bf C }}
\nc{\bD}{ { \bf D }}
\nc{\bd}{ { \bf d }}
\nc{\be}{ { \bf e }}
\nc{\bF}{ { \bf F }}
\nc{\bG}{ { \bf G }}
\nc{\bh}{ { \bf h }}
\nc{\bH}{ { \bf H }}
\nc{\bI}{ { \bf I }}
\nc{\bJ}{ { \bf J }}
\nc{\bK}{ { \bf K }}
\nc{\bL}{ { \bf L }}
\nc{\bM}{ { \bf M }}
\nc{\bn}{ { \bf n }}
\nc{\bO}{ { \bf O }}
\nc{\bP}{ { \bf P }}
\nc{\br}{ { \bf r }}
\nc{\bR}{ { \bf R }}
\nc{\bs}{ { \bf s }}
\nc{\bS}{ { \bf S }}
\nc{\bT}{ { \bf T }}
\nc{\bt}{ { \bf t }}
\nc{\bu}{ { \bf u }}
\nc{\bU}{ { \bf U }}
\nc{\bv}{ { \bf v }}
\nc{\bV}{ { \bf V }}
\nc{\bW}{ { \bf W }}
\nc{\bw}{ { \bf w }}
\nc{\bx}{ { \bf x }}
\nc{\bX}{ { \bf X }}
\nc{\by}{ { \bf y }}
\nc{\bY}{ { \bf Y }}
\nc{\bz}{ { \bf z }}
\nc{\bZ}{ { \bf Z }}

\nc{\YR}{[\bY,R]}
\nc{\YgivenR}{[\bY \mid R]}
\nc{\RgivenY}{[R \mid \bY]}
\nc{\Y}{[\bY]}
\nc{\R}{[R]}

\nc{\dio}{d_i^o}
\nc{\timi}{t_{i,m_i}}
\nc{\betahat}{\hat{\bbet}}
\nc{\mui}{\bmu_{\rm I}}
\nc{\mue}{\bmu^{\rm E}}
\nc{\mup}{\bmu^{\rm P}}
\nc{\muihat}{\hat{\bmu}_{\rm I}}
\nc{\muehat}{\hat{\bmu}^{\rm E}}
\nc{\muphat}{\hat{\bmu}^{\rm P}}
\nc{\delhat}{\hat{\bdel}}
\nc{\muhat}{\hat{\bmu}}

\nc{\iid}{\stackrel{\rm iid}{\sim}}
\nc{\law}{\stackrel{\scl}{=}}

\nc{\phiij}{ \phi_{ij}( \Delta_0) }
\nc{\phiiprmj}{ \phi_{i'j}( \Delta_0) }
\nc{\phiijprm}{ \phi_{ij'}( \Delta_0) }
\nc{\phixy}{ \phi( X_i(S_{ik}), Y_j(T_{jl}) ) }
\nc{\phixydo}{ \phi( X_i(S_{ik}), Y_j(T_{jl})-\Delta_0 ) }
\nc{\phixyd}{ \phi( X_i(S_{ik}), Y_j(T_{jl})-\Delta) }
\nc{\phixydstar}{ \phi^*( X_i(S_{ik}), Y_j(T_{jl})-\Delta) }
\nc{\phixystdttil}{ \tilde{\phi}( X_i(s), Y_j(t)-\Delta, \theta) }
\nc{\phixydttil}{ \tilde{\phi}( X_i(S_{ik}), Y_j(T_{jl})-\Delta, \theta) }
\nc{\Nmn}{{\sqrt{N} \over mn}}

\nc{\Xis}{X_i(s)}
\nc{\Yjt}{Y_j(t)}

\nc{\bthehat}{\hat{\bthe}}

\nc{\Ritil}{\tilde{R}_i}

\nc{\Ybar}{\overline{Y}}
\nc{\Rbar}{\overline{R}}
\nc{\Nbar}{\overline{N}}
\nc{\intzeroinf}{\int_0^\infty}

\nc{\Fhat}{\hat{F}}
\nc{\Ghat}{\hat{G}}

\nc{\FhatS}{\hat{F}(S_{ik})}
\nc{\GhatT}{\hat{G}(T_{jl})}

\nc{\Fhatik}{\hat{F}_{ik}}
\nc{\Ghatjl}{\hat{G}_{jl}}
\nc{\Fik}{F_{ik}}
\nc{\Gjl}{G_{jl}}
\nc{\phiijkl}{\phi_{ik,jl}(\Delta)}
\nc{\phiijkltil}{\tilde{\phi}_{ik,jl}(\Delta_0,\theta_0)}
\nc{\ord}{N^{-3/2}}           
\nc{\sumijkl}{\sum_{ijkl}}

\nc{\Citil}{\tilde{C}_i}
\nc{\Crtil}{\tilde{C}_r}
\nc{\Djtil}{\tilde{D}_j}
\nc{\Ditil}{\tilde{D}_i}

\nc{\Cithe}{\tilde{C}^{\theta}_i}
\nc{\Djthe}{\tilde{D}^{\theta}_j}

\nc{\Sikthe}{S_{ik}^{\theta}}
\nc{\Tjlthe}{T_{jl}^{\theta}}

\nc{\Zi}{ \bZ_{-i}}
\nc{\zic}{ \lbc z(\bs_j) \: : \: i \neq j \rbc }
\nc{\zkap}{ \bz_{\kappa} }
\nc{\sumi}{ \sum_i }
\nc{\sumj}{ \sum_j }
\nc{\sumij}{ \sum_{i < j} }
\nc{\sumiandj}{ \sum_{i, j} }
\nc{\zsi}{ z(\bs_i) }
\nc{\Zsi}{ Z(\bs_i) }
\nc{\zsj}{ z(\bs_j) }
\nc{\zsn}{ z(\bs_n) }
\nc{\zsone}{ z(\bs_1) }
\nc{\pZ}{ \Pr \lbc \bZ \rbc }
\nc{\qz}{ Q( \bz ) }
\nc{\qZ}{ Q( \bZ ) }

\nc{\thetaYD}{\theta_{Y\mid D}}
\nc{\thetaD}{\theta_D}
\nc{\psiDY}{\psi_{D\mid Y}}
\nc{\psiY}{\psi_Y}

\nc{\tn}{\Theta^{\nu}}
\nc{\Etn}{E_{\theta^{\nu}}}
\nc{\tnone}{\Theta^{\nu+1}}
\nc{\Lm}{L_{\text{m}}}
\nc{\Lo}{L_{\text{o}}}
\nc{\Ym}{Y_{\text{m}}}
\nc{\Yo}{Y_{\text{o}}}
\nc{\ym}{y_{\text{m}}}
\nc{\yo}{y_{\text{o}}}

\nc{\vijb}{v_{ij} - \bX_{i(j)}  \bbet}
\nc{\vikb}{v_{ik} - \bX_{i(k)}  \bbet}
\nc{\vilb}{v_{il} - \bX_{i(l)}  \bbet}
\nc{\betart}{ \bbet^{(r)}_{t_i} }
\nc{\betarj}{ \bbet^{(r)}_j }
\nc{\yij}{y_{ij}}
\nc{\Xmisi}{ {\bX_{ i{\rm (mis)} }} }
\nc{\Xobsi}{ {\bX_{ i{\rm (obs)} }} }
\nc{\Zobsi}{ {\bZ_{ i{\rm (obs)} }} }
\nc{\bSigobs}{ \bSig_{  {\rm obs} } }
\nc{\bSigmis}{ \bSig_{  {\rm mis} } }
\nc{\bSigmo}{ \bSig_{  {\rm mis,obs} } }
\nc{\bSigom}{ \bSig_{  {\rm obs,mis} } }
\nc{\Xil}{{\bX}_{il}}
\nc{\Zil}{{\bZ}_{il} }
\nc{\omilr}{\omega_{il}^{(r)}}
\nc{\delio}{\bdel_i^{{\rm obs}} }

\nc{\yio}{ {y_i^{\rm o} }}
\nc{\Yio}{ {Y_i^{\rm o }} }
\nc{\Yim}{ {Y_i^{\rm m} }}
\nc{\yim}{ {y_i^{\rm m} }}
\nc{\Yc}{Y^{\rm c}}
\nc{\Yic}{Y_i^{\rm c}}
\nc{\yc}{y^{\rm c}}
\nc{\yic}{y_i^{\rm c}}

\nc{\yi}{y_i}
\nc{\Yi}{Y_i}

\nc{\fyic}{f ( \yic ; \; \psiY )}
\nc{\fyi}{f ( y_i ; \; \psiY ) }
\nc{\fdigivenyic}{f ( d_i  \mid  \yic ; \; \psiDY )}
\nc{\fditilgivenyic}{f ( \tilde{d}_i  \mid  \yic ; \; \psiDY )}
\nc{\fditilgivenyi}{f ( \tilde{d}_i  \mid  \yi ; \; \psiDY )}
\nc{\Fditilgivenyic}{F ( \tilde{d}_i  \mid  \yic ; \; \psiDY )}
\nc{\Fditilgivenyi}{F ( \tilde{d}_i  \mid  \yi ; \; \psiDY )}
\nc{\fdigivenyi}{f (d_i \mid y_i ; \;  \psiDY  )}
\nc{\fyicdi}{f \left( \yic, d_i \right)}
\nc{\fyidi}{f \left( \yi, d_i \right)}

\nc{\fymidr}{f_{Y \mid R}}
\nc{\fyr}{f_{Y,R}}
\nc{\frmidy}{f_{R \mid Y}}
\nc{\fy}{f_Y}
\nc{\fr}{f_R}

\nc{\fyicgivendi}{f (\yic \mid d_i; \; \thetaYD )}
\nc{\fyigivendi}{f (\yi \mid d_i; \; \thetaYD )}
\nc{\fyicgivens}{f (\yic \mid s; \; \thetaYD )}
\nc{\fyigivens}{f (\yi \mid s; \; \thetaYD )}
\nc{\fdi}{f ( d_i; \; \thetaD )}

\nc{\fyicX}{f ( \yic \mid X_i; \; \psiY )}
\nc{\fyiX}{f ( y_i \mid X_i; \; \psiY ) }
\nc{\fdigivenyicX}{f ( d_i  \mid  \yic, X_i ; \; \psiDY )}
\nc{\fdigivenyiX}{f (d_i \mid y_i, X_i ; \;  \psiDY  )}
\nc{\fyicdiX}{f \left( \yic, d_i \mid X_i \right)}

\nc{\fyicgivendiX}{f (\yic \mid d_i, X_i; \; \thetaYD )}
\nc{\fyigivendiX}{f (y_i \mid d_i, X_i; \; \thetaYD )}
\nc{\fdiX}{f ( d_i \mid X_i; \; \thetaD )}

\nc{\Yistar}{\bY_i^*}

\nc{\Dio}{D_i^{\rm obs}}
\nc{\bdelio}{\bdel_{ i \, {\rm (obs)}} }

\nc{\fygivend}{f_{Y \mid \delta}}
\nc{\fyd}{f_{Y, \delta}}
\nc{\fd}{f_\delta}
\nc{\FD}{F_D}
\nc{\fygivenbd}{f_{Y\mid b, \delta}}

\nc{\alphahat}{\hat{\bal}}
\nc{\phihat}{\hat{\bphi}}
\nc{\thetahat}{\hat{\bthe}}
\nc{\thetatilde}{\tilde{\bthe}}
\nc{\scoretheta}{\bS(\bthe; \, \scc)}
\nc{\hesstheta}{\bH(\bthe; \, \scc)}
\nc{\infotheta}{\sci(\bthe; \, \scc)}
\nc{\sitheta}{\bs_i(\bthe; \, \scc_i)}
\nc{\sithetahat}{\bs_i(\thetahat; \, \scc_i)}

\nc{\loglikobs}{\ell_{{\rm o}}(\bthe; \, \sco)}
\nc{\scoreobs}{\bS_{{\rm o}}(\bthe; \, \sco)}
\nc{\hessobs}{\bH_{{\rm o}}(\bthe; \, \sco)}
\nc{\infoobs}{\scj_{{\rm o}}(\bthe; \, \sco)}

\nc{\Cil}{\scc_{il}}
\nc{\olog}{\lambda^*(\bthe, \Xobs)}
\nc{\LthetaC}{\scl(\bthe; \, \scc)}
\nc{\LthetaCi}{\scl_i(\bthe; \, \scc_i)}
\nc{\LthetaCil}{\scl_i (\bthe; \, \scc_{il}) }
\nc{\lthetaC}{\ell(\bthe; \, \scc)}
\nc{\lthetaCi}{\ell_i(\bthe; \, \scc_i)}
\nc{\lthetaCil}{\ell_i (\bthe; \, \scc_{il}) }
\nc{\Qtheta}{\scq \left( \bthe \, \left| \,  \bthe^{(r)} \right. \right)}
\nc{\thetar}{\bthe^{(r)}}
\nc{\thetas}{\bthe^{(s)}}
\nc{\alphas}{\bal^{(s)}}
\nc{\psis}{\psi^{(s)}}
\nc{\alphasplusone}{\bal^{(s+1)}}
\nc{\psisplusone}{\bpsi^{(s+1)}}
\nc{\alphapsis}{\left( \alphas, \psis \right)}

\nc{\thetarplusone}{\bthe^{(r+1)}}
\nc{\ologi}{\lambda^*_i(\bthe, \Xobs)}
\nc{\llogi}{\lambda_i \left( \bthe, \tilde{\Xcom}_{il} \right) }

\nc{\scxil}{\tilde{\Xcom}_{il}}
\nc{\siginv}{\bSig_i^{-1}}

\nc{\fofym}{ f \left( \by_i \mid \bbet_m, \bSig \right) }
\nc{\mphim}{ \phi_M \lsq \bSig^{-1/2}(\by_i - \bX_i \bbet_m) \rsq }
\nc{\mphit}{ \phi_M \lsq \bSig^{-1/2}(\by_i - \bX_i \bbet_{t_i}) \rsq }
\nc{\mphij}{ \phi_M \lsq \bSig^{-1/2}(\by_i - \bX_i \bbet_j) \rsq }
\nc{\mphik}{ \phi_M \lsq \bSig^{-1/2}(\by_i - \bX_i \bbet_k) \rsq }
\nc{\expkerm}{ \exp  \lbc -\half \bu_i(\bbet_m)' \bSig^{-1} \bu_i(\bbet_m)
  \rbc }
\nc{\expkerk}{ \exp  \lbc -\half \bu_i(\bbet_k)' \bSig^{-1} \bu_i(\bbet_k)
  \rbc }
\nc{\expkerj}{ \exp  \lbc -\half \bu_i(\bbet_j)' \bSig^{-1} \bu_i(\bbet_j)
  \rbc }
\nc{\normscorem}{\left( \bX_i' \bSig^{-1} \bX_i \bbet_m - \bX_i' \bSig^{-1}
  \by_i \right) }
\nc{\normscorej}{\left( \bX_i' \bSig^{-1} \bX_i \bbet_j - \bX_i' \bSig^{-1}
  \by_i \right) }
\nc{\piti}{ \pi \left( t_i, \bal, \bZ_i\bgam \right) }
\nc{\omij}{ \om_{ij} \left( t_i, \bal, \bZ_i\bgam \right) }
\nc{\phibetak}{ \phi_M(\bbet_k) }
\nc{\phibetaj}{ \phi_M(\bbet_j) }
\nc{\dphidbetak}{ \left. \dpl \phibetak \right/ \dpl \bbet_k }
\nc{\dphidbetakf}{ \frac{ \dpl \phibetak }{ \dpl \bbet_k } }
\nc{\uik}{\bu_i \left( \bbet_k  \right)}
\nc{\mset}{ \{ 0, 1, \ldots, M \} }
\nc{\betasigma}{ \left( \lbc \bbet^{(r)}_t \rbc, \bSig^{(r)} \right) }
\nc{\Thetar}{ \bThe^{(r)} }

\nc{\shatkm}{\hat{S}_{\rm KM}}

\nc{\ds}{\displaystyle}

\nc{\beq}{\begin{eqnarray*}}
\nc{\eeq}{\end{eqnarray*}}

\nc{\beqna}{\begin{eqnarray}}
\nc{\eeqna}{\end{eqnarray}}

\nc{\bct}{\begin{center}}
\nc{\ect}{\end{center}}

\nc{\bds}{\begin{description}}
\nc{\eds}{\end{description}}

\nc{\bit}{\begin{itemize}}
\nc{\eit}{\end{itemize}}

\nc{\bnu}{\begin{enumerate}}
\nc{\enu}{\end{enumerate}}

\nc{\bgt}{\begin{table}}
\nc{\bgtb}{\begin{center} \begin{tabular}}
\nc{\entb}{\end{tabular} \end{center} }
\nc{\ent}{\end{table}}

\nc{\ts}{\textstyle}

\nc{\bgl}{\begin{letter}}
\nc{\op}{\opening}
\nc{\incl}{\input{sendout.ltr} \closing{Best Regards,} \end{letter} }

\nc{\trd}[1]{\textcolor[rgb]{1,0.00,0.00}{#1}}
\nc{\tp}[1]{\textcolor[rgb]{1,0.00,1}{#1}}

\title{Inference procedures in sequential trial emulation with survival outcomes: comparing confidence intervals based on the sandwich variance estimator, bootstrap and jackknife}

\author{ \href{https://orcid.org/0000-0002-1140-2350}{\includegraphics[scale=0.06]{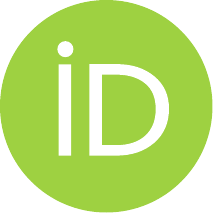}\hspace{1mm}Juliette M. Limozin}\\
	MRC Biostatistics Unit\\
	University of Cambridge\\
	\texttt{juliette.limozin@mrc-bsu.cam.ac.uk} \\
	\And
 \href{https://orcid.org/0000-0003-3726-5937}{\includegraphics[scale=0.06]{orcid.pdf}\hspace{1mm}Shaun R. Seaman} \\
	MRC Biostatistics Unit\\
	University of Cambridge\\
	\texttt{shaun.seaman@mrc-bsu.cam.ac.uk} \\
 \And
	\href{https://orcid.org/0000-0003-0919-3462}{\includegraphics[scale=0.06]{orcid.pdf}\hspace{1mm}Li Su} \\
	MRC Biostatistics Unit\\
	University of Cambridge\\
	\texttt{li.su@mrc-bsu.cam.ac.uk} \\
}

\renewcommand{\shorttitle}{Inference procedures in sequential trial emulation with survival outcomes}

\begin{document}
\maketitle
\begin{abstract}
\small{
Sequential trial emulation (STE) is an approach to estimating causal treatment effects by emulating a sequence of target trials from observational data. In STE, inverse probability weighting is commonly utilised to address time-varying confounding and/or dependent censoring. Then structural models for potential outcomes are applied to the weighted data to estimate treatment effects. For inference, the simple sandwich variance estimator is popular but conservative, while nonparametric bootstrap is computationally expensive, and a more efficient alternative, linearised estimating function (LEF) bootstrap, has not been adapted to STE. We evaluated the performance of various methods for constructing confidence intervals (CIs) of marginal risk differences in STE with survival outcomes by comparing the coverage of CIs based on nonparametric/LEF bootstrap, jackknife, and the sandwich variance estimator through simulations. LEF bootstrap CIs demonstrated the best coverage with small/moderate sample sizes, low event rates and low treatment prevalence, which were the motivating scenarios for STE. They were less affected by treatment group imbalance and faster to compute than nonparametric bootstrap CIs. With large sample sizes and medium/high event rates, the sandwich-variance-estimator-based CIs had the best coverage and were the fastest to compute. These findings offer guidance in constructing CIs in causal survival analysis using STE.
}
\end{abstract}


\section{Introduction} \label{intro}

\subsection{Target trial emulation with survival outcomes}
 Target Trial Emulation (TTE) has become a popular approach for causal inference using observational longitudinal data \cite{hernan_using_2016,hansford_reporting_2023}. The goal of TTE is to estimate and make inferences about causal treatment effects that are comparable to those that would be obtained from a target Randomised Controlled Trial (RCT) \cite{hernan_using_2016,matthews_comparing_2021}. TTE can be helpful when it is not possible to conduct this RCT because of time, budget and ethical constraints. Hernán and Robins (2016) have proposed a formal framework for TTE \cite{hernan_using_2016}, which highlights the need to specify the target trial's protocol, i.e., the protocol of the RCT that would have ideally been conducted, in order to guide the design and analysis of the emulated trial using data extracted from observational databases such as disease registries or electronic health records. The protocol should include the eligibility criteria, treatment strategies, assignment procedures, outcome(s) of interest, follow-up periods, causal contrast of interest and an analysis plan; see some step-by-step guides to TTE in \cite{hernan_target_2022}, \cite{matthews_target_2022}, \cite{maringe_reflection_2020}.

In TTE there are various sources of bias that must be addressed. Firstly, unlike in an RCT, non-random assignment of treatment at baseline must be accounted for when estimating the causal effect (intention-to-treat or per-protocol) of a treatment in an emulated trial. 
Secondly, similar to an RCT, it is necessary to account for censoring caused by loss to follow-up in an emulated trial. Thirdly, 
when the per-protocol effect is the causal effect of interest, it is also necessary to handle non-adherence to assigned treatments. To address these issues in TTE with survival outcomes,
a useful approach is to fit a marginal structural Cox model (MSCM) using inverse probability weighting (IPW) \cite{robins_marginal_2000,robins_correcting_2000, hernan_marginal_2000, hernan_marginal_2001, cain_inverse_2009, seaman_review_2013,clare_causal_2019}, after first artificially censoring the patients' follow-up at the time of treatment non-adherence \cite{hernan_observational_2008, daniel_methods_2013,murray_causal_2021}.
Baseline confounders are included in this MSCM as covariates to adjust for the non-random treatment assignment at baseline using regression. 
The inverse probability weights are the product of two sets of time-varying weights: one to address selection bias from censoring due to loss to follow-up, and one to address selection bias from the artificial censoring due to treatment non-adherence.
Counterfactual hazard ratios can be estimated from the fitted MSCM using weighted data. A modification of this approach is to discretise the survival time and replace the MSCM with a pooled logistic regression model \cite{
hernan_marginal_2000,hernan_marginal_2001}.
Provided that the probability of failure between the discrete times is small, this pooled logistic model well approximates the MSCM \cite{dagostino_relation_1990}.

The counterfactual hazard ratio has been criticised as lacking a causal interpretation, 
and it has been proposed that other estimands be used instead, e.g., the marginal risk difference (MRD) \cite{hernan_hazards_2010, buchanan_worth_2014, keogh_causal_2023}.
The MRD over time can be estimated by first using the counterfactual hazard ratio estimates from a marginal structural model (MSM) 
together with an estimate of the baseline hazard to predict the survival probabilities of the patients in the emulated trial under two scenarios: when all are treated and when none are treated.
For each scenario, the predicted survival probabilities are averaged over all enrolled patients.
The estimate of the MRD is then calculated as the difference between these two averages \cite{murray_causal_2021, keogh_causal_2023}.

\subsection{Constructing confidence intervals in sequential trial emulation}

A potential problem when emulating a trial is that the number of treated and/or untreated patients eligible for inclusion in a trial that begins at any \textit{given} time may be small.
This can be addressed by sequential trial emulation (STE) \cite{hernan_observational_2008,gran_sequential_2010}, which takes advantage of the fact that patients may meet the eligibility criteria for the target trial multiple times during their follow-up in an observational database.
In STE, a sequence of target trials is emulated, each starting at a different time. The data from these sequential trials are pooled and analysed to produce an overall estimate of the treatment effect \cite{keogh_causal_2023}. This approach was first proposed by Hernán et al. (2008) \cite{hernan_observational_2008} and Gran et al. (2010) \cite{gran_sequential_2010} as a simple way to improve the efficiency of treatment effect estimation relative to emulating a \textit{single} trial.
There have been several applications of STE; see Keogh et al. (2023) \cite{keogh_causal_2023} for a list.

Despite the increasing popularity of TTE, there is a lack of research on different methods for constructing confidence intervals (CIs) of treatment effects in STE. 
The sandwich variance estimator, bootstrap or jackknife can be used to obtain a variance estimate of the parameter estimates in an MSM by accounting for correlations induced by the same patient being eligible for multiple
trials.
In causal survival analysis with IPW to adjust for baseline confounding of point treatments \cite{austin_variance_2016,shu_variance_2021}, the sandwich variance estimator of Lin and Wei (1989) \cite{lin_robust_1989} is frequently used. However, this estimator does not account for the uncertainty due to weight estimation, and can consequently overestimate the true variance \cite{austin_variance_2016,shu_variance_2021}. 
More complex sandwich variance estimators that account for this uncertainty are available. 
Shu et al. (2021) \cite{shu_variance_2021} proposed a variance estimator for the hazard ratio of a point treatment in an MSCM with IPW used for baseline confounding only. 
 Enders et al. (2018) \cite{enders_robust_2018} developed a sandwich variance estimator for the hazard ratios in an MSCM when IPW is used to deal with treatment switching and censoring due to loss to follow-up. They found no substantial differences in the performance of the simple sandwich variance estimator and their sandwich variance estimator, and the latter performed comparatively poorly in scenarios with small sample sizes and many confounders. No off-the-shelf software has implemented the variance estimator by Enders et al. (2018) \cite{enders_robust_2018}.

Bootstrap has been recommended as an alternative to the sandwich variance estimator because 
it accounts for uncertainty in weight estimation. In the simple setting where IPW is used to estimate the effect of a point treatment on a survival outcome, Austin (2016) found that bootstrap CIs performed better than the sandwich-variance-estimator-based CIs when the sample size was moderate (1000) \cite{austin_variance_2016}. In the setting with continuous and binary outcomes, 
Austin (2022) found that when sample sizes were small (250 or 500) to moderate (1000), bootstrap resulted in more accurate estimates of standard errors than sandwich variance estimators. However, bootstrap CIs did not achieve nominal coverage when the sample sizes were small to moderate and the treatment prevalence was either very low or very high \cite{austin_bootstrap_2022}. 
Mao et al. (2018) observed similar results when constructing CIs of hazard ratios for a binary point treatment using IPW: with small (500) and moderate (1000) sample sizes and strong associations between confounders and treatment assignment, both the sandwich variance estimator and bootstrap resulted in under-coverage of CIs \cite{mao_propensity_2018}. For the longitudinal setting with binary time-varying treatments, Seaman and Keogh (2023) found that with moderate (1000) sample sizes, bootstrap and the sandwich variance estimator both led to slightly under-coverage of CIs for hazard ratios in an MSCM but with no notable difference between the two methods \cite{seaman_simulating_2023}. With small (250, 500) sample sizes, the coverage of the CIs deteriorated, but bootstrap CIs had coverage closer to the nominal level than the sandwich-estimator-based CIs \cite{seaman_simulating_2023}.

Jackknife resampling 
has been used in TTE to construct CIs of hazard ratio and risk difference (see \cite{serdarevic_emergency_2023} and \cite{virtanen_antidepressant_2024} for recent examples). Gran et al. (2010) also used jackknife to construct the CI of the hazard ratio of a binary treatment in the STE setting because in their analysis bootstrap led to non-convergence problems due to the large number of covariates used \cite{gran_sequential_2010}. Jackknife could be advantageous when the sample size is small because it is computationally faster than bootstrap and it is less likely to lead to non-convergence problems since only one patient's data are left out in each jackknife sample.

 The works mentioned above all focused on variance estimation and CIs for counterfactual hazard ratios, which are often chosen as the estimand in the literature of TTE with survival outcomes \cite{hernan_observational_2008, danaei_observational_2013}. While these works were not researched specifically for TTE with survival outcomes, they could be easily applied to such a setting. However, in the more complex setting of STE, less attention appears to be paid to the development and evaluation of CI construction methods. There is a lack of research on comparing the nonparametric bootstrap and the sandwich variance estimator for constructing CIs of the MRD in various settings of STE. It is also desirable to develop computationally more efficient CI methods than nonparametric bootstrap. 

\subsection{The contribution of this article}

To fill this gap, we carry out an extensive simulation study to compare different methods for constructing a CI of the MRD in STE with a survival outcome.
The first method uses the sandwich variance estimator that ignores the uncertainty caused by weight estimation.
The second method is nonparametric bootstrap.
This has the drawback of being computationally expensive.
For this reason, the third method that we investigate is the computationally less intensive Linearised Estimating Function (LEF) bootstrap. Hu and Kalbfleisch (2000) \cite{hu_estimating_2000} first developed the estimating function bootstrap approach. In settings with cross-sectional/longitudinal survey data with design weights, Rao and Tausi \cite{rao_estimating_2004} and Binder et al. \cite{binder_design-based_2004} proposed the LEF bootstrap to improve computational efficiency and to avoid ill-conditioned matrices when fitting logistic models to bootstrap samples.
We develop two forms of the LEF bootstrap for the STE setting.
The fourth method is jackknife resampling.
In our simulation study,
we consider scenarios with varying sample sizes, treatment prevalence, outcome event rates, and strength of time-varying confounding.
Our results provide some guidance to practitioners on which methods could perform better in different settings. 

The article is organised as follows. In Section \ref{motivating example} we introduce the HIV Epidemiology Research Study (HERS) data as a motivating example and describe a protocol of STE based on the HERS data. Section \ref{estimation} describes the notation, causal estimand, causal assumptions and MRD estimation procedure in STE. In Section \ref{CI methods}, we describe the CI construction methods that we compare in this article, including our proposed LEF bootstrap CIs.
Section \ref{simulation} presents our simulation study.
 In Section \ref{application}, we apply STE to the HERS data.
We conclude in Section \ref{discussion} with a discussion.

\section{HIV Epidemiology Research Study: a motivating example} \label{motivating example}

The HIV Epidemiology Research Study (HERS) included 1310 women with, or at high risk of, HIV infection at four sites (Baltimore, Detroit, New York, Providence) enrolled between 1993-1995 and followed up to 2000 \cite{ko_estimating_2003}. 
 The HERS had 12 approximately six-monthly scheduled visits, where clinical, behavioural, and sociological outcomes and (self-reported) treatment were recorded.

Following Ko et al. (2003) \cite{ko_estimating_2003} and Yiu and Su (2022) \cite{yiu_joint_2022}, we aim to estimate the causal effect of (self-reported) Highly Active AntiRetroviral Treatment (HAART) on all-cause mortality among HIV-infected patients in the HERS cohort. Clinical and demographic variables related to treatment assignment and disease progression were available, including CD4 cell count, HIV viral load, self-reported HIV symptoms, race, and the site in which a patient was enrolled. Following Yiu and Su (2022) \cite{yiu_joint_2022}, we treat visit 8 in 1996 as the baseline of the observational cohort, as HAART was more widely used and recorded in the HERS by then. There were 584 women assessed at visit 8.
Time of death during follow-up was recorded exactly and there were 24 deaths in total.
Some patients were also lost to follow-up, with 179 patients assessed at the last follow-up visit 12. 
Yiu and Su (2022) conducted their analysis with standard MSMs with IPW by defining the time-varying treatment as ordinal with 3 levels: ``no treatment'', ``antiretroviral therapy other than HAART", and ``HAART'' \cite{yiu_joint_2022}. In our analysis we consider a binary treatment: ``HAART'' versus ``no HAART''. 
A hypothetical RCT (the target trial) to estimate the per-protocol effect of HAART (vs. other or no treatment) on all-cause mortality could be emulated using the HERS data. The target trial protocol can be found in Table 16 of the Supplementary Materials.

 As mentioned in Section~\ref{intro}, a practical problem for TTE is that if we only emulate a single trial, the number of patients who initiate (i.e. start to receive) the treatment at baseline and the number of outcome events among them could be small. In the HERS, only 76 patients initiated the HAART when baseline is defined as visit 8.
By emulating a \textit{sequence} of target trials and combining their analyses, more efficient estimates of treatment effects can be obtained.
For example, an additional 62 women initiated HAART at visit 9 of the HERS, and so would be in the treatment arm of an emulated trial with baseline defined as visit 9. 

In Section \ref{application} we emulate 5 sequential trials from the HERS data labelled from 0 to 4 with sequential enrollment periods so that the trials start at visits 8, 9, 10, 11 and 12, respectively. The trial protocol, and more specifically the eligibility criteria, remain the same across all 5 trials, which in our example means that patients must have no prior use of HAART before the baseline of the trial. The study horizons differ: trial 0 has 4 follow-up assessments at visits 9 to 12; trial 1 has 3 follow-up assessments at visits 10 to 12; and so on. 
Trial 4 only has a baseline assessment at visit 12 and no further follow-up.
This approach means that we can use data from patients who started receiving the HAART later in the HERS cohort. Table \ref{HERS data summary} presents tabulation of the HERS data prepared for STE, where we note that the total number of patients in the treatment arm is increased from 76 (in a single trial with baseline at visit 8) to 234 by using the STE approach. 
A patient can be eligible for multiple trials. For example, a patient who had not been receiving HAART at visits 8 and 9 but started to receive HAART from visit 10 will be eligible as a member of the control arm in trials 0 and 1, and as a member of the treatment arm in trial 2. Moreover, this patient's follow-up in trials 0 and 1 will be artificially censored at visit 10. 
Figure 1 of the Supplementary Materials provides a schematic illustration of the STE approach.

\begin{table}[htbp]
 \centering
 \caption{Data tabulation of the HERS data prepared for a per-protocol analysis in sequential trial emulation. The numbers in a column represent the number of patients enrolled in an emulated sequential trial by their assigned treatment strategies, outcome and censoring status by the end of the emulated trial. Treatment: assigned treatment strategies; 0, never treated with HAART; 1, always treated with HAART. Outcome: indicator of all-cause mortality. Censoring: indicator of censoring due to loss to follow-up.}
 \begin{tabular}{*{3}{l}*{5}{r}}
 \toprule
 & & \multicolumn{1}{r}{\: \: \: \:\:Trial} & \multicolumn{1}{c}{0} & \multicolumn{1}{c}{1} & \multicolumn{1}{c}{2} & \multicolumn{1}{c}{3} & \multicolumn{1}{c}{4} \\
 Treatment & Outcome & Censoring & &&&&\\
 \midrule
 0 & 0 & 0 & 390 & 315 & 250 & 175 & 155 \\
 & & 1 & 14 & 16 & 15 & 49 & 0 \\ \addlinespace[3pt]
 & 1 & 0 & 11 & 8 & 5 & 5 & 4 \\ \addlinespace[6pt]
 1 & 0 & 0 & 73 & 54 & 49 & 25 & 19 \\
 & & 1 & 1 & 5 & 0 & 1 & 0 \\ \addlinespace[3pt]
 & 1 & 0 & 2 & 3 & 1 & 0 & 1 \\
 \bottomrule
 &&Total eligible in trial & 491 & 401& 320 &255 & 179 
 \end{tabular}
 
 \label{HERS data summary}
\end{table}

\section{Estimation of the per-protocol effect in sequentially emulated trials} \label{estimation}

\subsection{Setting and notation}

Consider an observational study in which $n$ patients are followed up from time $t_0$ until the earliest of the event of interest, loss to follow-up, and the end of the study.
For each patient, time-independent variables are measured at time $t_0$, and time-varying variables are measured at regular times $t_0 < t_1 < \ldots < 
t_{n_v-1}$ during follow-up, where $t_{n_v}$ denotes the time of the end of the study ($ t_{{n_v}-1} <t_{n_v} $) and $n_v$ is therefore the maximum number of study visits before $t_{n_v}$. 
Data on each patient are assumed to be independent and identically distributed.
Data from this study will be used to create data for a sequence of ${n_v}$ trials.
The $m^{\text{th}}$ sequential trial (i.e., trial $m$, $m=0, \ldots, {n_v}-1$) begins at time $t_m$, includes patients who are eligible for enrolment at this time, and ends at time $t_{n_v}$.
Hence, within this trial, the time-varying variables are measured at times $t_m, \ldots, t_{{n_v}-1}$.
We shall refer to these ${n_v}-m$ measurement times as the trial visits for trial $m$.
For example, trial visit 0 in trial $m$ takes place at time $t_m$ and trial visit ${n_v}-m-1$ takes place at time $t_{{n_v}-1}$.
For each of the $n$ patients in the observational study, we define the following variables.

\begin{itemize} 
\item $E_m$ is an indicator of whether the patient is eligible ($E_m = 1$) or not ($E_m = 0$) for trial $m$.
\item $Y_{m,k} = 1$ ($k=0, 1, \ldots$) if the patient experiences the event of interest in time interval $[t_{m}, t_{m+k+1})$, and $Y_{m,k} = 0$ otherwise.
\item $V$ denotes the patient's vector of time-independent covariates measured at time $t_0$.
\item $A_{m,k}$ and $L_{m,k}$ denote, respectively, the patient's treatment and time-varying covariates measured at time $t_{m+k}$.
\item $C_{m,k}$ is an indicator that the patient is censored due to loss to follow-up in the interval $[t_{m+k}, t_{m+k+1})$.
So, if $C_{m,k} = 1$ then $Y_{m,k+1}, Y_{m,k+2},..., Y_{m+1,k}, Y_{m+1,k+1},...$ are not observed.
\end{itemize}

$Y_{m,k}$, $A_{m,k}$, $L_{m,k}$ and $C_{m,k}$ will serve as, respectively, the outcome indicator, the binary treatment indicator, the time-varying covariates and the censoring indicator at trial visit $k$ ($k=0, 1, \ldots$) in trial $m$ ($m=0, \ldots, {n_v}-1$).
Also, we shall use the overbar
to denote variable history, for example, $\overline{A}_{m,k} = (A_{m,0}, \ldots, A_{m,k})$ denotes the patient's history of treatment up to trial visit $k$ in trial $m$, and define $\overline{0} = (0, \dots, 0)$ and $\overline{1} = (1, \ldots, 1)$.
We assume the temporal ordering $(L_{m,k},A_{m,k},Y_{m,k},C_{m,k})$ within $[t_{m+k}, t_{m+k+1})$, $\forall m,k$.

For a patient enrolled in trial $m$ (i.e., with $E_m=1$) and for ${\overline a_k}$ equal to either $\overline{0}$ or $\overline{1}$, we define the potential variable $Y_{m,k}^{\overline a_k}$ to be a binary indicator that the patient would have experienced the event of interest during the time interval $[t_m, t_{m+k+1})$ if he/she/they had, possibly contrary to fact, received treatment $a \in \{0,1\} $ since the baseline of trial $m$, i.e., from time $t_m$ up to time $t_{m+k}$.
Analogously, $L_{m,k}^{\overline a_k}$ denotes the potential time-varying covariates the patient would have if he/she/they had received this treatment since $t_m$ up to time $t_{m+k}$. Note that $Y_{m,k}^{\overline a_k}$ and $L_{m,k}^{\overline a_k}$ are not defined for patients ineligible for trial $m$, i.e.\ patients with $E_m=0$.
We shall omit the explicit conditioning on $E_m=1$ when describing the causal estimand in the next section. 

\subsection{Causal estimand and assumptions} \label{causal estimand and assumptions}
We define the per-protocol effect in trial $m$ in terms of the MRD. 
The MRD at trial visit $k$ in trial $m$ is the difference between the marginal cumulative incidence at time $t_{m+k}$ if all patients in the population eligible for trial $m$ were always treated by time $t_{m+k}$ and the marginal cumulative incidence if they were not treated at all by time $t_{m+k}$ \cite{keogh_causal_2023}.
That is,
\begin{equation}\label{mrdformula}
 \mbox{MRD}_m(k) = \Pr(Y_{m,k}^{\overline a_k= \overline 1} = 1) - \Pr(Y_{m,k}^{\overline a_k= \overline 0} = 1)
\end{equation}

Identification of \eqref{mrdformula} requires standard causal assumptions of no interference, consistency, positivity, no unmeasured confounding of treatment assignment at the trial baseline, and positivity and sequential ignorability of treatment adherence \cite{keogh_causal_2023}. Positivity and sequential ignorability of censoring must also hold. Details of these assumptions are provided in Section 2 of the Supplementary Materials. 
Equation~\eqref{mrdformula} can be written equivalently as
 \begin{eqnarray}
 \mbox{MRD}_m(k)
 & = &
 \Pr(Y_{m,k}^{\overline a_k= \overline 0} = 0) - \Pr(Y_{m,k}^{\overline a_k= \overline 1} = 0)
 \nonumber \\
 & = &
 {\rm E}_{ V, L_{m,0}}\left\{ \Pr(Y_{m,k}^{\overline a_k= \overline 0} = 0 \mid V, L_{m,0}) - \Pr(Y_{m,k}^{\overline a_k= \overline 1} = 0 \mid V, L_{m,0}) \right\} 
 \label{mrd formula counterfactuals}
 \end{eqnarray}

 
The counterfactual survival probabilities in \eqref{mrd formula counterfactuals} can be written in terms of counterfactual discrete-time hazards as follows:
\begin{align} 
 \Pr(Y^{\overline a_k= \bar{a}}_{m,k} = 0) & \nonumber \\
 = {\rm E}_{ V, L_{m,0}} & \left[\left\{1-\mbox{Pr}\left(Y_{m,0}^{a_0=a} =1 \mid V, L_{m,0}\right)\right\} \prod_{j=1}^k \left\{1-\mbox{Pr}\left(Y_{m,j}^{\overline a_j= \bar a} =1 \mid Y_{m,j-1}^{\overline a_{j-1}= \bar a}=0, V, L_{m,0}\right) \right\}\right], \nonumber \\
 ~~~~~~& ~~~~~~~~~~~~~~~~~~~~~~~~~~~~~~~~~~~~~~~~~~~~~~~~~~~~~~~~~~~~~~~~~~~~~~~~~~~~~~~~~~~~~~~~a \in \{0,1\} \nonumber
\end{align} 

\subsection{Marginal structural model with inverse probability weighting}\label{perprotocol}

We assume the following MSM in the form of a pooled logistic model with regression parameters $\boldsymbol \beta$: 
\begin{align} \label{outcomemodelmultiple}
 \text{logit} \left\{ \mbox{Pr}\left(Y_{m,k}^{\overline a_k = \overline a} =1 \mid {Y}_{m,k-1}^{\overline a_{k-1} = \overline a}=0, V, L_{m,0}\right) \right\} =\beta_{0}(m) + \beta_{1}(k) + \beta_2\cdot a + \boldsymbol \beta_3\trans V + \boldsymbol \beta_4\trans L_{m,0}, \\ ~~~~~~~~~~~~~~a \in \{0,1\}, \nonumber
\end{align}
where $\beta_{0} (m)$ is a trial-specific intercept and $\beta_{1}(k)$ is the baseline hazard at trial visit $k$. This MSM could be fitted separately in each emulated trial. However, a combined analysis can be more efficient and may be necessary when the number of treated patients for some trials are small. This involves making modelling assumptions about how the MSM parameters vary across trials. For example, Danaei et al. (2013) allowed a trial-specific intercept term $\beta_0(m)$ but assumed that the coefficient for treatment $\beta_2$ was the same in all trials \cite{danaei_observational_2013}, thus allowing for borrowing strength across trials for treatment effect estimation. Parametric forms or splines can be used for $\beta_{0} (m)$ and $\beta_{1}(k)$. In~\eqref{outcomemodelmultiple} it is also assumed that the baseline hazard and the coefficients of $V$ and $L_{m,0}$ do not vary across trials. Interactions between trials, trial follow-up visits, treatment and baseline covariates can be included as well. Keogh et al. (2023) discussed the MSM specification in STE and recommended that formal tests be performed for the inclusion of any covariate interaction \cite{keogh_causal_2023}.

Given the baseline covariates $V$ and $L_{m,0}$, we assume no unmeasured confounding of baseline treatment assignment.
Thus if all patients adhere to the treatment assigned and no censoring occurs, the MSM parameters can be estimated by fitting the pooled logistic model in~\eqref{outcomemodelmultiple}
to the observed data in the emulated trials. In practice, not all eligible patients for trial $m$ would adhere to their assigned treatments. We artificially censor patients' follow-up in trial $m$ at the time at which they cease to adhere to the treatment $A_{m,0}$ received at the baseline of trial $m$ \cite{hernan_observational_2008, gran_sequential_2010}. We use IPW as mentioned in Section \ref{intro} to account for the artificial censoring and censoring due to loss-to-follow-up.

To address artificial censoring due to treatment switching and censoring due to loss to follow-up, we calculate each patient's stabilised inverse probability of treatment weight $sw^A_{m,k}$ and stabilised inverse probability of censoring weight $sw^C_{m,k}$ at trial visit $k$ in trial $m$. The formulae of these weights are provided Section 3 of the Supplementary Materials. Each patient's \textit{stabilised inverse probability of treatment and censoring weight} (SIPTCW) \cite{danaei_observational_2013, daniel_methods_2013,murray_causal_2021, robins_marginal_2000} at trial visit $k$ in trial $m$ is therefore $sw_{m,k}^{AC} = sw_{m,k}^A \times sw_{m,k}^C$.

We follow the method of Danaei et al. (2013) \cite{danaei_observational_2013}, who fitted logistic models to the treatment and censoring data from the original observational study to estimate the conditional probabilities used for calculating SIPTCWs. 
They used observed treatment and censoring data of each patient from the visits that correspond to the baselines of the eligible trials until the trial visits where the patient stopped adhering to the assigned treatments or the last trial visits. If patients were eligible for multiple trials, duplicates of the treatment and censoring data within patients were discarded. 

We pool the observed data from the ${n_v}$ trials and fit the MSM in~\eqref{outcomemodelmultiple} to the pooled, artificially censored and weighted data to obtain a point estimate $\bm{\hat\beta}$ of the MSM parameters $\bm\beta$ in~\eqref{outcomemodelmultiple}.

\subsection{Estimating the causal estimand}\label{mrdest}

The MRD at trial visit $k$ in trial $m$ can be estimated using the parameter estimates $\hat{\boldsymbol \beta}$ from the MSM by the empirical standardization formula:
\begin{eqnarray}
 \label{mrd estimate}
 \widehat{\mbox{MRD}}_m(k)
 & =&
 \frac{1}{n_m} \sum_{i=1}^n E_{m,i} \prod_{j = 0}^{k}\left\{1-\text{logit}^{-1}\left\{\mu(j , m, a = 0 , V_i, L_{m,0,i}; \hat{\boldsymbol \beta})\right\}\right\} \nonumber
 \\
 && -
 \frac{1}{n_m} \sum_{i=1}^n E_{m,i} \prod_{j = 0}^{k}\left\{1-\text{logit}^{-1}\{\mu(j , m, a = 1 , V_i, L_{m,0,i}; \hat{\boldsymbol \beta})\}\right\}, 
\end{eqnarray}
where $i$ indexes the patient in the original observational study, $n_m = \sum_{i=1}^n E_{m,i}$ is the total number of patients enrolled in trial $m$, $\text{logit}^{-1}(\cdot) = \exp(\cdot)/\{1+\exp(\cdot)\}$ and $\mu(j, m, a, L_{m,0},V; $ $\bm{\hat\beta})$ $= \hat \beta_{0}(m) + \hat\beta_{1}(j) + \hat\beta_2\cdot a + \bm{\hat\beta}_3\trans V + \bm{\hat \beta}_4\trans L_{m,0} $ for $j=0, \ldots, k$.
$\widehat{\mbox{MRD}}_m(k)$ is an estimate of the MRD at trial visit $k$ in the population of patients eligible for trial $m$.
Alternatively, we could standardise to a population characterised by a different distribution of $(V, L_{m,0})$.

To summarise, the MRD in \eqref{mrd formula counterfactuals} is estimated by the following steps: 
\begin{enumerate}
 \item Estimate the SIPTCWs using data from the observational study.
 
 \item Expand the observational data by assigning patients to eligible sequential trials and artificially censor patients' trial follow-up when they were no longer adhering to the treatment assigned at the baseline of each sequential trial.
 
 \item Estimate the MSM for the sequential trials using the expanded, artificially censored data with the estimated weights in Step 1.
 
 \item Create two datasets containing patients from a target population with their baseline covariates data, setting their treatment assignment to either treatment arm, and calculate the estimated counterfactual survival probabilities of each patient in each of these two datasets.
 
 \item Estimate the MRD by averaging the survival probabilities in each of the two datasets and taking the difference between these two averages, as is done in equation~\eqref{mrd estimate}.
 \end{enumerate}

In this article, we use the \texttt{R} package \texttt{TrialEmulation} \cite{rezvani_trialemulation_2023} (see Section 4 of Supplementary Materials) to implement 
these steps. 

\section{Constructing confidence intervals of the marginal risk difference} \label{CI methods}

In this section, we describe the methods for constructing CIs of the MRD based on the simple sandwich variance estimator, nonparametric bootstrap, LEF bootstrap and jackknife resampling.

\subsection{Sandwich variance estimator} \label{sandwich CI}
The simple sandwich variance estimator accounts for the inverse probability weights and the correlation induced by patients being eligible to multiple trials \cite{hernan_marginal_2000,danaei_observational_2013}. Specifically, for the pooled logistic regression model in~\eqref{outcomemodelmultiple},
the simple sandwich variance estimator of $\hat{\bm \beta}$ is
$$
 \hat{\bm{\Sigma}} =
 \left\{\sum_{i=1}^n
 \frac{\partial \bU_i(\bm{\beta})}{\partial \bm{\beta}^{\text{T}}} \right\}_{\bm \beta = \hat{\bm \beta}}^{-1}
 \left\{ \sum_{i=1}^n \bU_i(\hat{\bm{\beta}}) \bU_i(\hat{\bm{\beta}})^{\text{T}}\right\}
 \left\{ \frac{\partial \bU_i(\bm{\beta})}{\partial \bm{\beta}^{\text{T}}} \right\}_{\bm \beta = \hat{\bm \beta}}^{-1}
$$
where $\bU_i(\hat{\bm{\beta}})$ is the weighted score function of the pooled logistic model evaluated at $\hat{\bm{\beta}}$ for patient $i$.

We follow the parametric bootstrap algorithm of Mandel (2013) to construct simulation-based CIs of the MRD as follows \cite{mandel_simulation-based_2013}:
\begin{enumerate}
 \item[1)] Obtain the parameter estimate $\hat{\boldsymbol \beta}$ and the sandwich variance estimate $\hat{\bm{\Sigma}}$ of the MSM in~\eqref{outcomemodelmultiple}.
 \item[2)] Draw an i.i.d.\ sample $\boldsymbol \beta^{(1)},...,\boldsymbol \beta^{(S)}$ of size $S$ (say $S=500$) from the multivariate Normal distribution with mean $\hat{\boldsymbol \beta}$ and variance $\hat{\bm{\Sigma}}$.
 \item[3)] For each vector $\boldsymbol \beta^{(s)}$ ($s = 1,...,S$), estimate the MRD at each trial visit by setting $\boldsymbol \beta^{(s)}$ as the MSM parameters.
 \item[4)] Use the $2.5^\text{th}$ and $97.5^\text{th}$ percentiles of these $S$ MRD estimates at each trial visit as the lower and upper bounds of the 95\% CI.
\end{enumerate}

This is currently the only CI method implemented in the \texttt{TrialEmulation} package. 

\subsection{Nonparametric bootstrap with the pivot method}\label{nonpboot}

We use the non-Studentized pivot method \cite{carpenter_bootstrap_2000} to construct CIs based on nonparametric bootstrap.
Specifically, we follow the steps described below:
\begin{enumerate}
 \item[1)] Draw $B$ bootstrap samples from the observational data, treating the $n$ patients as the resampling units.
 \item[2)] For each bootstrap sample $b$ ($b = 1,\ldots,B$), obtain the bootstrap parameter estimate $ \hat {\boldsymbol \beta}^{(b)}$ and estimate the MRD at trial visit $k$ in trial $m$, ${\mbox{MRD}}_m(k)$, using the method in Section~\ref{estimation}.
 \item[3)] Define the lower and upper bounds of the 95\% CI for the MRD at each trial visit $k$ in trial $m$ as, respectively, 
 $
 2\widehat{\mbox{MRD}}_m(k)- \widehat{\mbox{MRD}}_m(k)^*_{(0.975)}
 $
 and
 $
 2\widehat{\mbox{MRD}}_m(k) - \widehat{\mbox{MRD}}_m(k) ^*_{(0.025)}, 
$
where $\widehat{\mbox{MRD}}_m(k)$ is the point estimate of the MRD at trial visit $k$ in trial $m$ estimated from the original dataset, and $\widehat{\mbox{MRD}}_m(k) ^*_{(0.025)}$, $\widehat{\mbox{MRD}}_m(k) ^*_{(0.975)}$ are the $2.5^\text{th}$ and $97.5^\text{th}$ percentiles of the $B$ bootstrap MRD estimates at trial visit $k$ in trial $m$, respectively.
\end{enumerate}

\subsection{Linearised Estimating Function bootstrap}

The main advantages of LEF bootstrap over nonparametric bootstrap are reduced computational time and non-convergence issues \cite{binder_design-based_2004,hu_estimating_2000}. In terms of computational time, unlike the nonparametric bootstrap, which involves fitting a regression model to each bootstrap sample, LEF bootstrap requires fitting this model only once to the original dataset. This can reduce computational time considerably, especially when iterative procedures are used to fit the model.
 In terms of non-convergence issues, Binder et al. (2004) found that when using nonparametric bootstrap for logistic regression it was possible to have several bootstrap samples for which the parameter estimation algorithm would not converge, due to ill-conditioned matrices that were not invertible \cite{binder_design-based_2004}. For both reasons, LEF bootstrap may have advantages in our STE setting, where logistic regression models are used both for estimating the SIPTCWs and for fitting the MSM.
We now explain how LEF bootstrap works in the general situation where the goal is to construct a CI for some function of a parameter vector $\btheta$.
In Section 4.3.1, we shall describe how to apply this general method to the specific setting of STE.

Let $\bU(\btheta)$ denote the estimating function for $\btheta$ (note that $\bU$ here is different from $\bU_i$ in Section~\ref{sandwich CI}).
Let $\bU^{\rm org} (\btheta)$ denote the sum of $\bU(\btheta)$ over the $n$ patients in the original dataset.
Then $\bU^{\rm org} (\btheta) = \mathbf{0}$ is the estimating equation for $\btheta$ based on the original dataset.
Let $\hat{\btheta}$ denote the estimate of $\btheta$ obtained by solving this equation.
Now suppose that $B$ bootstrap samples have been generated by resampling with replacement from this original dataset.
Let $\bU^{(b)} (\btheta) = \mathbf{0}$ denote the estimating equation for $\btheta$ based on the $b^{\text{th}}$ bootstrap sample, and let $\hat{\btheta}^{(b)}$ denote the corresponding estimate of $\btheta$ ($b=1, \ldots, B$).
If we apply Taylor linearisation to the function $\bU^{(b)} (\btheta)$ around $\hat{\btheta}$, we obtain 
\[
\bU^{(b)}({\btheta}) \approx \bU^{(b)}(\hat{\btheta})+\left\{\frac{\partial \bU^{(b)} ({\btheta})}{\partial \btheta} \right\}_{\btheta=\hat{\btheta}}(\btheta-\hat{\btheta}).
\]
From this and the fact that $\bU^{(b)} (\hat{\btheta}^{(b)}) = \mathbf{0}$ by definition, we obtain
\[
\mathbf{0} \approx \bU^{(b)}(\hat{\btheta})+\left\{\frac{\partial \bU^{(b)} ({\btheta})}{\partial \btheta} \right\}_{\btheta=\hat{\btheta}}(\hat{\btheta}^{(b)} - \hat{\btheta}) 
\]
Rearranging the terms, we get
\[
\hat{\btheta}^{(b)} \approx \hat{\btheta}- \left\{\frac{\partial \bU^{(b)} ({\btheta})}{\partial \btheta} \right\}_{\btheta=\hat{\btheta}}^{-1} \bU^{(b)}(\hat{\btheta}). 
\]
Replacing the matrix $\left\{ \partial \bU^{(b)} (\btheta) / \partial \btheta \right\}_{\btheta=\hat{\btheta}}^{-1}$ by
$\left\{ \partial \bU^{\rm org} (\btheta) / \partial \btheta \right\}_{\btheta=\hat{\btheta}}^{-1}$, we obtain the following approximation of $\hat{\bm \theta}^{(b)}$:
\begin{eqnarray}\label{lef approx}
\hat{\btheta}^{(b)}_{\rm{LEF}} &\approx & \hat{\btheta}- \left\{\frac{\partial \bU^{\rm org} ({\btheta})}{\partial \btheta} \right\}_{\btheta=\hat{\btheta}}^{-1} \bU^{(b)}(\hat{\btheta}) =\hat{\btheta} + \text{vcov}({\hat{\bm\theta}}) \bU^{(b)}(\hat{\btheta}), 
\end{eqnarray} where $\text{vcov}({\hat{\bm\theta}})$ is the model-based variance matrix based on the original dataset.

We propose the following two ways of using LEF bootstrap to construct CIs for the MRD.

\subsubsection{Approach 1: LEF bootstrap for the MSM parameters}\label{approach1}

This approach applies Taylor linearisation only to the estimating function of the pooled logistic regression for the MSM in~\eqref{outcomemodelmultiple}, with the SIPTCWs first being estimated from each bootstrap sample by fitting the corresponding logistic models to that bootstrap sample, as is done in nonparametric bootstrap.
We shall use $w_{m,k,i}$ to denote the estimated SIPTCW for patient $i$ ($i=1, \ldots, n$) in the original dataset at trial visit $k$ in trial $m$ (if patient $i$ is not enrolled in trial $m$, i.e.\ $E_{m,i}= 0$, then $w_{m,k,i}=0$). 
 Analogously, $w^{(b)}_{m,k,i}$ will denote the estimated SIPTCW for patient $i$ at trial visit $k$ in trial $m$ in the $b^{\text{th}}$ bootstrap sample.
The procedure is as follows.

\begin{enumerate}
 
\item[(1)] Estimate $w_{m,k,i}$ using the original dataset.
Using $w_{m,k,i}$, fit the weighted pooled logistic model to the original dataset and obtain the point estimate $\hat{\boldsymbol \beta}$ of the MSM parameters.

\item[(2)] Create $B$ bootstrap samples using the $n$ patients as resampling units.
Using the $b^{\text{th}}$ bootstrap sample, estimate $w^{(b)}_{m,k,i}$ for the patients in the $b^{\text{th}}$ bootstrap sample ($b = 1,...,B$).

\item[(3)] Calculate approximate bootstrap parameter estimate $\hat{\boldsymbol \beta}_{\rm LEF}^{(b)}$ from the $b^{\text{th}}$ bootstrap sample using the weights $w^{(b)}_{m,k,i}$ according to the formula in equation \eqref{lef approx}.

For our case with a pooled logistic regression for MSM in~\eqref{outcomemodelmultiple}, this formula can be written as
\begin{eqnarray}\label{lefformula}
 \hat{\boldsymbol \beta}_{\rm LEF}^{(b)} = \hat{\boldsymbol \beta} &-& 
 \left(\sum_{m=0}^{{n_v}-1} \sum_{i=1}^n E_{m,i} \sum_{k=0}^{q_{m,i}-1} \; w_{m,k,i} \; \mbox{logit}^{-1}\{\mu_{m,k,i}(\hat{\boldsymbol \beta})\}\right. \nonumber\\
 & & ~~~ \left. \left[1 - \mbox{logit}^{-1}\{\mu_{m,k,i}(\hat{\boldsymbol \beta})\} \right] \bX_{m,k,i}\bX_{m,k,i}\trans\right)^{-1} \nonumber\\
& &
\times \sum_{m=0}^{{n_v}-1} \sum_{i=1}^n E^{(b)}_{m,i} \sum_{k=0}^{q_{m,i}-1} w^{(b)}_{m,k,i} \; \bX^{(b)}_{m,k,i} \; \left[Y^{(b)}_{m,k,i} - \mbox{logit}^{-1}\{\mu^{(b)}_{m,k,i}(\hat{\boldsymbol \beta})\}\right], 
\end{eqnarray}
where $E_{m,i}$ and $E^{(b)}_{m,i}$ are the eligibility indicators of patient $i$ for trial $m$ in the original dataset and 
in the $b^{\text{th}}$ bootstrap sample, respectively; $q_{m,i}$ is the total number of trial visits made by patient $i$ eligible in trial $m$ before being artificially censored, loss to follow-up, the occurrence of outcome event or reaching the end of trial $m$; $\mbox{logit}^{-1}\{\mu_{m,k,i}(\hat{\boldsymbol \beta})\}$ and $\mbox{logit}^{-1}\{\mu^{(b)}_{m,k,i}(\hat{\boldsymbol \beta})\}$ are the estimated discrete-time hazards at trial visit $k$ for patient $i$ in trial $m$ evaluated at $\hat{\boldsymbol \beta}$ using 
the original dataset and the $b^{\text{th}}$ bootstrap sample, respectively; $\bX_{m,k,i}$ and $\bX^{(b)}_{m,k,i}$ are the design vector for the discrete-time hazard MSM at trial visit $k$ for patient $i$ in trial $m$ 
in the original dataset and in the $b^{\text{th}}$ bootstrap sample, respectively; 
$Y^{(b)}_{m,k,i}$ is the observed outcome indicator at trial visit $k$ for patient $i$ in trial $m$ in the $b^{\text{th}}$ bootstrap sample. 

\item[(4)] For the $b^{\text{th}}$ bootstrap sample, estimate the MRD at each trial visit by setting $\hat{\boldsymbol \beta}_{\rm LEF}^{(b)}$ as the MSM parameters.

\item[(5)] Construct the pivot CI using the $2.5^\text{th}$ and $97.5^\text{th}$ percentiles of the $B$ LEF bootstrap estimates of the MRD at each trial visit, as in Section~\ref{nonpboot}.
\end{enumerate}

\subsubsection{Approach 2: LEF bootstrap for the model parameters for SIPTCWs and the MSM parameters}

In the second approach, the Taylor linearisation is applied to the estimating functions of both the models for estimating SIPTCWs and the pooled logistic regression for the MSM~\eqref{outcomemodelmultiple}. Approach 2 should be even more computationally efficient than Approach 1, because it avoids fitting the models for the SIPTCWs to each bootstrap sample. Moreover, Approach 2 could be useful when there are many covariates (relative to the sample size) in the models for the SIPTCWs, in which case ill-conditioned matrices could arise when fitting these models to some of the bootstrap samples.

Because multiple models have to be fitted to obtain the SIPTCWs, we explain the steps for implementing the LEF bootstrap using the model for the denominator term of the stabilised inverse probability of treatment weights (IPTWs, see equation (1) of the Supplementary Materials) given $\overline A_{m,k-1} = \overline 1$ as an illustration. Implementing the LEF bootstrap is analogous for all other models involved in equations (1) and (2) of the Supplementary Materials.

Let $p_{m, k}=\Pr(A_{m,k} = 1 \mid \overline A_{m,k-1} = \overline 1, V, L_{m,k}, E_m = 1, Y_{m,k-1} = 0, C_{m,k-1} = 0)$, the conditional probability that the patient remains treated at trial visit $k$ in trial $m$ given they received treatment up to trial visit $k-1$, conditional on their observed variables up to trial visit $k$. Suppose that a logistic regression model is assumed for $p_{m,k}$, 
\begin{align}\label{pmodel}
 \logit(p_{m,k}) = \bZ_{m, k}\trans {\boldsymbol \gamma},
\end{align} where $\bZ_{m,k}$ is the design vector and ${\boldsymbol \gamma}$ is the regression parameter vector. Note that $\bZ_{m,k}$ only contains rows for patients who were always treated in trial $m$ up until the previous trial visit $k-1$, i.e. $\overline A_{m,k-1} = \overline 1$ .
The procedure is as follows.
\begin{enumerate}

 \item[(1)] Fit the weighted pooled logistic model to the original dataset and obtain the point estimate $\hat{\boldsymbol \beta}$ of the MSM parameters, using the estimated weights $w_{m,k,i}$ based on the original dataset.

\item[(2)] Create $B$ bootstrap samples with patients as resampling units, and let $s^{(b)}_i$ denote the number of times that patient $i$ is sampled in the $b^{\text{th}}$ bootstrap sample ($b = 1,\dots,B$). Note that $s^{(b)}_i=0$ if patient $i$ is not sampled in the $b^{\text{th}}$ bootstrap sample.

\item[(3)] Calculate the LEF bootstrap parameter estimates of the models for estimating SIPTCWs for each bootstrap sample according to \eqref{lef approx}. For example, the LEF bootstrap estimates $\hat {\boldsymbol \gamma}^{(b)}$ of $\boldsymbol \gamma$ in~\eqref{pmodel} can be obtained by 
\begin{eqnarray}\label{gammaformula}
 \hat {\boldsymbol \gamma}^{(b)} = \hat {\boldsymbol \gamma} &-& \left[\sum_{m = 0}^{n_v-1}\sum_{i = 1}^n E_{m,i} \sum_{k = 0}^{q_{m,i}-1} \mathds{1}_{\{\overline A_{m,k-1,i} = \overline 1\}}p_{m,k,i}(\hat {\boldsymbol \gamma})\left\{1 - p_{m,k,i}(\hat{\boldsymbol \gamma})\right\} \bZ_{m,k,i}\bZ_{m,k,i}\trans \right]^{-1} \nonumber\\
 &&\sum_{m = 0}^{n_v-1}\sum_{i = 1}^n s^{(b)}_i E_{m,i} \sum_{k = 0}^{q_{m,i}-1}\mathds{1}_{\{\overline A_{m,k-1,i} = \overline 1\}}\bZ_{m,k,i} \left\{A_{m,k,i} - p_{m,k,i}(\hat{\boldsymbol \gamma})\right\}, 
\end{eqnarray} 
where $\mathds{1}_{\{\overline A_{m,k-1,i} = \overline 1\}}$ is the indicator for whether patient $i$ has always been treated up to trial visit $k-1$ in trial $m$,
$p_{m,k,i}(\hat {\boldsymbol \gamma})$ is the estimated probability of treatment adherence given previous treatment at trial visit $k$ for patient $i$ in trial $m$ evaluated at $\hat{\boldsymbol \gamma}$,
$\bZ_{m,k,i}$ is the corresponding design vector for the logistic regression model~\eqref{pmodel}, and $A_{m,k,i}$ is the observed treatment indicator at trial visit $k$ for patient $i$ in trial $m$.

\item[(4)] Based on LEF bootstrap parameter estimates for the models for estimating SIPTCWs, calculate a new set of weights $\Tilde{w}^{(b)}_{m,k,i}$ for the $b^{\text{th}}$ bootstrap sample.

\item[(5)] Construct the pivot CI using Steps (3)-(5) in Approach 1 in Section~\ref{approach1}, replacing ${w}^{(b)}_{m,k,i}$ with $\Tilde{w}^{(b)}_{m,k,i}$ in~\eqref{lefformula}. 
\end{enumerate}

\subsection{Jackknife}
\label{Jackknife}
We use jackknife resampling to construct two types of jackknife-based CIs: 1) using a jackknife estimate of the MRD standard error \cite{tibshirani_introduction_1994}, and 2) using the jackknife variance estimator \cite{lipsitz_using_1990, friedl_jackknife_2006} of $\hat{\boldsymbol \beta}$. The resampling units are the patients.

\subsubsection{Approach 1: Wald-type CI using the jackknife estimate of the MRD standard error}
\label{Jackknife approach 1}
We use the jackknife estimator of the standard error of the MRD estimate and a Normal approximation to obtain CIs as follows:
\begin{enumerate}
\item[1)] Obtain the parameter estimate $\hat{\boldsymbol \beta}$ of the MSM in~\eqref{outcomemodelmultiple} and the MRD estimate $\widehat{\mbox{MRD}}_m(k)$ at each trial visit $k$ in trial $m$ using the method in Section \ref{estimation}.
 \item[2)] For $i=1, \ldots, n$, obtain parameter estimates $\hat{\bm{\beta}}^{(-i)}$ and $\widehat{\mbox{MRD}}^{(-i)}_m(k)$ using the method in Section \ref{estimation} after leaving out data from patient $i$. 
 \item[3)] Obtain the jackknife standard error estimate $\widehat{\mbox{SE}}^J_m(k)$ of the MRD at trial visit $k$ in trial $m$:
 \begin{align}
 \widehat{\mbox{SE}}^J_m(k) = \left[ \frac{n-1}{n}\sum_{i = 1}^n\left(\widehat{\mbox{MRD}}^{(-i)}_m(k) - \widehat{\mbox{MRD}}_m(k)\right)^2\right]^{1/2}
 \end{align}
 \item[4)] Define the lower and upper bounds of the 95\% CI for the MRD at trial visit $k$ in trial $m$, respectively, 
 $
 \widehat{\mbox{MRD}}_m(k)- 1.96 \cdot \widehat{\mbox{SE}}^J_m(k)
 $
 and
 $
 \widehat{\mbox{MRD}}_m(k) + 1.96 \cdot \widehat{\mbox{SE}}^J_m(k).
$
\end{enumerate}

\subsubsection{Approach 2: Multivariate Normal sampling using the jackknife variance estimator of the MSM parameters}
\label{Jackknife approach 2}
We use the jackknife variance estimator \cite{lipsitz_using_1990, friedl_jackknife_2006} of $\hat{\boldsymbol \beta}$ to construct simulation-based CIs of the MRD in a manner similar to the sandwich-variance-estimator-based CIs as follows:
\begin{enumerate}
\item[1)] Obtain the parameter estimate $\hat{\boldsymbol \beta}$ of the MSM in~\eqref{outcomemodelmultiple}.
 \item[2)] For $i=1, \ldots, n$, obtain parameter estimates $\hat{\bm{\beta}}^{(-i)}$ using the method in Section \ref{estimation} after leaving out data from patient $i$.
 \item[3)] Obtain the jackknife variance estimate $\hat{\bm{V_J}}$ of the MSM parameters:
\begin{align}
 \hat{\bm{V_J}} = \frac{1}{n(n-1)}\sum_{i = 1}^n(\widetilde{\bm{\beta}}_i - \bar {\bm{\beta}})(\widetilde{\bm{\beta}}_i - \bar{\bm{\beta}})^{\rm T}
\end{align}
where $\widetilde{\bm{\beta}}_i = n\hat{\bm{\beta}} - (n-1)\hat{\bm{\beta}}^{(-i)}$ for $i = 1,...,n$ and $\bar\beta = n^{-1}\sum_{i = 1}^n\widetilde{\bm{\beta}}_i$.
 \item[4)] Construct a 95\% CI by repeating Steps (2)-(4) of Section~\ref{sandwich CI} and replacing the sandwich variance matrix estimate $\hat{\bm{\Sigma}}$ with $\hat{\bm{V_J}}$.
\end{enumerate}

\section{Simulation study} \label{simulation}

We conducted an extensive simulation study to compare the performance of CIs obtained using the sandwich variance estimator, nonparametric bootstrap, the proposed LEF bootstrap and jackknife methods. For simplicity, we assumed there was no loss to follow-up. 

\subsection{Study setup}
\subsubsection{Data generating mechanism}
We used the algorithm described in Young and Tchetgen Tchetgen (2014) \cite{young_simulation_2014} to simulate data.
This algorithm ensure that previous treatments affect time-varying variables (confounders) that are associated with both current treatment and hazard of the outcome event. 
The data generating mechanism is described in Table~\ref{data simu mechanism}.

\begin{table}[h]
\caption{Summary of data generating mechanism of the simulation study.}
\label{data simu mechanism}
\centering
\small
\begin{tabular}{p{0.35\linewidth} p{0.55\linewidth}} 
 \toprule
 Data simulation setting specifications & $n$: number of patients \\
 & $n_v= 5$: number of visits\\
 & $t_j = 0,..., n_v-1$: visit time for visit $j$ \\
 & $\alpha_a$: intercept in the treatment model, representing the baseline rate of treatment initiation \\
 & $\alpha_c$: coefficient that describes the strength of confounding due to time-varying variable $X_{1,t_j}$ \\
 & $\alpha_y$: intercept term in the discrete-time hazard model, representing the baseline hazard \\
 & \\
 Time-varying confounder & $ X_{1,t_j} \sim N(Z_{t_j} - 0.3A_{t_{j-1}}, 1)$, where $A_{t_{-1}} \equiv 0$ and $Z_{t_j} \sim N(0,1)$.
 \\
 Time-invariant confounder
 & $X_2 \sim N(0,1)$. \\
 & \\
 Treatment & $\text{logit}\{\Pr(A_{t_j} = 1\mid A_{t_{j-1}},X_{1,t_j}, X_2, Y_{t_{j-1}} = 0)\} $ \\
 & \quad $= \alpha_a + 0.05 A_{t_{j-1}} + \alpha_cX_{1,t_j} + 0.2X_2 $, where $Y_{t_{-1}} \equiv 0$. \\
 & \\
 & \\
 Discrete-time hazard of the outcome event & $\text{logit}\{\Pr(Y_{t_j} = 1 \mid A_{t_j}, X_{1,t_j},X_2, Y_{t_{j-1}} = 0) \}$ \\
 & \quad $ = \alpha_y - 0.5 A_{t_j} + \alpha_cX_{1,t_j} + X_2$. \\
 &\\
 Trial Eligibility & $E_{t_j} = 1$ if patient has not received treatment before $t_j$ \\
 & and has not experienced the outcome event before $t_j$; \\
 & $E_{t_j} =0$ otherwise \\
 \bottomrule
\end{tabular}

\end{table}

\subsubsection{Monte Carlo simulation settings}

We considered three settings with different outcome event rates, by setting the baseline hazards $\alpha_y$ as $-4.7$, $-3.8$ and $-3$ to allow low ($5-6.5\%$), medium ($10-14\%$) and high ($20-25\%$) percentage of patients experiencing the event during follow-up in the simulated data, respectively. In total, we investigated 81 scenarios for generating the simulated data using the mechanism in Table~\ref{data simu mechanism}, by considering the combinations of the specifications presented in Table \ref{simu scenarios}. We varied the number of patients, confounding strength of a time-varying confounder and treatment prevalence (i.e., percentage of patients who ever received treatments during follow-up). By varying the intercept term of the treatment model $\alpha_a$, we could generate a low ($25-30\%$), medium ($50-60\%$) and high ($75-80\%$) treatment prevalence in the simulated data. For each scenario, we generated 1000 simulated datasets.

\begin{table}[ht]
\centering
\caption{Summary of the specifications for the 81 scenarios considered in the simulations.}
\label{simu scenarios}\small
\begin{tabular}{cccc } 
\toprule
 Outcome event rate & Sample size & Confounding strength & Treatment prevalence \\
 \midrule
 Low: $\alpha_y=-4.7$& Small: $n=200$ & $\alpha_c=0.1$ & Low: $\alpha_a=-1$ \\
 Medium: $\alpha_y=-3.8$ & Medium: $n=1000$ & $\alpha_c=0.5$ & Medium: $\alpha_a=0$\\
 High: $\alpha_y=-3$ & Large: $n=5000$ & $\alpha_c=0.9$ & High: $\alpha_a=1$\\
 \bottomrule
\end{tabular}
\end{table}


\subsubsection{Estimation and inference}

For each simulated dataset, we emulated 5 trials ($m=0, \ldots,4$), with trial 0 including trial visits $k=0, \ldots, 4$, trial 1 including trial visits $k=0, \ldots, 3$, and so on. 
Our estimand of interest is the MRDs at trial visits $k = 0,\ldots,4$ for patients eligible in trial $0$. We chose trial 0 patients as the target population because they had the longest follow-up and so we can assess the methods for estimation and inference of the MRDs at later visits.

 Correctly specified logistic regression models were fitted to the simulated data to estimate the denominator terms of the stabilised IPTWs in equation (1) of the Supplementary Materials, while the numerator terms were estimated by fitting an intercept-only logistic model stratified by treatments received at the immediately previous visit $A_{t_{j-1}}$. 
 
Then the following weighted pooled logistic model was fitted to the artificially censored data in the five sequential trials to estimate the counterfactual discrete-time hazard:
\begin{eqnarray} 
\text{logit}\left\{\mbox{Pr}\left(Y_{m,k}^{\overline a_k = \overline a} =1 \mid {Y}_{m,k-1}^{\overline a_{k-1} = \overline a}=0, X_{1m,0}, X_2\right)\right\} = \beta_{0,k} + \beta_{1,k} \cdot a + \beta_{2,k} X_{1,m,0}+ \beta_{3,k} X_2 \label{outcome model multiple int}
\end{eqnarray}
 where $X_{1,m,0}$ is the value of the time-varying confounder at the baseline of trial $m$ ($m=0, \ldots, 4$) and $ \beta_{0,k}$, $ \beta_{1,k}$, $ \beta_{2,k}$ $ \beta_{3,k}$ are regression coefficients that vary by trial visit ($k=0, \ldots, 4$) but are assumed to be the same across trials.
 Note that we were not able to correctly specify the MSM for counterfactual discrete-time hazard via pooled logistic regression models, due to the non-collapsibility of the logistic model used in the data-generating mechanism \cite{keogh_causal_2023}. It was to minimise this misspecification (and potential bias caused by it) that we chose the model in~(\ref{outcome model multiple int}), which is a rich model that allows the coefficients of treatment and confounders to vary by trial visit. 

We applied the methods for constructing 95\% CIs of the MRD in Section~\ref{CI methods} to each simulated dataset, where we set \textit{$S = 500$} for the CIs based on the sandwich variance estimator and \textit{$B = 500$} for the CIs based on the bootstrap methods. We applied the two jackknife methods only when $n = 200$, because it would be computationally inefficient relative to bootstrap to use the jackknife CI methods for larger sample sizes. The jackknife method is a linear approximation of the bootstrap \cite{efron_bootstrap_1979}, and so we would not expect jackknife CIs to perform any better than bootstrap with larger sample sizes.
A pseudo-code algorithm for the simulation study is presented in Section 5 of the Supplementary Materials.

\subsubsection{True values}

True values of the MRDs for each simulation scenario were obtained by generating data for a very large randomized controlled trial, as proposed by Keogh et al. (2023) \cite{keogh_causal_2023}. The true marginal risks in trial 0 when all patients were always treated or all patients were never treated were approximated by Kaplan-Meier estimates from two extremely large datasets ($n= 1,000,000$).
For the first dataset, the treatment strategy was set to `always treated' (by setting $A_{0,k}=1$ for $k=0, \ldots, 4$).
For the second, it was set to `never treated' (by setting $A_{0,k}=0$ for $k=0, \ldots, 4$).

\subsubsection{Performance measures}

\textit{Empirical coverage} of the CIs was calculated for each trial visit in each simulation scenario. This was done by dividing the number of times that a CI of the MRD contained the true MRD by 1000 minus the number of times there was an error in the CI construction. Such errors include issues with the sandwich variance estimation or jackknife variance estimation that led to non-positive definite variance matrices. We also considered \textit{bias-eliminated CI coverage} to adjust for the impact of bias in the coverage results \cite{morris_using_2019}. This involved calculating the proportion of the 1000 CIs that contained the sample mean of the 1000 MRD estimates for each scenario.

Morris et al (2019) \cite{morris_using_2019} state four potential reasons for CI under- or over-coverage when examining simulation results: (i) MRD estimation bias, (ii) the standard error estimates from a CI method underestimate the empirical standard deviation (SD) of the MRD estimator, (iii) the distribution of the MRD estimates is not normal and CIs have been constructed assuming normality, and (iv) the standard error estimates from a CI method are too variable.
Therefore, for each simulation scenario, we calculated the \textit{empirical bias} of the MRD estimates at each trial visit to check reason (i) for CI under-coverage. For each CI method, we calculated the SD of the MRD estimates (from resampling), denoted as $\widehat{\rm SE}$. This provided us with 1,000 standard error estimates for each CI method.
We then computed $\frac{\widehat{\rm SE}}{{\rm SD}_{\widehat{\rm MRD}}}$, the \textit{ratio of standard error estimate for each CI method to the SD of the MRD empirical distribution} ${\rm SD}_{\widehat{\rm MRD}}$, where the ${\rm SD}_{\widehat{\rm MRD}}$ was obtained by taking the SD of the 1000 MRD point estimates across simulations. We referred to this ratio as the `Standard Error (SE) ratio'. The summary statistics of the SE ratio across 1000 simulations would allow us to assess reasons (ii) and (iv) for potential CI under-coverage or over-coverage.

We recorded the number of times that each CI method encountered an error (a \textit{`construction failure'}). We also calculated the \textit{Monte Carlo standard error} of the empirical coverage (due to using a finite number of simulated datasets) \cite{morris_using_2019}:
\begin{align}
 \mbox{SE}_{\rm MC} = \sqrt{\frac{\widehat{\text{Coverage}}(1- \widehat{\text{Coverage}})}{n_{\rm sim}}}, 
\end{align} where $\widehat{\text{Coverage}}$ is the empirical coverage and ${n_{\rm sim}}$ is the number of simulations performed (${n_{\rm sim}}= 1000$ in our case). 

Finally, we reported the relative computation time that it took to construct CIs based on nonparametric bootstrap, LEF bootstrap and jackknife compared to CIs based on the sandwich variance estimator.
\subsubsection{Computational resources} 

The simulations were conducted using \texttt{R} (version 4.1.3) \cite{r_core_team_r_2023}. All R scripts are available at \url{https://github.com/juliettelimozin/Multiple-trial-emulation-IPTW-MSM-CIs}.
Point estimation and CI construction based on the sandwich variance estimator were performed using 
 the R package \texttt{TrialEmulation} \cite{rezvani_trialemulation_2023} (version 0.0.3.9), and some of the key functions in the \texttt{R} package are summarised in Section 5 of the Supplementary Materials.
We used packages \texttt{doParallel} and \texttt{doRNG} to parallelise the simulation, which took 12 hours to finish by using 67 cores of the University of Cambridge high-performance computing cluster.

\subsection{Results} \label{results}
In this section, we focus on the discussion of the results in low event rate scenarios; results for medium and high event rate scenarios can be found in Section 6 of the Supplementary Materials. The reason for this focus is two-fold: 1) The HERS data example had low event rates; 2) the CI methods performed the worst in low event rate scenarios and yet we still observed interesting patterns for their relative performance, which were not drastically different from those in medium and high event rate scenarios.

\begin{sidewaysfigure}
 \centering
 \includegraphics[width = \textwidth]{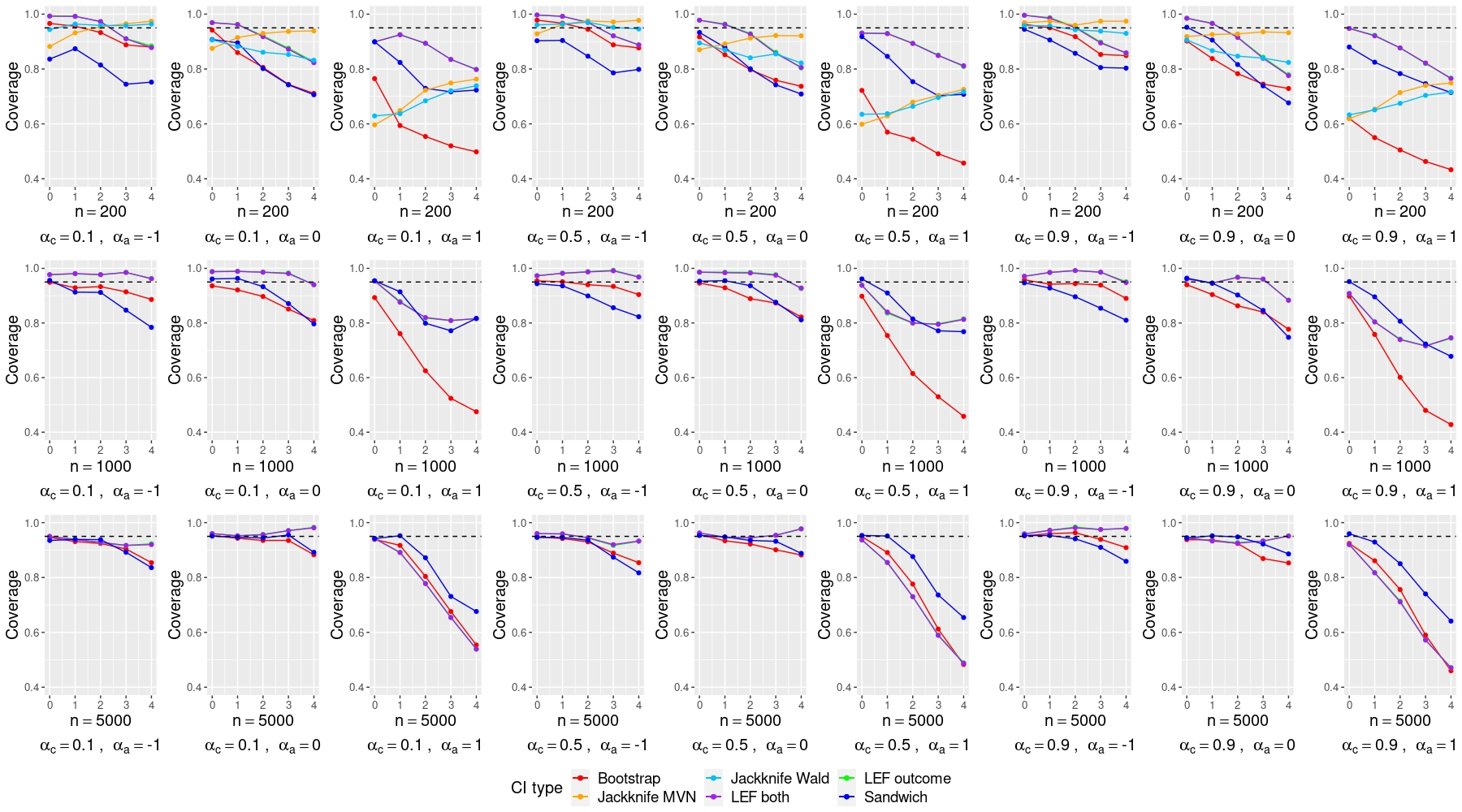}
 \caption{Empirical coverage of the CIs in the scenarios with \textbf{low event rates}. Bootstrap: CIs constructed by nonparametric bootstrap; LEF both: CIs constructed by applying Approach 2 of LEF bootstrap; LEF outcome: CIs constructed by applying Approach 1 of LEF bootstrap; Jackknife Wald: CIs constructed by applying Approach 1 of jackknife resampling; jackknife MVN: CIs constructed by applying Approach 2 of jackknife resampling; Sandwich: CIs based on the sandwich variance estimator. Note that the results for applying Approaches 1 and 2 of LEF bootstrap were very similar so that the purple and green lines overlapped.}
 \label{fig:coverage_low}
\end{sidewaysfigure}

Figure~\ref{fig:coverage_low} shows the empirical coverage rates of the CIs when the event rate was low; see Figures 2 and 3 of the Supplementary Materials for results with medium and high event rates.
Differences between these coverage rates and the corresponding bias-eliminated coverage rates were negligible (see Figures 4, 5 and 6 of the Supplementary Materials).

We will initially examine the bias of the MRD estimates, followed by the SE ratio results for the CI methods, to explain the under-coverage or over-coverage of the CIs and the relative performance of these CI methods in various simulation scenarios. Afterwards, we will discuss computation time for the CI methods. Results for Monte Carlo standard errors of the CI coverage estimates and CI construction failure rates can be found in Section 6.6 and 6.7 of the Supplementary Materials. Since we observe very little difference in the performance of the two LEF bootstrap CI methods, we will refer to them as one when discussing the results. We refer to the CIs constructed using the sandwich variance estimator as `Sandwich CIs', those using Approach 1 from jackknife resampling as `Jackknife Wald CIs' and those using Approach 2 as `Jackknife Multivariate Normal (MVN) CIs'.

\subsubsection{Empirical bias and its impact on CI coverage}

\begin{sidewaysfigure}
 \centering
 \includegraphics[width = \textwidth]{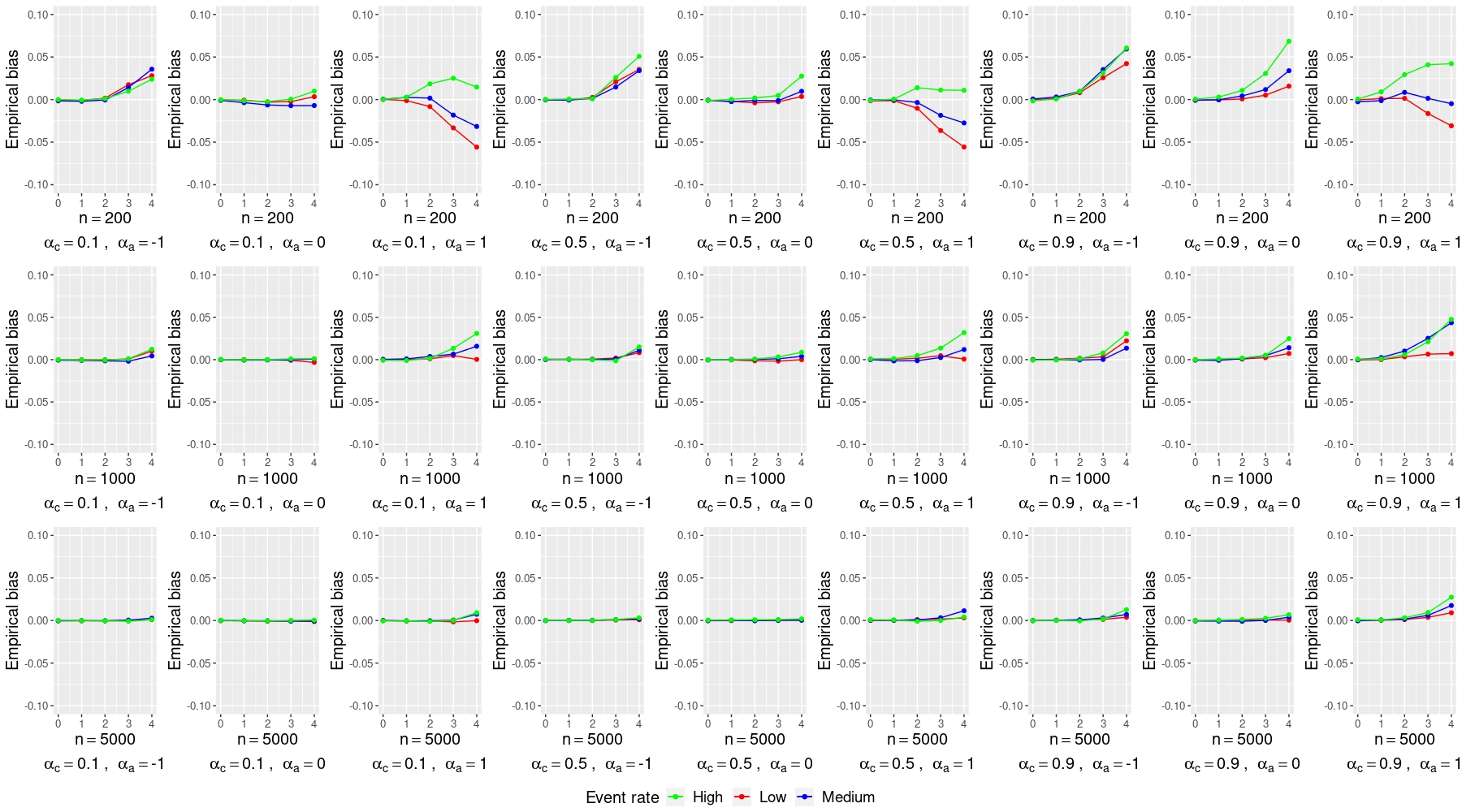}
 \caption{Empirical biases of the MRD estimates in various simulation scenarios.}
 \label{bias}
\end{sidewaysfigure}

Figure~\ref{bias} presents the empirical biases of the MRD estimates.

Minimal biases were observed at \textit{earlier visits} ($k < 2$), which approached zero as sample sizes increased. The increasing absolute bias at \textit{later visits} ($k >2$) could be explained by the increasing data sparsity and treatment arm imbalances at later visits. Very few events occurred after trial visit $2$ in most scenarios, even in those with large sample sizes ($n = 5000$) (see Table 3 of Supplementary Materials for an example).

The absolute biases in \textit{low or high treatment prevalence} scenarios were larger than those in medium treatment prevalence scenarios. This likely stemmed from data sparsity issues at later trial visits that were aggravated by treatment arm imbalance caused by low or high treatment prevalence. Table 1 of the Supplementary Materials shows an example of this imbalance.

\textit{Confounding strength} had a negligible effect on absolute bias, perhaps because the specified values of confounding strength $\alpha_c$ had relatively small influences on the data generating mechanism.

Increasing the \textit{outcome event rate} tended to reduce absolute bias, because it reduced data sparsity.

These bias results explain the general trends of CI coverage in Figure \ref{fig:coverage_low}. As the sample size increased, the CI coverage at earlier visit approached the nominal 95\% level, which was not surprising given that all the methods rely on asymptotic approximations. The increasing absolute bias at \textit{later visits} ($k >2$) coincided with the deterioration in CI coverage that we observed across all simulation scenarios at later visits (reason (i) of under-coverage according to Morris et al. \cite{morris_using_2019}).

 In scenarios with low treatment prevalence ($\alpha_a = -1$), CI coverage did not achieve the nominal 95\% level at later visits ($k >2$) even with larger sample sizes. Similarly, in scenarios with high treatment prevalence ($\alpha_a = 1$), nominal coverage was rarely achieved, except at the baseline visit ($k=0$), and coverage decayed considerably at later visits for Sandwich CIs, nonparametric and LEF bootstrap CIs.

We observed minimal impact of confounding strength on CI coverage. Increasing the event rate improved the CI coverage for all methods, as shown in Figures 2 and 3 of the Supplementary Materials.

The empirical SD and the root-Mean Squared Error (MSE) of the MRD estimates largely followed the patterns of the bias (see Figures 7 and 8 of the Supplementary Materials): scenarios with larger biases also exhibited larger SDs and MSEs. 

\subsubsection{SE ratio and its impact on CI coverage}

\begin{figure}[!ht]
 \centering
 \includegraphics[width = \textwidth]{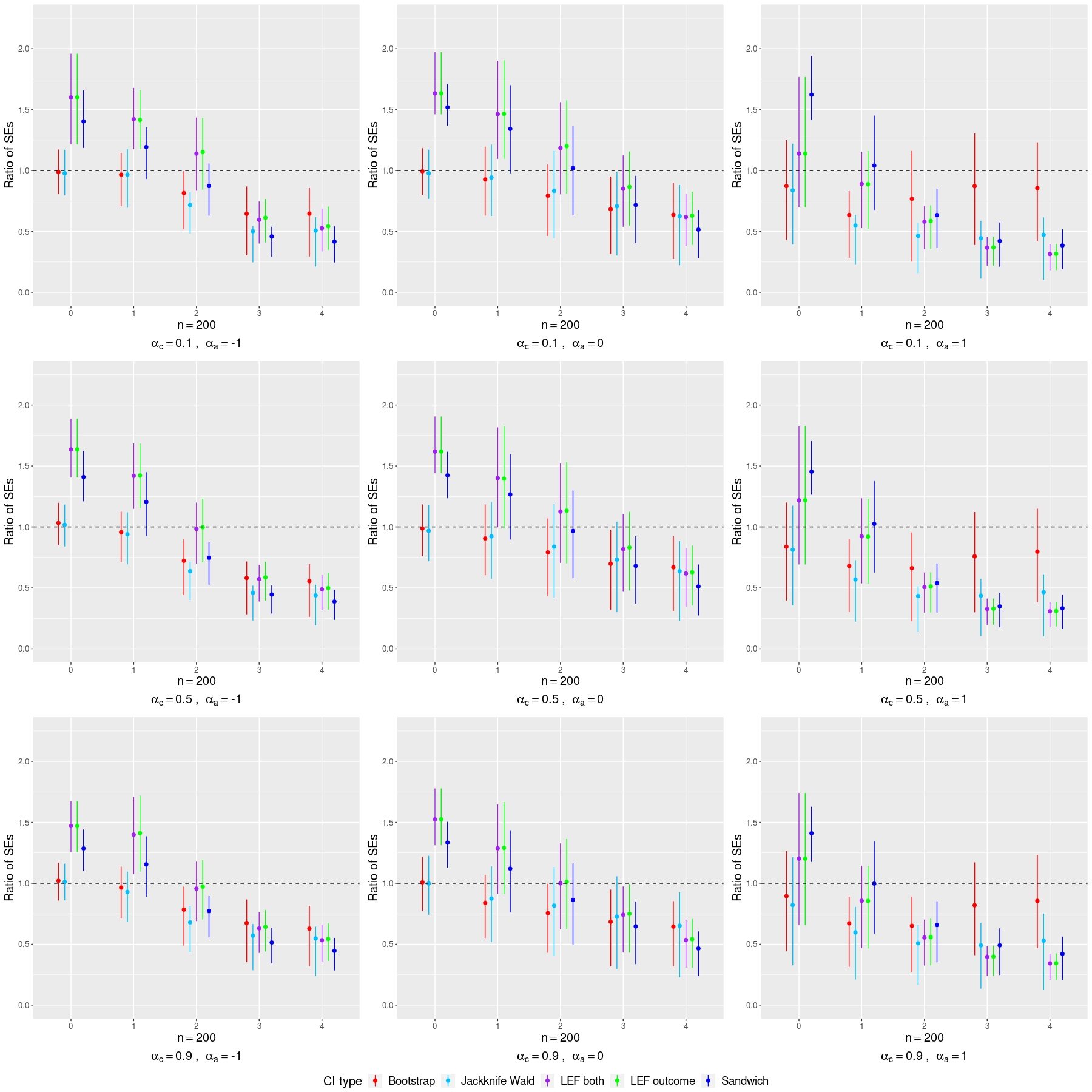}
 \caption{Ratio of the estimated standard error to empirical standard deviation of the MRD estimator (SE ratio) in \textbf{low event rate and small sample size scenarios}. The dots represent the averages of the ratio, with the bottom and top of the bar being the 1st and 3rd quartile of this ratio, respectively. Bootstrap: CIs constructed by nonparametric bootstrap; LEF both: CIs constructed by applying Approach 2 of LEF bootstrap; LEF outcome: CIs constructed by applying Approach 1 of LEF bootstrap; Jackknife Wald: CIs constructed by applying Approach 1 of jackknife resampling; Sandwich: CIs based on the sandwich variance estimator. }
 \label{fig:se_ratio_low}
\end{figure}

\begin{figure}[!ht]
 \centering
 \includegraphics[width = \textwidth]{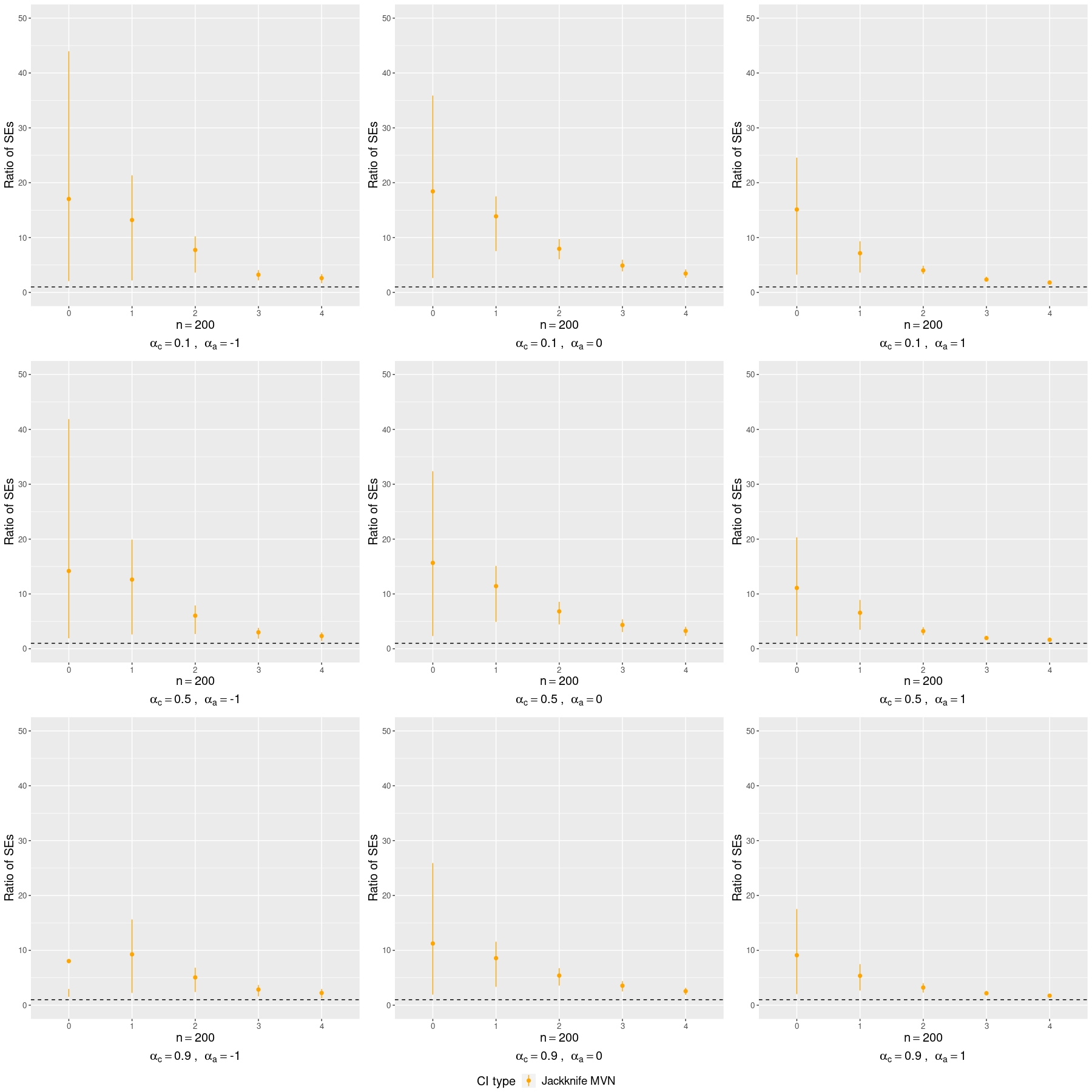}
 \caption{Ratio of the estimated standard error to empirical standard deviation of the MRD estimator (SE ratio) in \textbf{low event rate and small sample size scenarios}. The dots represent the averages of the ratio, with the bottom and top of the bar being the 1st and 3rd quartile of this ratio respectively. Jackknife MVN: CIs constructed by applying Approach 2 of jackknife resampling.}
 \label{fig:se_ratio_low_jackknifeMVN}
\end{figure}

Figures~\ref{fig:se_ratio_low} and \ref{fig:se_ratio_low_jackknifeMVN} present the summary statistics of the SE ratio, $\frac{\widehat{\rm SE}}{{\rm SD}_{\widehat{\rm MRD}}}$, for each CI method in scenarios with low event rates and small sample size ($n=200$); see Figures 9--15 of the Supplementary Materials for scenarios with low event rates and moderate/large sample sizes and for scenarios with medium and high event rates.

We found that the SE ratio results largely explained the differences in CI coverage of the various CI methods. The comparative performance of the CI methods in terms of coverage generally followed the patterns of the SE ratios: when the average SE ratios were above 1, we observed over-coverage; when the average SE ratios were about 1, we observed close to nominal coverage; when the average SE ratios were below 1, we observed under-coverage. Higher variability of the SE ratios also coincided with lower CI coverage. Notably, in scenarios with \textit{high event rate, large sample size, medium treatment prevalence and weak confounding}, all CI methods achieved nominal coverage, which corresponded to all CI methods having SE ratios close to 1 on average and with low variability.

\textbf{\textit{Low event rate and small sample size scenarios} } From Figure~\ref{fig:se_ratio_low}, we can see that the SE ratios for nonparametric bootstrap CIs were on average lower than 1 at later visits, and they were also more variable, especially with \textit{ high treatment prevalence}. This is consistent with the coverage results in Figure \ref{fig:coverage_low}, where we expect nonparametric bootstrap CIs to have low coverage if the variability of the MRD estimator was underestimated and the variability estimates were highly variable across simulations \cite{morris_using_2019}. The nonparametric bootstrap CIs had lower coverage compared to LEF bootstrap CIs in scenarios with low/medium treatment prevalence, and lower coverage than Sandwich CIs when treatment prevalence was high. This performance is mirrored in the SE ratio results for nonparametric bootstrap CIs in such scenarios. It was likely that the data sparsity issues at later visits increased the instability of parameter estimation in bootstrap samples and thus resulted in the lower SE ratios and large variability of SE ratios for nonparametric bootstrap CIs. In contrast, LEF bootstrap reduced instability of parameter estimation in bootstrap samples by design, and thus led to smaller variability of the SE ratios at later visits.
Similar findings for nonparametric bootstrap CIs were observed by Austin (2022) \cite{austin_bootstrap_2022}.

In Figure~\ref{fig:se_ratio_low}, the SE ratios for jackknife Wald CIs were also lower than 1 at later visits, but in low and high treatment prevalence scenarios, they were less variable compared to those for nonparametric bootstrap CIs, which might partly explain the better coverage of jackknife Wald CIs in these scenarios. 
However, in Figure~\ref{fig:se_ratio_low_jackknifeMVN}, the SE ratios for jackknife MVN CIs were much larger than 1 in all scenarios with small sample sizes, which may explain the over-coverage or closer-to-nominal coverage for jackknife MVN CIs in some scenarios.

In Figure~\ref{fig:se_ratio_low}, the SE ratios for LEF bootstrap CIs and Sandwich CIs were larger than 1 at \textit{early visits} ($k<2$), but they gradually dropped below 1 at later visits. When the \textit{treatment prevalence was high} ($\alpha_a = 1$), their SE ratios were much lower but less variable than those of nonparametric bootstrap CIs at the later visits. 
LEF bootstrap CIs also provided better coverage than Sandwich CIs at later visits in small sample size
scenarios, possibly because the SE ratios for LEF bootstrap CIs were on average closer to 1 or less variable than the Sandwich CIs' SE ratios.
In such scenarios, Sandwich CIs might have suffered from finite-sample bias of variance matrix estimation that 
affected the multivariate normal sampling for the MRD 
estimation, as well as the construction failures due to non-positive definite variance matrices (see Table 13 of the 
Supplementary Materials).

\textbf{\textit{Low event rate and medium/large sample size scenarios} } 
In Figure 9 of the Supplementary Materials, with \textit{larger sample sizes} ($n = 1000, 5000$), the SE ratio results for all CI methods were improved in low/medium treatment prevalence scenarios, which was reflected correspondingly by the better coverage. 
The SE ratios of the LEF bootstrap CIs were above 1 at most visits in low/medium treatment prevalence scenarios, which could explain some of the CI over-coverage we observed and the better coverage compared with Sandwich CIs. With high treatment prevalence, the SE ratios of the LEF bootstrap sample at later visits approached to 1 on average as sample sizes increased, but they became more variable as well. This might explain the corresponding decreased coverage of LEF bootstrap CIs, which was now similar to the coverage of nonparametric bootstrap CIs. Similar patterns in high treatment prevalence scenarios were also observed for the SE ratios of Sandwich CIs. However, they were slightly less variable than those of LEF bootstrap CIs and closer to 1 than those of nonparametric bootstrap CIs, which might be due to the much reduced number of construction failures for Sandwich CIs in large sample size and high treatment prevalence scenarios. Overall, this resulted in better coverage of 
of Sandwich CIs in these scenarios.

\textbf{\textit{Medium/high event rate scenarios} }
From Figures 10--15 in the Supplementary Materials, increasing the outcome event rate also improved the SE ratio results for all CI methods, with the SE ratios approaching 1 and becoming less variable. This is consistent with the coverage results in Figures 2--3 in the Supplementary Materials, where an increase of outcome event rate resulted in the various CI methods converging to similar closer-to-nominal coverage. However, in scenarios with large sample sizes and low or high treatment prevalence, we still observed SE ratios to be lower than 1 and highly variable at later visits. Together with the larger empirical bias, this could explain the low coverage at later visits for all CI methods in these scenarios. 

With increased event rates, we observed that differences in CI coverage among the methods were primarily due to SE ratio variability. In most scenarios with medium/high event rates and medium/large sample sizes, the average SE ratios of Sandwich CI and LEF bootstrap CIs were very similar but the SE ratios of Sandwich CIs tended to be less variable, which might lead to their better coverage. With medium/large sample sizes, nonparametric bootstrap CIs and LEF bootstrap CIs 
 had similar SE ratios, resulting in similar coverage, whereas with small sample sizes, the SE ratios of nonparametric bootstrap CIs tended to be much more variable than those for LEF bootstrap CIs, leading to worse coverage.

\textbf{\textit{Confounding strength scenarios} } Similarly to the empirical bias results, the confounding strength appeared to have minimal impact on the SE ratio results. This also translated to minimal impact of confounding strength on CI coverage.

\subsubsection{Computation time}

Table~\ref{computation time} summarises the average computation time (across 1000 simulated data sets and the scenarios of treatment prevalence and confounding strength) for constructing a CI of the MRD using each method, relative to the time for constructing Sandwich CI. 
On average, nonparametric bootstrap CIs took about 3 to 8 times longer to compute than Sandwich CIs. The computation time increased exponentially as sample sizes increased.
LEF bootstrap CIs, using both Approaches 1 and 2, took about 1.6 to 2.5 times longer to compute compared to Sandwich CIs. LEF bootstrap CIs were not as affected by sample sizes in terms of computation time compared to nonparametric bootstrap CIs. 

For small sample sizes ($n = 200$), jackknife Wald CIs had very similar computation times to LEF bootstrap CIs. Jackknife MVN CIs took 3 times longer to construct compared to Sandwich CIs, most likely because the former method required two sampling steps: the jackknife resampling step and the MVN sampling step, while Sandwich CIs only involved the MVN sampling step.

The large gain in computational efficiency through LEF bootstrap is very much of interest; with large sample sizes ($n = 5000$), LEF bootstrap was on average 3 times faster than nonparametric bootstrap. However, the sandwich variance estimator-based method remains the fastest for CI construction, especially if it is not possible to parallelise the computing of the bootstrap-based CIs.

\begin{table}[ht]
\caption{Summary of average computation time of the CI construction methods in simulations with the sandwich-variance-estimator-based CIs as the reference. Bootstrap: CIs constructed by nonparametric bootstrap; LEF both: CIs constructed by applying Approach 2 of LEF bootstrap; LEF outcome: CIs constructed by applying Approach 1 of LEF bootstrap; Jackknife Wald: CIs constructed by applying Approach 1 of jackknife resampling; Jackknife MVN: CIs constructed by applying Approach 2 of jackknife resampling; Sandwich: CIs based on the sandwich variance estimator.}
\label{computation time}
\centering
\resizebox{\textwidth}{!}{
\begin{tabular}{llcccccc}
 \hline
 & & Bootstrap & LEF & LEF & Jackknife & Jackknife & Sandwich \\ 
Outcome event rate& Sample size&&outcome&both& Wald&MVN & \\ 
 \hline
 Low & 200 & 2.94 & 1.83 & 1.62 & 1.78 & 3.00 & 1 \\ 
 & 1000 & 4.28 & 1.92 & 1.74 & & & 1 \\ 
 & 5000 & 8.39 & 2.66 & 2.50 & & & 1 \\ 
 Medium & 200 & 2.99 & 1.88 & 1.66 & 1.81 & 3.04 & 1 \\ 
 & 1000 & 4.31 & 1.99 & 1.77 & & & 1 \\ 
 & 5000 & 8.39 & 2.68 & 2.51 & & & 1 \\ 
 High & 200 & 2.97 & 1.88 & 1.66 & 1.80 & 3.02 & 1 \\ 
 & 1000 & 4.23 & 1.96 & 1.77 & & & 1 \\ 
 & 5000 & 7.99 & 2.65 & 2.48 & & & 1 \\ 
 \hline
\end{tabular}
}
\end{table}

\subsubsection{Summary}

Our simulation results suggest that LEF bootstrap CIs provided better coverage compared to Sandwich CIs in scenarios with small/medium sample sizes, low/medium outcome event rates and low/medium treatment prevalence. This might be attributed to the finite sample bias of the sandwich variance estimator and high frequency of the construction failure of Sandwich CIs in such scenarios. The performance of nonparametric bootstrap CIs was considerably affected by treatment arm imbalance and data sparsity, particularly in scenarios with low event rates and small/moderate sample sizes. In these scenarios, the LEF bootstrap method not only performed better but was also more computationally efficient than nonparametric bootstrap. While jackknife Wald CIs achieved nominal coverage for small sample sizes and low treatment prevalence, they were sensitive to treatment prevalence. Due to data sparsity and finite-sample bias, all methods performed poorly when treatment prevalence was high, and jackknife MVN CIs faced numerous construction failures, making them impractical for use.

Since STE is particularly useful for data scenarios with small numbers of patients initiating treatments at any given time (low treatment prevalence) and with low event rates, LEF bootstrap offers a useful alternative to the sandwich variance estimator and the nonparametric bootstrap for CI construction, especially in small/medium sample sizes. Although jackknife Wald CIs also provided good coverage, they become computationally inefficient as the number of patients exceeds the number of bootstrap samples. For large sample sizes and medium/high event rates, Sandwich CIs has the advantage of being computationally more efficient than the bootstrap CI methods. Overall, our investigation underscores the importance of considering sample size, outcome event rate, and treatment prevalence when selecting a CI construction method in STE.

\section{Application to the HERS data} \label{application}

In this section, we applied the methods described in Sections~\ref{estimation}--\ref{CI methods} to the HERS data. 
The treatment process for the denominator term of the stabilised IPTWs was modelled by two logistic regressions stratified by treatment received at the immediately previous visit $A_{t_{j-1}}$ ($j=1, \ldots, 4$). 
We included the following covariates: CD4 count (after square root transformation and standardization) and HIV viral load (after $\log_{10}$ transformation and standardization) measured at the previous two visits (since HAART treatment was a self-reported status between the last visit and the current visit), HIV symptoms at the previous visit, ethnicity (Caucasian, Black, other) and study sites. The numerator term of the stabilised IPTWs was estimated as the marginal probability of receiving HAART in the stratum defined by previous treatment $A_{t_{j-1}}$. 

Similarly, the censoring process for the denominator term of the stabilised inverse probability of censoring weights (IPCWs, see equation (2) of the Supplementary Materials) was modelled by two logistic regressions stratified by the previous treatment $A_{t_{j-1}}$. 
The following covariates were included: CD4 count, HIV viral load and HIV symptoms measured at the current and previous visits, ethnicity and study sites. The numerator term of the stabilised IPCWs was estimated by the marginal probabilities of remaining in the HERS cohort, stratified by the previous treatment $A_{t_{j-1}}$. 

\begin{table}[!ht]
\centering
\caption{Results for the fitted MSM for the sequentially emulated trials using the HERS data. Robust standard error: standard errors based on the sandwich variance estimate.}
\begin{tabular}{lrr}
 \hline
 & Estimate & Robust standard error \\ 
 \hline
Intercept & -5.159 & 1.131 \\ 
 Assigned treatment: HAART vs. non-HAART& -0.015 & 0.549 \\ 
 CD4 count at 1 visit before trial baseline & 0.146 & 0.380 \\ 
 CD4 count at 2 visits before trial baseline & -0.277 & 0.372 \\ 
 Viral load at 1 visit before trial baseline & 0.015 & 0.354 \\ 
 Viral load at 2 visits before trial baseline & 0.615 & 0.275 \\ 
 Site 2 vs. Site 1 & 0.503 & 1.148 \\ 
 Site 3 vs. Site 1 & 0.071 & 0.936 \\ 
 Caucasian vs. Black & -0.182 & 0.757 \\ 
 Other ethnicity vs. Black & 0.977 & 1.078 \\ 
Interaction between assigned treatment and & -0.004 & 0.554 \\ 
CD4 count at 1 visit before trial baseline & & \\
 \hline
\end{tabular}
\label{HERS msm}
\end{table}

The pooled logistic regression for the MSM included an intercept term, main effect of treatment, main effects of CD4 count and HIV viral load at previous two visits before the trial baseline, main effects of ethnicity and study sites, and the interaction between treatment and CD4 count at the visit before trial baseline. 
A summary of the fitted MSM is provided in Table~\ref{HERS msm}.

 We used 500 bootstrap samples for Nonparametric and LEF bootstrap CIs, and drew a sample of size $S = 500$ of MSM parameters for constructing Sandwich CIs. 

\begin{figure}[ht!]
 \centering
 \includegraphics[width = \textwidth]{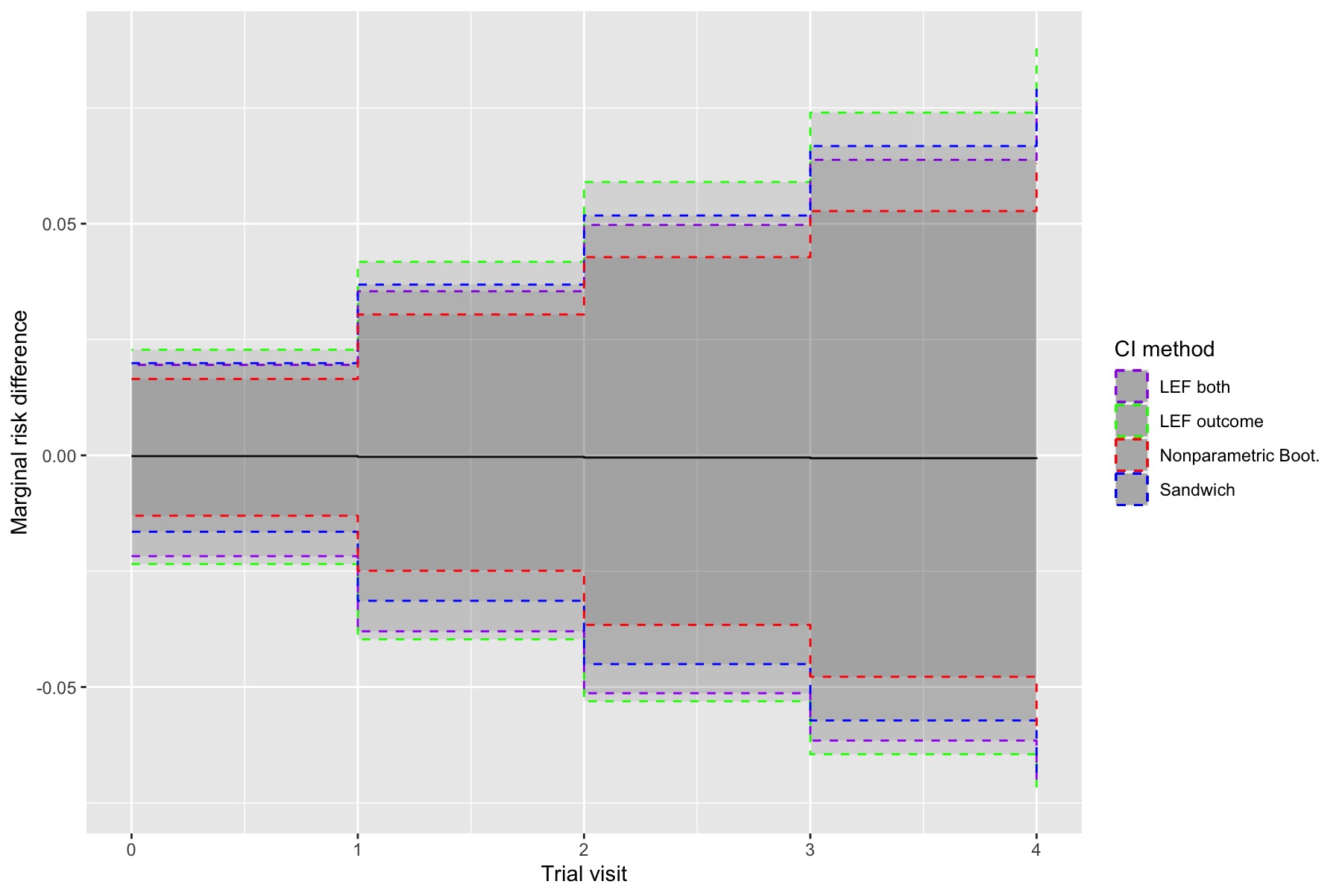}
 \caption{Estimates and 95\% CIs of the MRD of all-cause mortality under HAART treatment strategy and the non-HAART strategy for patients enrolled in the first emulated trial (trial 0) of the HERS data. Nonparametric Boot.: CIs constructed by nonparametric bootstrap; LEF both: CIs constructed by applying Approach 2 of LEF bootstrap; LEF outcome: CIs constructed by applying Approach 1 of LEF bootstrap; Sandwich: CIs based on the sandwich variance estimator.}
 \label{fig:HERS results}
\end{figure}

Figure \ref{fig:HERS results} presents the estimated MRD, which is the difference in risk of all-cause mortality when all patients eligible to trial 0 adhere to the treatment strategies assigned (HAART vs. no HAART). Figure \ref{fig:HERS results} also includes 95\% CIs of the MRD obtained by the four methods described in Section~\ref{CI methods}. We note that the results are not statistically significant given that all four CIs include zero. 

\section{Discussion} \label{discussion}

In this article, we focused on the application of STE to estimate and make inferences about the per-protocol effect of treatments on a survival outcome in terms of counterfactual MRDs over time. We conducted a simulation study to compare the relative performance of four CI construction methods for the MRD, based on the sandwich variance estimator, nonparametric bootstrap, LEF bootstrap (which previously had not been extended to STE) and jackknife resampling. In scenarios with small/moderate sample sizes, low event rates and low treatment prevalence, we observed relatively better coverage for LEF bootstrap CIs than for nonparametric bootstrap and Sandwich CIs. These results align with previous findings on the limitations of the sandwich variance estimator when the sample size is small \cite{pan_small-sample_2002, rogers_assessment_2021}. 
Since STE is particularly useful in scenarios with small sample sizes, low event rates and low treatment prevalence, the LEF bootstrap offers a valuable alternative to the sandwich variance estimator and the nonparametric bootstrap for constructing CIs of the MRD. With large sample sizes and medium/high event rates, Sandwich CIs exhibited relatively better performance in our simulations and were also computationally more efficient, therefore one could choose to use them in such scenarios. 
Our simulation results also highlighted how data sparsity issues which are inherent when implementing STE can greatly affect CI performance, meaning that one should carefully consider features of their data when choosing a CI method.

Although dependent censoring was not included in the data generating mechanism of our simulation study, we expect that the relative performance of CIs constructed using LEF bootstrap and the sandwich variance estimator observed in our simulation study would also hold if there were dependent censoring and IPCWs were used to handle it. Inference would still face the same issues, but with the additional uncertainty caused by the estimation of IPCWs and possible exacerbation of the finite-sample bias as a result of loss of information. 

We primarily focused on pooled logistic regressions for fitting MSMs and did not explore alternative survival time models, such as additive hazards models \cite{keogh_causal_2023}. 
Additive hazard models can be used in moderate-to-high event rates settings, while in low event rate settings they might result in negative hazard estimates. Inference procedures for STE based on additive hazard models warrant further research. 

 We have used parametric models for estimation of the SIPTCWs. Recent developments in nonparametric data-adaptive methods have led to their widespread use in biomedical research. This is partly because these models have become more accessible due to increased automation in their implementation across programming languages.
Because the consistency of the weighted estimators of the MSM parameters relies on correct specification of treatment and censoring models, data-adaptive methods are attractive. However, the subsequent inference can be challenging \cite{schuler_targeted_2017,chernozhukov_doubledebiased_2017,cai_nonparametric_2020, petersen_targeted_2014,zheng_doubly_2016,ertefaie_nonparametric_2023}.

Throughout this article, it was assumed that there were no unmeasured confounders. However, in practice, this assumption is unlikely to hold, as it is challenging to identify and measure all potential confounders in observational databases. Instrumental variable approaches and sensitivity analysis have been proposed to deal with such unmeasured confounding for point treatments. Recently, Tan (2023) proposed a sensitivity analysis approach in general longitudinal settings \cite{tan_sensitivity_2023}. In the specific setting of STE, sensitivity analysis methods for unmeasured confounding warrant further research. 

\section*{Acknowledgements}

The authors would like to thank Dr Brian Tom and Dr Pantelis Samartsidis for helpful comments and suggestions. Data from the HERS were collected under grant U64-CCU10675 from the U.S. Centers for Disease Control and Prevention. This work is supported by the U.K. Medical Research Council [grants MC\_UU\_00002/15, MC\_UU\_00040/03 and MC\_UU\_00040/05] and the MRC Biostatistics Unit Core Studentship.


\bibliographystyle{SageV}
\bibliography{references}

\end{document}


\maketitle
\section{Schematic illustration of sequential trial emulation} \label{appendix: sequential explanation}

\begin{figure}[h]
 \centering
 \includegraphics[width = \columnwidth]{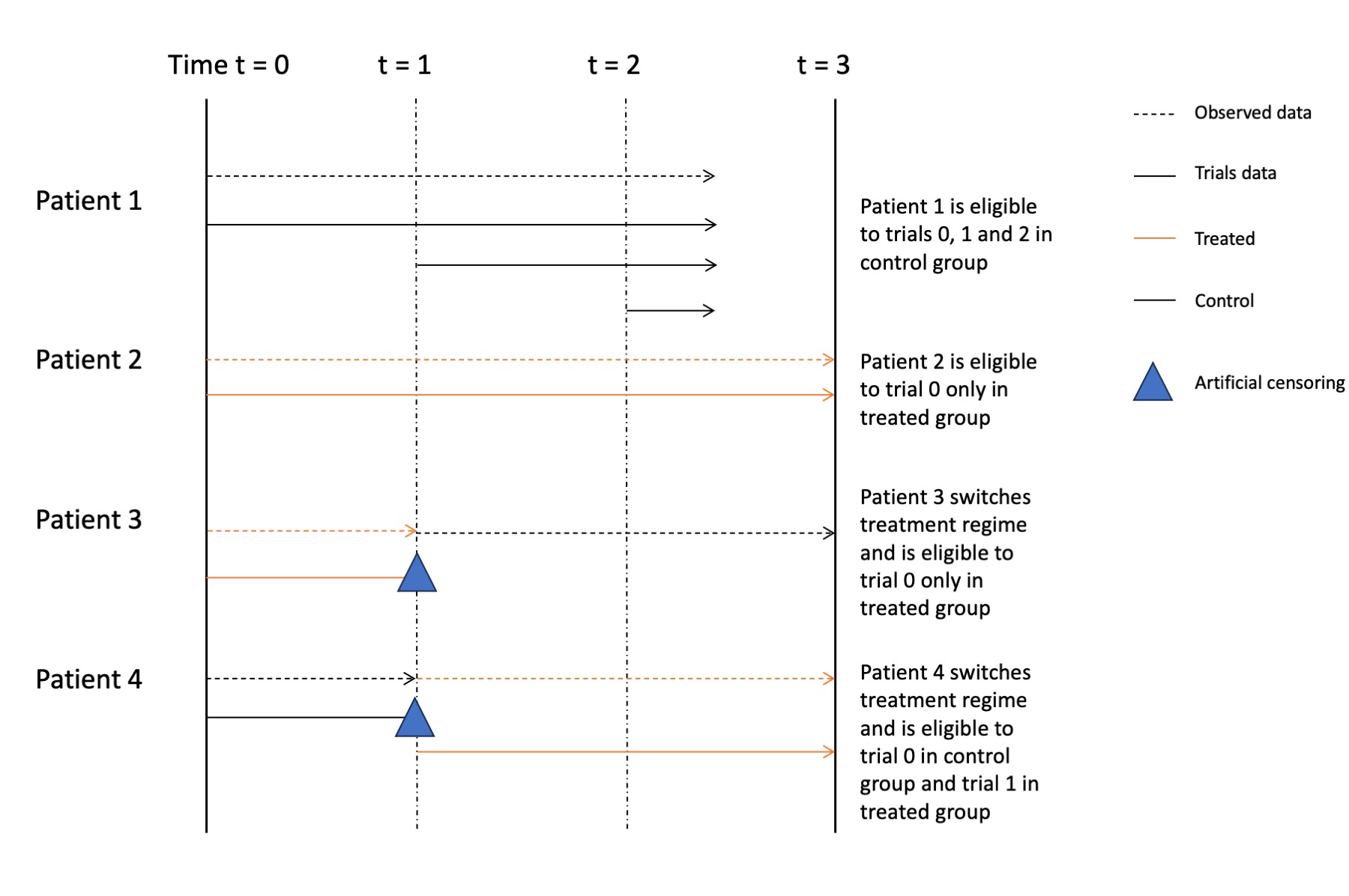}
 \caption{A schematic illustration of how patients'  data are  utilised in sequential trial emulation.}
 \label{appendix:trials_temp}
\end{figure}

Figure~\ref{appendix:trials_temp} provides a schematic illustration of how patients'  data are  utilised in  sequential trial emulation (STE). 
Three target trials, starting at visits at time $t = 0,1,2$, were sequentially emulated, which were referred  as `trial 0', `trial 1', and `trial 2'. Patient 1 never initiated treatment for the duration of the study until he/she was lost to follow-up after $t = 2$, so he/she/they was eligible as a control (marked by the black arrow) in all three trials. Patient 2 initiated treatment (marked by the orange arrow) at $t= 0$ and kept taking the treatment until the end of follow-up so he/she/they was in the treated group for trial 0 starting at $t = 0$, but not eligible for trials 1 and 2, starting at $ t = 1, 2$, respectively, under the criterion that patients were only eligible if there was no prior history of treatment. Patient 3 initiated treatment at $t = 0$ but stopped from $t = 1$, so his/her/their follow-up was artificially censored from $t = 1$. Patient 3 was included in the treatment group for trial 0, but his/her data was not included in other trials because he/she/they had a prior history of treatment. Patient 4 was not taking treatment at $t = 0$ but initiated treatment from $t = 1$ onwards: hence he/she/they was eligible as a control in trial 0 and artificially censored at the time he/she started treatment, $t = 1$, and he/she was eligible as a treated patient in trial 1 but not eligible in trial 2.

\section{Causal Assumptions} \label{appendix:causal assumptions}

\textit{No Interference.}

For $i \ne j$, patient $i$'s received treatment has no effect on patient $j$'s potential outcomes \cite{daniel_methods_2013}, i.e., a patient's event time is not affected by other patient's treatments.\\

\textit{Consistency.}

$\forall m, \:k, \:\overline Y_{m,k} = \overline Y_{m,k}^{\overline A_{m,k}}$ and $\overline L_{m,k} = \overline L_{m,k} ^{\overline A_{m,k}}$, meaning that the observed outcomes $\overline Y_{m,k}$ and covariates $\overline L_{m,k}$ are equal to their potential outcomes and covariates, under the treatment assignment which they actually received, for every trial visit $k$ in trial $m$ \cite{daniel_methods_2013,young_simulation_2014}.\\

\textit{Positivity.}

$\forall m, \:k, \: \Pr\{A_{m,k} = a \mid \overline A_{m,k-1} = \overline a_{k-1}, V, \overline L_{m,k}, Y_{m,k-1} = 0, C_{m,k-1} = 0\} >0$ and $\Pr\{C_{m,k} = 0 \mid C_{m,k-1} = 0, \overline A_{m,k} = \overline a_{k}, V, \overline L_{m,k}, Y_{m,k-1} = 0\} >0$, i.e., a patient  has non-zero probability of adhering to the treatment assigned and  remaining in the study at all time conditional on the observed histories of treatment and covariates \cite{young_simulation_2014}.\\

\textit{Sequentially ignorable treatment assignment.}

$\forall m, \:k, \: Y_{m,k}^{\overline a_k}
\indep A_{m,k} \mid V, \overline L_{m,k}, \overline A_{m,k-1}= \overline a_{k-1}, Y_{m,k-1} = 0, C_{m,k-1} = 0 $ for $ a =  0,  1$, i.e., at a given time, conditional on past treatment assignment and covariate history, there is no unmeasured confounding between the potential outcome and the current treatment received \cite{daniel_methods_2013}.\\

\textit{Sequentially ignorable loss to follow up.}

$\forall m, \:k, \: \overline C_{m,k} \indep Y_{m,k+1}^{\overline a_k}, Y_{m,k+2}^{\overline a_k}, \ldots \mid \overline C_{m,k-1}=0$, $Y_{m,k}=0$, $\overline A_{m,k}= \overline a_{k}$, $V$, $\overline L_{m,k}$, for $ a = 0, 1$ \cite{daniel_methods_2013}. 
 In other words, at a given time, a patient's probability of being under follow-up does not depend on their future risk of event, conditional on the treatment and covariate history up to that time.

\section{Inverse probability weighting}

For addressing artificial censoring due to treatment non-adherence, a patient's \textit{stabilised inverse probability of treatment weight} \cite{danaei_observational_2013, daniel_methods_2013,murray_causal_2021, robins_marginal_2000} at trial visit $k$ ($k>0$) in trial $m$ is defined as 
\begin{align}\label{SIPTW_m}
 sw^A_{m,k} = \frac{\prod_{j = 1}^k \Pr( A_{m,j}=a\mid\overline A_{m,j-1}= \overline a_{j-1},V, L_{m,0},E_m = 1, Y_{m,j-1} = 0, C_{m,j-1} = 0)}{\prod_{j = 1}^k \Pr( A_{m,j}= a\mid\overline A_{m, j-1}=\overline a_{j-1}, V, \overline L_{m,j},E_m = 1,Y_{m,j-1} = 0, C_{m,j-1} = 0)},   ~~~~~~~~~~a \in \{0,1\}.
\end{align} 
Here, $\Pr(A_{m,j}=a\mid\overline A_{m,j-1}=\overline a_{j-1}, V, \overline L_{m,j}, E_m = 1,Y_{m,j-1} = 0, C_{m,j-1} = 0)$ is the conditional probability that the patient’s treatment received at trial visit $j$ in trial $m$ remained the same as their treatment received up to trial visit $j-1$, conditional on their observed variables up to trial visit $j$. The baseline covariates $V$ and $L_{m,0}$ in the numerator of \eqref{SIPTW_m} 
are also included as covariates in the MSM of equation (3) of the main text \cite{daniel_methods_2013}. 

To address censoring due to loss to follow-up, the \textit{stabilised inverse probability of censoring weight} \cite{danaei_observational_2013, daniel_methods_2013,murray_causal_2021, robins_marginal_2000} for a patient at trial visit $k$ ($k>0$) in trial $m$ is 
\begin{align}\label{SIPCW_m}
 sw^C_{m,k} = \frac{\prod_{j = 0}^{k-1}\Pr(C_{m,j} =0\mid C_{m,j-1} = 0, \overline A_{m,j}= \overline a_{j}, V, L_{m,0}, E_m = 1, Y_{m,j} = 0)}{\Pi_{j = 0}^{k-1}\Pr(C_{m,j} =0 \mid C_{m,j-1} = 0,\overline A_{m,j}= \overline a_{j}, V, \overline L_{m,j}, E_m = 1, Y_{m,j} = 0)},  ~~~~~~~~~~~~~~~~~a \in \{0,1\}.
\end{align}
Here, $\Pr(C_{m,j} =0\mid\overline C_{m,j-1} = 0,\overline A_{m,j} = \overline a_{j}, V, \overline L_{m,j}, E_m = 1, Y_{m,j} = 0)$ is the conditional probability that the patient remains in the trial during $[t_{m,j+1}, t_{m,j+2})$ given that they had been in the trial in $[0, t_{m,j+1})$ and their observed covariate history up to time $t_{m,j}$.
Again, whenever we condition on certain variables in the numerator of~\eqref{SIPCW_m}, these variables are also be included as covariates in the MSM of equation (3) of the main text.

\section{More details for  key functions of the \texttt{TrialEmulation} package} \label{appendix: TrialEmulation}

\begin{itemize}
 \item \texttt{data\_preparation}: takes longitudinal data and the specifications  of the treatment and censoring models as input, and returns an expanded dataset with patients' data spread across their eligible sequentially emulated trials. This dataset includes a column containing the patients' SIPTCWs as calculated from the formulae (4) and (5), according to whether we are conducting an intention-to-treat or per-protocol analysis. The modelling fitting required for calculating SIPTCWs was performed using the original longitudinal data with the variables specified in the function input. For the per-protocol analysis, the  expanded data have been artificially censored whenever patients stop adhering to their assigned treatment strategies in each emulated trial. This function implements Steps 1 and 2 of MRD estimation  described at the end of Section 3 of the main text.
 \item \texttt{trial\_msm}: takes the expanded data with SIPTCWs such as one returned from \texttt{data\_preparation} and the MSM specification such as in equation (3) in the main text as inputs and produces a \texttt{fit.glm} type output of the MSM with estimated hazard ratios and the sandwich variance matrix estimate. This function implements Step 3 of MRD estimation  described at the end of Section 3 of the main text. We can use \texttt{predict.glm} on the output of this function to estimate marginal risks for a target population with given baseline covariate data, which is what we use to carry out Step 5 of the MRD estimation. 
 \item \texttt{initiators}: a wrapper function combining the two functions above.
\end{itemize}

The \texttt{data\_preparation} function in the \texttt{TrialEmulation}  package was used to estimate the stabilised IPTWs from the simulated datasets, and this function outputs the expanded and artificially censored data in  the sequential trials as described in Section 3 of the main text, with each patient's inverse probability weights at each trial's follow-up visit. The MSM for the discrete-time counterfactual hazard was fitted using the  \texttt{trial\_msm} function.

The sandwich variance matrix estimate is provided in the output of \texttt{trial\_msm} and was used to obtain Sandwich CIs. The code for implementing  nonparametric bootstrap CIs,   LEF bootstrap CIs and jackknife CIs, were not included in the \texttt{TrialEmulation} package so they were created for the purpose of this article and can be found in the GitHub repository \url{https://github.com/juliettelimozin/Multiple-trial-emulation-IPTW-MSM-CIs}.
More details on the \texttt{TrialEmulation} package can be found at \url{https://cran.r-project.org/web/packages/TrialEmulation/TrialEmulation.pdf}

\clearpage
\section{Pseudo code for the simulation study} \label{appendix: pseudo code}

Algorithm~\ref{alg:cap} presents a pseudo-code of the simulation study in Section 5 of the main text. The R code can be found at \url{https://github.com/juliettelimozin/Multiple-trial-emulation-IPTW-MSM-CIs}.

\begin{algorithm}[ht]
\caption{Pseudo-code of Monte Carlo simulation algorithm.}\label{alg:cap}
\begin{algorithmic}
\For{MC iteration $m= 1,2,...,1000$}
\For{Simulation scenario $l = 1,2,...,81$}
\State Data A $\gets$ Generate data with scenario $l$
\State Select eligible patients 
\State Fit  models for the denominator and numerator terms of  stabilised IPTWs  to eligible patients' data up to the first visit they switched treatment after trial baseline.
\State Data B $\gets$ the original dataset expanded into data for sequentially emulated trials and assign patients to eligible trials
\For{Each trial} 
\State Artificially censor patients' follow-up when they no longer adhere to the  treatment assigned at the trial baseline
\State Estimate each patient's stabilised IPTW based on covariate history in this trial. At trial baseline, the weights are always one.
\EndFor
\State Fit weighted pooled logistic regression to Data B 
\State $v \gets $ MRD estimate for trial 0 patients
\State $[B_1,B_2] \gets$ lower- and upper-bound of nonparametric bootstrap pivot CI of MRD
\State $[LEF^{(1)}_1,LEF^{(1)}_2] \gets$ lower- and upper-bound of LEF bootstrap pivot CI of MRD using approach 1
\State $[LEF^{(2)}_1,LEF^{(2)}_2] \gets$ lower- and upper-bound of LEF bootstrap pivot CI of MRD using approach 2
\State $[S_1,S_2] \gets$ lower- and upper-bound of the sandwich-variance-estimator-based CI of MRD
\State $[J^{(1)}_1,J^{(1)}_2] \gets$ lower- and upper-bound of jackknife Wald CI of MRD
\State $[J^{(2)}_1,J^{(2)}_2] \gets$ lower- and upper-bound of jackknife MVN CI of MRD
\EndFor
\EndFor
\State \Return{$v, [B_1,B_2],[LEF^{(1)}_1,LEF^{(1)}_2],[LEF^{(2)}_1,LEF^{(2)}_2],[S_1,S_2],[J^{(1)}_1,J^{(1)}_2],[J^{(2)}_1,J^{(2)}_2]$}
\end{algorithmic}
\end{algorithm}

\clearpage

\section{Additional simulation results}

\subsection{Data tabulation of sequentially emulated trials in simulations  }

Tables~\ref{data tabulation low 200}--\ref{data tabulation high 5000} provide examples of data  prepared for sequentially emulated trials using simulated data in Section 3 of the  main text.  Data are aggregated across trials and tabulated by treatment arm, trial visit and outcome status in various simulation scenarios.  

\begin{table}[htbp]
  \centering
  \caption{Data tabulation by treatment arm and outcome status when generating one dataset using the data generating algorithm in Table 2 of the Main Text for \textbf{small sample sizes ($n = 200$) under low event rate}.}
  \resizebox{\textwidth}{!}{%
  \begin{tabular}{*{5}{l}*{5}{r}}
    \toprule
     &  & &  &   \hspace{2cm} Trial visit &  & &  &  &  \\
    Sample size & Confounding strength & Treatment prevalence & Assigned treatment & Outcome \hspace{2cm} & \multicolumn{1}{c}{0} & \multicolumn{1}{c}{1} & \multicolumn{1}{c}{2} & \multicolumn{1}{c}{3} & \multicolumn{1}{c}{4} \\
    \midrule
    200         & 0.1  & -1    & 0                  & 0                              & 409 & 269 & 172 &  98 &  43 \\ 
              &      &       &                    & 1                              &   5 &   4 &   1 &   0 &   0 \\ 
              &      &       & 1                  & 0                              & 148 &  34 &  11 &   2 &   1 \\ 
              &      &       &                    & 1                              &   4 &   2 &   0 &   0 &   0 \\ 
              &      & 0     & 0                  & 0                              & 162 &  77 &  37 &  16 &   5 \\ 
              &      &       &                    & 1                              &   5 &   2 &   1 &   0 &   0 \\ 
              &      &       & 1                  & 0                              & 190 &  89 &  45 &  25 &   7 \\ 
              &      &       &                    & 1                              &   0 &   0 &   0 &   1 &   0 \\ 
              &      & 1     & 0                  & 0                              &  85 &  29 &   8 &   3 &   1 \\ 
              &      &       &                    & 1                              &   0 &   0 &   0 &   0 &   0 \\ 
              &      &       & 1                  & 0                              & 198 & 149 & 108 &  74 &  42 \\ 
              &      &       &                    & 1                              &   1 &   2 &   0 &   0 &   0 \\ 
              & 0.5  & -1    & 0                  & 0                              & 400 & 254 & 156 &  84 &  34 \\ 
              &      &       &                    & 1                              &   3 &   2 &   0 &   0 &   0 \\ 
              &      &       & 1                  & 0                              & 162 &  40 &  11 &   2 &   0 \\ 
              &      &       &                    & 1                              &   1 &   2 &   1 &   0 &   0 \\ 
              &      & 0     & 0                  & 0                              & 200 &  98 &  47 &  21 &   8 \\ 
              &      &       &                    & 1                              &   2 &   1 &   0 &   0 &   0 \\ 
              &      &       & 1                  & 0                              & 187 & 101 &  47 &  24 &  12 \\ 
              &      &       &                    & 1                              &   3 &   0 &   1 &   0 &   0 \\ 
              &      & 1     & 0                  & 0                              &  81 &  29 &  11 &   5 &   2 \\ 
              &      &       &                    & 1                              &   0 &   0 &   0 &   0 &   0 \\ 
              &      &       & 1                  & 0                              & 196 & 130 &  95 &  64 &  36 \\ 
              &      &       &                    & 1                              &   2 &   1 &   0 &   2 &   0 \\ 
              & 0.9  & -1    & 0                  & 0                              & 397 & 257 & 156 &  86 &  37 \\ 
              &      &       &                    & 1                              &   1 &   1 &   0 &   0 &   0 \\ 
              &      &       & 1                  & 0                              & 160 &  34 &  11 &   2 &   0 \\ 
              &      &       &                    & 1                              &   2 &   1 &   0 &   0 &   0 \\ 
              &      & 0     & 0                  & 0                              & 194 &  99 &  49 &  25 &  10 \\ 
              &      &       &                    & 1                              &   2 &   2 &   1 &   0 &   0 \\ 
              &      &       & 1                  & 0                              & 185 &  80 &  36 &  16 &   5 \\ 
              &      &       &                    & 1                              &   3 &   2 &   1 &   0 &   0 \\ 
              &      & 1     & 0                  & 0                              &  97 &  30 &   9 &   4 &   2 \\ 
              &      &       &                    & 1                              &   0 &   0 &   0 &   0 &   0 \\ 
              &      &       & 1                  & 0                              & 197 & 150 &  97 &  58 &  28 \\ 
              &      &       &                    & 1                              &   1 &   2 &   1 &   0 &   0 \\
    \bottomrule
  \end{tabular}
  }
  
  \label{data tabulation low 200}
\end{table}

\begin{table}[htbp]
  \centering
  \caption{Data tabulation by treatment arm and outcome status when generating one dataset using the data generating algorithm in Table 2 of the Main Text for \textbf{medium sample sizes ($n = 1000$) under low event rate}.}
  \resizebox{\textwidth}{!}{%
  \begin{tabular}{*{5}{l}*{5}{r}}
    \toprule
     &  & &  &   \hspace{2cm} Trial visit &  & &  &  &  \\
    Sample size & Confounding strength & Treatment prevalence & Assigned treatment & Outcome \hspace{2cm} & \multicolumn{1}{c}{0} & \multicolumn{1}{c}{1} & \multicolumn{1}{c}{2} & \multicolumn{1}{c}{3} & \multicolumn{1}{c}{4} \\
    \midrule
    1000        & 0.1  & -1    & 0                  & 0                              & 2058 & 1357 &  839 &  462 &  193 \\ 
              &      &       &                    & 1                              &   35 &   24 &   16 &    9 &    5 \\ 
              &      &       & 1                  & 0                              &  765 &  188 &   41 &    7 &    2 \\ 
              &      &       &                    & 1                              &    7 &    2 &    2 &    0 &    0 \\ 
              &      & 0     & 0                  & 0                              &  952 &  480 &  239 &  111 &   42 \\ 
              &      &       &                    & 1                              &   13 &    6 &    3 &    0 &    0 \\ 
              &      &       & 1                  & 0                              &  937 &  473 &  231 &   91 &   42 \\ 
              &      &       &                    & 1                              &    8 &    5 &    3 &    2 &    0 \\ 
              &      & 1     & 0                  & 0                              &  357 &   81 &   13 &    2 &    0 \\ 
              &      &       &                    & 1                              &    6 &    0 &    0 &    0 &    0 \\ 
              &      &       & 1                  & 0                              &  991 &  727 &  536 &  358 &  222 \\ 
              &      &       &                    & 1                              &    3 &    7 &    5 &    5 &    0 \\ 
              & 0.5  & -1    & 0                  & 0                              & 1960 & 1265 &  758 &  412 &  173 \\ 
              &      &       &                    & 1                              &   27 &   20 &   10 &    6 &    1 \\ 
              &      &       & 1                  & 0                              &  790 &  196 &   45 &   10 &    2 \\ 
              &      &       &                    & 1                              &   10 &    5 &    0 &    0 &    0 \\ 
              &      & 0     & 0                  & 0                              &  977 &  464 &  211 &   83 &   29 \\ 
              &      &       &                    & 1                              &   12 &    6 &    2 &    2 &    1 \\ 
              &      &       & 1                  & 0                              &  951 &  450 &  183 &   82 &   26 \\ 
              &      &       &                    & 1                              &    8 &    4 &    4 &    0 &    1 \\ 
              &      & 1     & 0                  & 0                              &  398 &  114 &   31 &   10 &    2 \\ 
              &      &       &                    & 1                              &    6 &    2 &    1 &    0 &    0 \\ 
              &      &       & 1                  & 0                              &  978 &  687 &  462 &  304 &  168 \\ 
              &      &       &                    & 1                              &   14 &    3 &    6 &    2 &    2 \\ 
              & 0.9  & -1    & 0                  & 0                              & 1897 & 1215 &  730 &  394 &  161 \\ 
              &      &       &                    & 1                              &   24 &    9 &    8 &    4 &    0 \\ 
              &      &       & 1                  & 0                              &  799 &  189 &   49 &    8 &    1 \\ 
              &      &       &                    & 1                              &   16 &    2 &    0 &    0 &    0 \\ 
              &      & 0     & 0                  & 0                              &  968 &  475 &  219 &   83 &   26 \\ 
              &      &       &                    & 1                              &   17 &    6 &    2 &    0 &    0 \\ 
              &      &       & 1                  & 0                              &  939 &  414 &  170 &   68 &   19 \\ 
              &      &       &                    & 1                              &   18 &    7 &    1 &    1 &    0 \\ 
              &      & 1     & 0                  & 0                              &  436 &  134 &   43 &   13 &    2 \\ 
              &      &       &                    & 1                              &    0 &    0 &    0 &    0 &    0 \\ 
              &      &       & 1                  & 0                              &  987 &  633 &  406 &  250 &  131 \\ 
              &      &       &                    & 1                              &   11 &    5 &    3 &    8 &    2 \\ 
    \bottomrule
  \end{tabular}
  }
  
  \label{data tabulation low 1000}
\end{table}
\begin{table}[htbp]
  \centering
  \caption{Data tabulation by treatment arms and outcome status when generating one dataset using the data generating algorithm in Table 2 of the Main Text for \textbf{large sample sizes ($n = 5000$) under low event rate}.}
  \resizebox{\textwidth}{!}{%
  \begin{tabular}{*{5}{l}*{5}{r}}
    \toprule
     &  & &  &   \hspace{2cm} Trial visit &  & &  &  &  \\
    Sample size & Confounding strength & Treatment prevalence & Assigned treatment & Outcome \hspace{2cm} & \multicolumn{1}{c}{0} & \multicolumn{1}{c}{1} & \multicolumn{1}{c}{2} & \multicolumn{1}{c}{3} & \multicolumn{1}{c}{4} \\
    \midrule
    5000        & 0.1  & -1    & 0                  & 0                              & 10181 &  6601 &  4075 &  2273 &   961 \\ 
              &      &       &                    & 1                              &   119 &    70 &    43 &    23 &    13 \\ 
              &      &       & 1                  & 0                              &  3891 &   976 &   250 &    49 &    12 \\ 
              &      &       &                    & 1                              &    29 &     6 &     2 &     2 &     1 \\ 
              &      & 0     & 0                  & 0                              &  4949 &  2446 &  1171 &   522 &   177 \\ 
              &      &       &                    & 1                              &    52 &    22 &    12 &     8 &     3 \\ 
              &      &       & 1                  & 0                              &  4732 &  2251 &  1052 &   477 &   145 \\ 
              &      &       &                    & 1                              &    39 &    18 &    13 &     2 &     2 \\ 
              &      & 1     & 0                  & 0                              &  1932 &   557 &   168 &    47 &    11 \\ 
              &      &       &                    & 1                              &    27 &     6 &     1 &     0 &     0 \\ 
              &      &       & 1                  & 0                              &  4912 &  3531 &  2545 &  1797 &  1028 \\ 
              &      &       &                    & 1                              &    50 &    20 &    31 &    11 &    14 \\ 
              & 0.5  & -1    & 0                  & 0                              & 10134 &  6539 &  3992 &  2193 &   916 \\ 
              &      &       &                    & 1                              &   135 &    85 &    50 &    24 &    11 \\ 
              &      &       & 1                  & 0                              &  3902 &   934 &   201 &    36 &     3 \\ 
              &      &       &                    & 1                              &    47 &     5 &     4 &     0 &     0 \\ 
              &      & 0     & 0                  & 0                              &  4858 &  2388 &  1144 &   517 &   185 \\ 
              &      &       &                    & 1                              &    65 &    32 &    16 &     6 &     2 \\ 
              &      &       & 1                  & 0                              &  4698 &  2151 &   973 &   406 &   127 \\ 
              &      &       &                    & 1                              &    52 &    16 &    17 &     7 &     5 \\ 
              &      & 1     & 0                  & 0                              &  1931 &   545 &   153 &    42 &    10 \\ 
              &      &       &                    & 1                              &    30 &     6 &     2 &     0 &     0 \\ 
              &      &       & 1                  & 0                              &  4911 &  3426 &  2365 &  1568 &   865 \\ 
              &      &       &                    & 1                              &    49 &    30 &    20 &    17 &    14 \\ 
              & 0.9  & -1    & 0                  & 0                              &  9673 &  6170 &  3729 &  2051 &   842 \\ 
              &      &       &                    & 1                              &   151 &   100 &    60 &    37 &    17 \\ 
              &      &       & 1                  & 0                              &  3922 &   925 &   205 &    50 &    10 \\ 
              &      &       &                    & 1                              &    85 &    10 &     2 &     0 &     0 \\ 
              &      & 0     & 0                  & 0                              &  4695 &  2267 &  1050 &   462 &   151 \\ 
              &      &       &                    & 1                              &    55 &    27 &     9 &     4 &     1 \\ 
              &      &       & 1                  & 0                              &  4727 &  1986 &   846 &   336 &   104 \\ 
              &      &       &                    & 1                              &    67 &    29 &     9 &     7 &     0 \\ 
              &      & 1     & 0                  & 0                              &  2183 &   644 &   196 &    53 &    15 \\ 
              &      &       &                    & 1                              &    20 &     7 &     0 &     0 &     0 \\ 
              &      &       & 1                  & 0                              &  4894 &  3147 &  1985 &  1231 &   603 \\ 
              &      &       &                    & 1                              &    71 &    42 &    34 &    18 &    15 \\ 
    \bottomrule
  \end{tabular}
  }
  
  \label{data tabulation low 5000}
\end{table}

\begin{table}[htbp]
  \centering
  \caption{Data tabulation by treatment arms and outcome status when generating one dataset using the data generating algorithm in Table 2 of the Main Text for \textbf{small sample sizes ($n = 200$) under medium event rate}.}
  \resizebox{\textwidth}{!}{%
  \begin{tabular}{*{5}{l}*{5}{r}}
    \toprule
     &  & &  &   \hspace{2cm} Trial visit &  & &  &  &  \\
    Sample size & Confounding strength & Treatment prevalence & Assigned treatment & Outcome \hspace{2cm} & \multicolumn{1}{c}{0} & \multicolumn{1}{c}{1} & \multicolumn{1}{c}{2} & \multicolumn{1}{c}{3} & \multicolumn{1}{c}{4} \\
    \midrule
    200         & 0.1  & -1    & 0                  & 0                              & 391 & 253 & 157 &  87 &  36 \\ 
              &      &       &                    & 1                              &  12 &   8 &   4 &   3 &   1 \\ \addlinespace[3pt]
              &      &       & 1                  & 0                              & 150 &  32 &   7 &   1 &   0 \\ 
              &      &       &                    & 1                              &   2 &   1 &   0 &   0 &   0 \\ \addlinespace[6pt]
              &      & 0     & 0                  & 0                              & 183 &  89 &  47 &  22 &   8 \\ 
              &      &       &                    & 1                              &   8 &   6 &   2 &   2 &   0 \\ \addlinespace[3pt]
              &      &       & 1                  & 0                              & 179 &  87 &  40 &  14 &   6 \\ 
              &      &       &                    & 1                              &   5 &   4 &   1 &   0 &   0 \\ \addlinespace[6pt]
              &      & 1     & 0                  & 0                              &  87 &  32 &  13 &   4 &   1 \\ 
              &      &       &                    & 1                              &   2 &   0 &   0 &   0 &   0 \\ \addlinespace[3pt]
              &      &       & 1                  & 0                              & 191 & 138 & 102 &  71 &  46 \\ 
              &      &       &                    & 1                              &   6 &   3 &   1 &   1 &   1 \\ \addlinespace[9pt]
              & 0.5  & -1    & 0                  & 0                              & 417 & 276 & 177 & 100 &  44 \\ 
              &      &       &                    & 1                              &  14 &   8 &   4 &   3 &   0 \\ \addlinespace[3pt]
              &      &       & 1                  & 0                              & 139 &  29 &   7 &   2 &   0 \\ 
              &      &       &                    & 1                              &   3 &   0 &   0 &   1 &   0 \\ \addlinespace[6pt]
              &      & 0     & 0                  & 0                              & 196 & 102 &  53 &  29 &  11 \\ 
              &      &       &                    & 1                              &   4 &   3 &   2 &   0 &   0 \\ \addlinespace[3pt]
              &      &       & 1                  & 0                              & 179 &  82 &  25 &   8 &   0 \\ 
              &      &       &                    & 1                              &   6 &   0 &   2 &   0 &   0 \\ \addlinespace[6pt]
              &      & 1     & 0                  & 0                              &  70 &  18 &   5 &   1 &   0 \\ 
              &      &       &                    & 1                              &   2 &   0 &   0 &   0 &   0 \\ \addlinespace[3pt]
              &      &       & 1                  & 0                              & 195 & 134 & 104 &  71 &  37 \\ 
              &      &       &                    & 1                              &   3 &   4 &   0 &   1 &   0 \\ \addlinespace[9pt]
              & 0.9  & -1    & 0                  & 0                              & 357 & 226 & 136 &  72 &  32 \\ 
              &      &       &                    & 1                              &   6 &   3 &   3 &   2 &   0 \\ \addlinespace[3pt]
              &      &       & 1                  & 0                              & 156 &  46 &   8 &   2 &   1 \\ 
              &      &       &                    & 1                              &   6 &   4 &   1 &   0 &   0 \\ \addlinespace[6pt]
              &      & 0     & 0                  & 0                              & 178 &  79 &  36 &  13 &   3 \\ 
              &      &       &                    & 1                              &   5 &   4 &   0 &   0 &   0 \\ \addlinespace[3pt]
              &      &       & 1                  & 0                              & 181 &  80 &  34 &  14 &   2 \\ 
              &      &       &                    & 1                              &  11 &   3 &   4 &   0 &   0 \\ \addlinespace[6pt]
              &      & 1     & 0                  & 0                              &  81 &  27 &   9 &   4 &   0 \\ 
              &      &       &                    & 1                              &   1 &   0 &   0 &   0 &   0 \\ \addlinespace[3pt]
              &      &       & 1                  & 0                              & 190 & 122 &  80 &  41 &  21 \\ 
              &      &       &                    & 1                              &   9 &   5 &   3 &   1 &   1 \\ 
    \bottomrule
  \end{tabular}
  }
  
  \label{data tabulation med 200}
\end{table}

\begin{table}[htbp]
  \centering
  \caption{Data tabulation by treatment arms and outcome status when generating one dataset using the data generating algorithm in Table 2 of the Main Text for \textbf{medium sample sizes ($n = 1000$) under medium event rate}.}
  \resizebox{\textwidth}{!}{%
  \begin{tabular}{*{5}{l}*{5}{r}}
    \toprule
     &  & &  &   \hspace{2cm} Trial visit &  & &  &  &  \\
    Sample size & Confounding strength & Treatment prevalence & Assigned treatment & Outcome \hspace{2cm} & \multicolumn{1}{c}{0} & \multicolumn{1}{c}{1} & \multicolumn{1}{c}{2} & \multicolumn{1}{c}{3} & \multicolumn{1}{c}{4} \\
    \midrule
  1000        & 0.1  & -1    & 0                  & 0                              & 2012 & 1321 &  825 &  460 &  194 \\ 
              &      &       &                    & 1                              &   66 &   40 &   20 &   11 &    3 \\ 
              &      &       & 1                  & 0                              &  722 &  176 &   37 &    6 &    1 \\ 
              &      &       &                    & 1                              &   18 &    8 &    1 &    0 &    0 \\ 
              &      & 0     & 0                  & 0                              & 1002 &  505 &  257 &  127 &   44 \\ 
              &      &       &                    & 1                              &   32 &   17 &    5 &    1 &    0 \\ 
              &      &       & 1                  & 0                              &  902 &  412 &  179 &   73 &   25 \\ 
              &      &       &                    & 1                              &   22 &    8 &    4 &    0 &    0 \\ 
              &      & 1     & 0                  & 0                              &  360 &   86 &   20 &    4 &    1 \\ 
              &      &       &                    & 1                              &    9 &    3 &    1 &    0 &    0 \\ 
              &      &       & 1                  & 0                              &  972 &  669 &  453 &  312 &  182 \\ 
              &      &       &                    & 1                              &   18 &   19 &   10 &    9 &    5 \\ 
              & 0.5  & -1    & 0                  & 0                              & 1997 & 1275 &  772 &  417 &  170 \\ 
              &      &       &                    & 1                              &   52 &   28 &   16 &    9 &    1 \\ 
              &      &       & 1                  & 0                              &  752 &  172 &   37 &   10 &    2 \\ 
              &      &       &                    & 1                              &   26 &    5 &    3 &    0 &    0 \\ 
              &      & 0     & 0                  & 0                              &  997 &  506 &  249 &  113 &   37 \\ 
              &      &       &                    & 1                              &   22 &    9 &    3 &    2 &    2 \\ 
              &      &       & 1                  & 0                              &  913 &  397 &  173 &   76 &   31 \\ 
              &      &       &                    & 1                              &   28 &   11 &    0 &    3 &    2 \\ 
              &      & 1     & 0                  & 0                              &  402 &  117 &   31 &    6 &    1 \\ 
              &      &       &                    & 1                              &   16 &    2 &    1 &    0 &    0 \\ 
              &      &       & 1                  & 0                              &  951 &  642 &  423 &  272 &  144 \\ 
              &      &       &                    & 1                              &   32 &   16 &   11 &    5 &    5 \\ 
              & 0.9  & -1    & 0                  & 0                              & 1742 & 1093 &  662 &  363 &  151 \\ 
              &      &       &                    & 1                              &   65 &   41 &   21 &   13 &    5 \\ 
              &      &       & 1                  & 0                              &  755 &  179 &   48 &   11 &    3 \\ 
              &      &       &                    & 1                              &   29 &   11 &    1 &    1 &    0 \\ 
              &      & 0     & 0                  & 0                              &  937 &  453 &  223 &  108 &   41 \\ 
              &      &       &                    & 1                              &   16 &    7 &    3 &    1 &    1 \\ 
              &      &       & 1                  & 0                              &  905 &  409 &  177 &   56 &   13 \\ 
              &      &       &                    & 1                              &   38 &   12 &    5 &    2 &    0 \\ 
              &      & 1     & 0                  & 0                              &  409 &  113 &   33 &    6 &    1 \\ 
              &      &       &                    & 1                              &    9 &    2 &    0 &    0 &    0 \\ 
              &      &       & 1                  & 0                              &  952 &  608 &  395 &  236 &  121 \\ 
              &      &       &                    & 1                              &   38 &   16 &   15 &    9 &    6 \\ 
    \bottomrule
  \end{tabular}
  }
  
  \label{data tabulation med 1000}
\end{table}

\begin{table}[htbp]
  \centering
  \caption{Data tabulation by treatment arms and outcome status when generating one dataset using the data generating algorithm in Table 2 of the Main Text for \textbf{large sample sizes ($n = 5000$) under medium event rate}.}
  \resizebox{\textwidth}{!}{%
  \begin{tabular}{*{5}{l}*{5}{r}}
    \toprule
     &  & &  &   \hspace{2cm} Trial visit &  & &  &  &  \\
    Sample size & Confounding strength & Treatment prevalence & Assigned treatment & Outcome \hspace{2cm} & \multicolumn{1}{c}{0} & \multicolumn{1}{c}{1} & \multicolumn{1}{c}{2} & \multicolumn{1}{c}{3} & \multicolumn{1}{c}{4} \\
    \midrule
5000        & 0.1  & -1    & 0                  & 0                              & 9935 & 6440 & 3976 & 2209 &  934 \\ 
              &      &       &                    & 1                              &  298 &  178 &   98 &   49 &   16 \\ 
              &      &       & 1                  & 0                              & 3679 &  911 &  215 &   42 &    7 \\ 
              &      &       &                    & 1                              &   89 &   24 &    7 &    1 &    0 \\ 
              &      & 0     & 0                  & 0                              & 4736 & 2298 & 1076 &  472 &  161 \\ 
              &      &       &                    & 1                              &  127 &   48 &   22 &   13 &    3 \\ 
              &      &       & 1                  & 0                              & 4624 & 2257 & 1037 &  445 &  132 \\ 
              &      &       &                    & 1                              &   88 &   55 &   18 &   15 &    1 \\ 
              &      & 1     & 0                  & 0                              & 1741 &  454 &  125 &   35 &    7 \\ 
              &      &       &                    & 1                              &   59 &   10 &    2 &    0 &    0 \\ 
              &      &       & 1                  & 0                              & 4826 & 3458 & 2427 & 1645 &  913 \\ 
              &      &       &                    & 1                              &  108 &   58 &   61 &   40 &   29 \\ 
              & 0.5  & -1    & 0                  & 0                              & 9471 & 6047 & 3655 & 1988 &  809 \\ 
              &      &       &                    & 1                              &  299 &  180 &  117 &   57 &   26 \\ 
              &      &       & 1                  & 0                              & 3785 &  904 &  224 &   48 &    5 \\ 
              &      &       &                    & 1                              &  107 &   21 &    6 &    1 &    1 \\ 
              &      & 0     & 0                  & 0                              & 4727 & 2298 & 1077 &  477 &  164 \\ 
              &      &       &                    & 1                              &  129 &   50 &   30 &   11 &    3 \\ 
              &      &       & 1                  & 0                              & 4570 & 2069 &  906 &  359 &  125 \\ 
              &      &       &                    & 1                              &  137 &   47 &   25 &   12 &    3 \\ 
              &      & 1     & 0                  & 0                              & 1846 &  497 &  124 &   32 &    9 \\ 
              &      &       &                    & 1                              &   47 &    8 &    2 &    0 &    0 \\ 
              &      &       & 1                  & 0                              & 4820 & 3247 & 2179 & 1380 &  748 \\ 
              &      &       &                    & 1                              &  124 &   64 &   45 &   32 &   20 \\ 
              & 0.9  & -1    & 0                  & 0                              & 8950 & 5602 & 3328 & 1769 &  710 \\ 
              &      &       &                    & 1                              &  297 &  165 &   93 &   55 &   24 \\ 
              &      &       & 1                  & 0                              & 3813 &  907 &  201 &   38 &    7 \\ 
              &      &       &                    & 1                              &  180 &   30 &    6 &    3 &    3 \\ 
              &      & 0     & 0                  & 0                              & 4795 & 2318 & 1124 &  485 &  166 \\ 
              &      &       &                    & 1                              &  124 &   47 &   19 &    7 &    1 \\ 
              &      &       & 1                  & 0                              & 4547 & 1917 &  805 &  304 &   84 \\ 
              &      &       &                    & 1                              &  163 &   80 &   14 &   13 &    8 \\ 
              &      & 1     & 0                  & 0                              & 2041 &  615 &  192 &   58 &   16 \\ 
              &      &       &                    & 1                              &   47 &   10 &    6 &    3 &    1 \\ 
              &      &       & 1                  & 0                              & 4775 & 3030 & 1941 & 1185 &  592 \\ 
              &      &       &                    & 1                              &  162 &   81 &   50 &   41 &   20 \\ 
    \bottomrule
  \end{tabular}
  }
  
  \label{data tabulation med 5000}
\end{table}

\begin{table}[htbp]
  \centering
  \caption{Data tabulation by treatment arms and outcome status when generating one dataset using the data generating algorithm in Table 2 of the Main Text for \textbf{small sample sizes ($n = 200$) under high event rate}.}
  \resizebox{\textwidth}{!}{%
  \begin{tabular}{*{5}{l}*{5}{r}}
    \toprule
     &  & &  &   \hspace{2cm} Trial visit &  & &  &  &  \\
    Sample size & Confounding strength & Treatment prevalence & Assigned treatment & Outcome \hspace{2cm} & \multicolumn{1}{c}{0} & \multicolumn{1}{c}{1} & \multicolumn{1}{c}{2} & \multicolumn{1}{c}{3} & \multicolumn{1}{c}{4} \\
    \midrule
200         & 0.1  & -1    & 0                  & 0                              & 345 & 213 & 126 &  67 &  28 \\ 
              &      &       &                    & 1                              &  17 &  13 &   7 &   4 &   1 \\ 
              &      &       & 1                  & 0                              & 148 &  32 &  14 &   3 &   1 \\ 
              &      &       &                    & 1                              &   7 &   2 &   1 &   1 &   0 \\ 
              &      & 0     & 0                  & 0                              & 187 &  95 &  46 &  20 &   5 \\ 
              &      &       &                    & 1                              &  12 &   2 &   2 &   0 &   0 \\ 
              &      &       & 1                  & 0                              & 180 &  83 &  33 &  14 &   5 \\ 
              &      &       &                    & 1                              &   3 &   5 &   2 &   2 &   1 \\ 
              &      & 1     & 0                  & 0                              &  71 &  20 &   6 &   1 &   0 \\ 
              &      &       &                    & 1                              &   3 &   0 &   0 &   0 &   0 \\ 
              &      &       & 1                  & 0                              & 188 & 121 &  83 &  49 &  29 \\ 
              &      &       &                    & 1                              &   9 &   8 &   4 &   3 &   1 \\ 
              & 0.5  & -1    & 0                  & 0                              & 354 & 226 & 140 &  80 &  34 \\ 
              &      &       &                    & 1                              &  23 &  14 &   9 &   5 &   2 \\ 
              &      &       & 1                  & 0                              & 135 &  31 &   5 &   2 &   0 \\ 
              &      &       &                    & 1                              &   8 &   1 &   1 &   0 &   0 \\ 
              &      & 0     & 0                  & 0                              & 192 &  96 &  45 &  19 &   7 \\ 
              &      &       &                    & 1                              &  13 &   4 &   1 &   1 &   0 \\ 
              &      &       & 1                  & 0                              & 173 &  81 &  36 &   7 &   2 \\ 
              &      &       &                    & 1                              &   7 &   4 &   1 &   1 &   0 \\ 
              &      & 1     & 0                  & 0                              &  55 &  12 &   5 &   1 &   0 \\ 
              &      &       &                    & 1                              &   1 &   0 &   0 &   0 &   0 \\ 
              &      &       & 1                  & 0                              & 188 & 127 &  82 &  55 &  36 \\ 
              &      &       &                    & 1                              &  11 &   9 &   4 &   3 &   0 \\ 
              & 0.9  & -1    & 0                  & 0                              & 377 & 238 & 145 &  75 &  31 \\ 
              &      &       &                    & 1                              &  24 &  14 &   8 &   6 &   4 \\ 
              &      &       & 1                  & 0                              & 136 &  24 &   6 &   1 &   0 \\ 
              &      &       &                    & 1                              &   9 &   1 &   0 &   0 &   0 \\ 
              &      & 0     & 0                  & 0                              & 187 &  91 &  46 &  20 &   6 \\ 
              &      &       &                    & 1                              &  10 &   4 &   3 &   2 &   0 \\ 
              &      &       & 1                  & 0                              & 172 &  72 &  30 &   6 &   2 \\ 
              &      &       &                    & 1                              &  12 &   2 &   2 &   3 &   0 \\ 
              &      & 1     & 0                  & 0                              &  74 &  18 &   0 &   0 &   0 \\ 
              &      &       &                    & 1                              &   4 &   2 &   1 &   0 &   0 \\ 
              &      &       & 1                  & 0                              & 188 & 125 &  74 &  49 &  24 \\ 
              &      &       &                    & 1                              &   8 &  10 &   6 &   1 &   0 \\ 
    \bottomrule
  \end{tabular}
  }
  
  \label{data tabulation high 200}
\end{table}

\begin{table}[htbp]
  \centering
  \caption{Data tabulation by treatment arms and outcome status when generating one dataset using the data generating algorithm in Table 2 of the Main Text for \textbf{medium sample sizes ($n = 1000$) under high event rate}.}
  \resizebox{\textwidth}{!}{%
  \begin{tabular}{*{5}{l}*{5}{r}}
    \toprule
     &  & &  &   \hspace{2cm} Trial visit &  & &  &  &  \\
    Sample size & Confounding strength & Treatment prevalence & Assigned treatment & Outcome \hspace{2cm} & \multicolumn{1}{c}{0} & \multicolumn{1}{c}{1} & \multicolumn{1}{c}{2} & \multicolumn{1}{c}{3} & \multicolumn{1}{c}{4} \\
    \midrule
1000        & 0.1  & -1    & 0                  & 0                              & 1865 & 1198 &  722 &  393 &  165 \\ 
              &      &       &                    & 1                              &  105 &   55 &   33 &   18 &    7 \\ 
              &      &       & 1                  & 0                              &  695 &  154 &   27 &    6 &    0 \\ 
              &      &       &                    & 1                              &   35 &    6 &    1 &    0 &    0 \\ 
              &      & 0     & 0                  & 0                              &  802 &  373 &  167 &   69 &   23 \\ 
              &      &       &                    & 1                              &   42 &   11 &    7 &    2 &    1 \\ 
              &      &       & 1                  & 0                              &  894 &  421 &  199 &   95 &   41 \\ 
              &      &       &                    & 1                              &   41 &   10 &    9 &    5 &    1 \\ 
              &      & 1     & 0                  & 0                              &  365 &  109 &   33 &   10 &    3 \\ 
              &      &       &                    & 1                              &   22 &    3 &    1 &    0 &    0 \\ 
              &      &       & 1                  & 0                              &  925 &  627 &  431 &  277 &  156 \\ 
              &      &       &                    & 1                              &   50 &   29 &   19 &   14 &   10 \\ 
              & 0.5  & -1    & 0                  & 0                              & 1871 & 1178 &  709 &  382 &  158 \\ 
              &      &       &                    & 1                              &   99 &   57 &   30 &   13 &    5 \\ 
              &      &       & 1                  & 0                              &  695 &  159 &   34 &    3 &    1 \\ 
              &      &       &                    & 1                              &   48 &    6 &    4 &    0 &    0 \\ 
              &      & 0     & 0                  & 0                              &  828 &  381 &  169 &   74 &   26 \\ 
              &      &       &                    & 1                              &   43 &   20 &    6 &    4 &    0 \\ 
              &      &       & 1                  & 0                              &  885 &  396 &  171 &   62 &   14 \\ 
              &      &       &                    & 1                              &   46 &   16 &    4 &    4 &    1 \\ 
              &      & 1     & 0                  & 0                              &  346 &   93 &   23 &    5 &    0 \\ 
              &      &       &                    & 1                              &   34 &   11 &    2 &    0 &    0 \\ 
              &      &       & 1                  & 0                              &  912 &  582 &  387 &  256 &  126 \\ 
              &      &       &                    & 1                              &   54 &   31 &    7 &    7 &    5 \\ 
              & 0.9  & -1    & 0                  & 0                              & 1698 & 1034 &  612 &  330 &  132 \\ 
              &      &       &                    & 1                              &  105 &   62 &   27 &   14 &    7 \\ 
              &      &       & 1                  & 0                              &  712 &  144 &   33 &   11 &    3 \\ 
              &      &       &                    & 1                              &   51 &   18 &    4 &    0 &    0 \\ 
              &      & 0     & 0                  & 0                              &  899 &  417 &  184 &   79 &   28 \\ 
              &      &       &                    & 1                              &   59 &   27 &   12 &    4 &    0 \\ 
              &      &       & 1                  & 0                              &  850 &  357 &  151 &   45 &   15 \\ 
              &      &       &                    & 1                              &   63 &   34 &    3 &    7 &    1 \\ 
              &      & 1     & 0                  & 0                              &  457 &  155 &   48 &   18 &    7 \\ 
              &      &       &                    & 1                              &   20 &    5 &    2 &    1 &    0 \\ 
              &      &       & 1                  & 0                              &  919 &  565 &  355 &  217 &  118 \\ 
              &      &       &                    & 1                              &   54 &   41 &   20 &   11 &    8 \\
    \bottomrule
  \end{tabular}
  }
  
  \label{data tabulation high 1000}
\end{table}

\begin{table}[htbp]
  \centering
  \caption{Data tabulation by treatment arms and outcome status when generating one dataset using the data generating algorithm in Table 2 of the Main Text for \textbf{large sample sizes ($n = 5000$) under high event rate}.}
  \resizebox{\textwidth}{!}{%
  \begin{tabular}{*{5}{l}*{5}{r}}
    \toprule
     &  & &  &   \hspace{2cm} Trial visit &  & &  &  &  \\
    Sample size & Confounding strength & Treatment prevalence & Assigned treatment & Outcome \hspace{2cm} & \multicolumn{1}{c}{0} & \multicolumn{1}{c}{1} & \multicolumn{1}{c}{2} & \multicolumn{1}{c}{3} & \multicolumn{1}{c}{4} \\
    \midrule
5000        & 0.1  & -1    & 0                  & 0                              & 9584 & 6122 & 3754 & 2069 &  851 \\ 
              &      &       &                    & 1                              &  593 &  356 &  188 &   85 &   40 \\ 
              &      &       & 1                  & 0                              & 3378 &  763 &  171 &   28 &    5 \\ 
              &      &       &                    & 1                              &  178 &   43 &    4 &    3 &    0 \\ 
              &      & 0     & 0                  & 0                              & 4440 & 2076 &  951 &  395 &  129 \\ 
              &      &       &                    & 1                              &  249 &  111 &   42 &   21 &    1 \\ 
              &      &       & 1                  & 0                              & 4426 & 2069 &  949 &  395 &  143 \\ 
              &      &       &                    & 1                              &  196 &   80 &   48 &   33 &    7 \\ 
              &      & 1     & 0                  & 0                              & 1739 &  471 &  125 &   34 &    9 \\ 
              &      &       &                    & 1                              &   89 &   25 &   10 &    1 &    0 \\ 
              &      &       & 1                  & 0                              & 4687 & 3258 & 2277 & 1508 &  845 \\ 
              &      &       &                    & 1                              &  215 &  146 &  110 &   78 &   35 \\ 
              & 0.5  & -1    & 0                  & 0                              & 9292 & 5884 & 3571 & 1958 &  808 \\ 
              &      &       &                    & 1                              &  574 &  341 &  193 &   93 &   39 \\ 
              &      &       & 1                  & 0                              & 3428 &  786 &  170 &   35 &    5 \\ 
              &      &       &                    & 1                              &  190 &   48 &   11 &    2 &    1 \\ 
              &      & 0     & 0                  & 0                              & 4482 & 2112 &  997 &  452 &  157 \\ 
              &      &       &                    & 1                              &  261 &   97 &   45 &   23 &    6 \\ 
              &      &       & 1                  & 0                              & 4382 & 1972 &  859 &  321 &  105 \\ 
              &      &       &                    & 1                              &  200 &   90 &   35 &   13 &    2 \\ 
              &      & 1     & 0                  & 0                              & 1772 &  484 &  131 &   29 &    6 \\ 
              &      &       &                    & 1                              &  106 &   20 &    5 &    1 &    0 \\ 
              &      &       & 1                  & 0                              & 4644 & 3119 & 2033 & 1328 &  746 \\ 
              &      &       &                    & 1                              &  244 &  134 &  122 &   51 &   34 \\ 
              & 0.9  & -1    & 0                  & 0                              & 8655 & 5343 & 3152 & 1676 &  672 \\ 
              &      &       &                    & 1                              &  545 &  317 &  172 &   94 &   45 \\ 
              &      &       & 1                  & 0                              & 3451 &  769 &  192 &   42 &    7 \\ 
              &      &       &                    & 1                              &  332 &   56 &   13 &    5 &    2 \\ 
              &      & 0     & 0                  & 0                              & 4517 & 2133 &  971 &  405 &  129 \\ 
              &      &       &                    & 1                              &  256 &  114 &   46 &   17 &    7 \\ 
              &      &       & 1                  & 0                              & 4295 & 1776 &  704 &  253 &   78 \\ 
              &      &       &                    & 1                              &  320 &   94 &   43 &   20 &    4 \\ 
              &      & 1     & 0                  & 0                              & 1920 &  528 &  161 &   44 &   10 \\ 
              &      &       &                    & 1                              &  107 &   19 &    9 &    2 &    0 \\ 
              &      &       & 1                  & 0                              & 4558 & 2802 & 1696 & 1013 &  475 \\ 
              &      &       &                    & 1                              &  325 &  167 &   90 &   54 &   27 \\ 
    \bottomrule
  \end{tabular}
  }
  
  \label{data tabulation high 5000}
\end{table}

\clearpage
\subsection{CI coverage in medium and high event rate scenarios}
\begin{sidewaysfigure}[h]
 \centering
 \includegraphics[width = \textwidth]{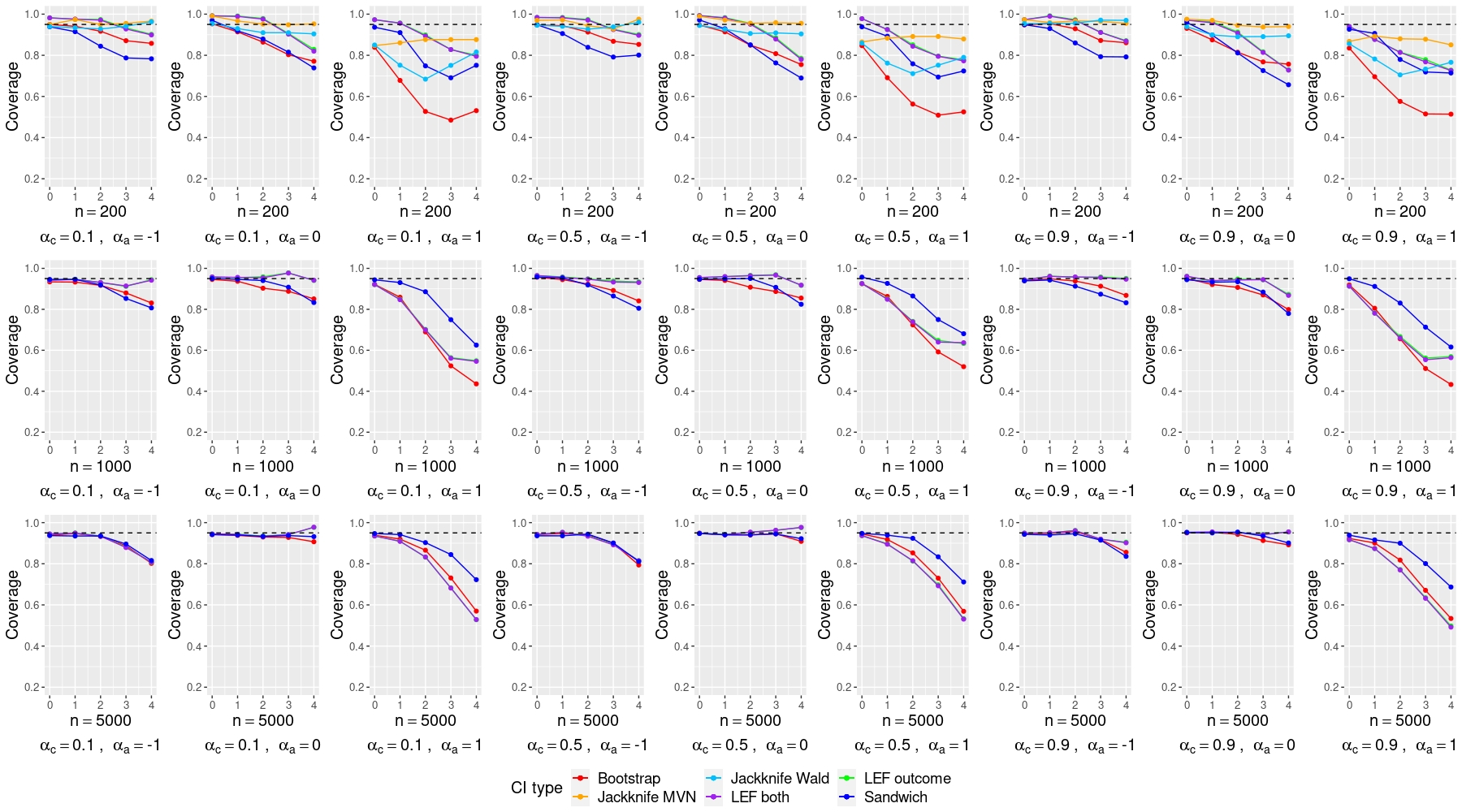}
 \caption{Coverage of the CIs under \textbf{medium event rates}. Bootstrap: CIs constructed by nonparametric bootstrap; LEF both: CIs constructed by applying Approach 2 of LEF bootstrap; LEF outcome: CIs constructed by applying Approach 1 of LEF bootstrap; Jackknife Wald: CIs constructed by applying Approach 1 of Jackknife resampling; Jackknife MVN: CIs constructed by applying Approach 2 of Jackknife resampling; Sandwich: CIs based on the sandwich variance estimator}
 \label{fig:coverage_med}
\end{sidewaysfigure}

\begin{sidewaysfigure}[h]
 \centering
 \includegraphics[width = \textwidth]{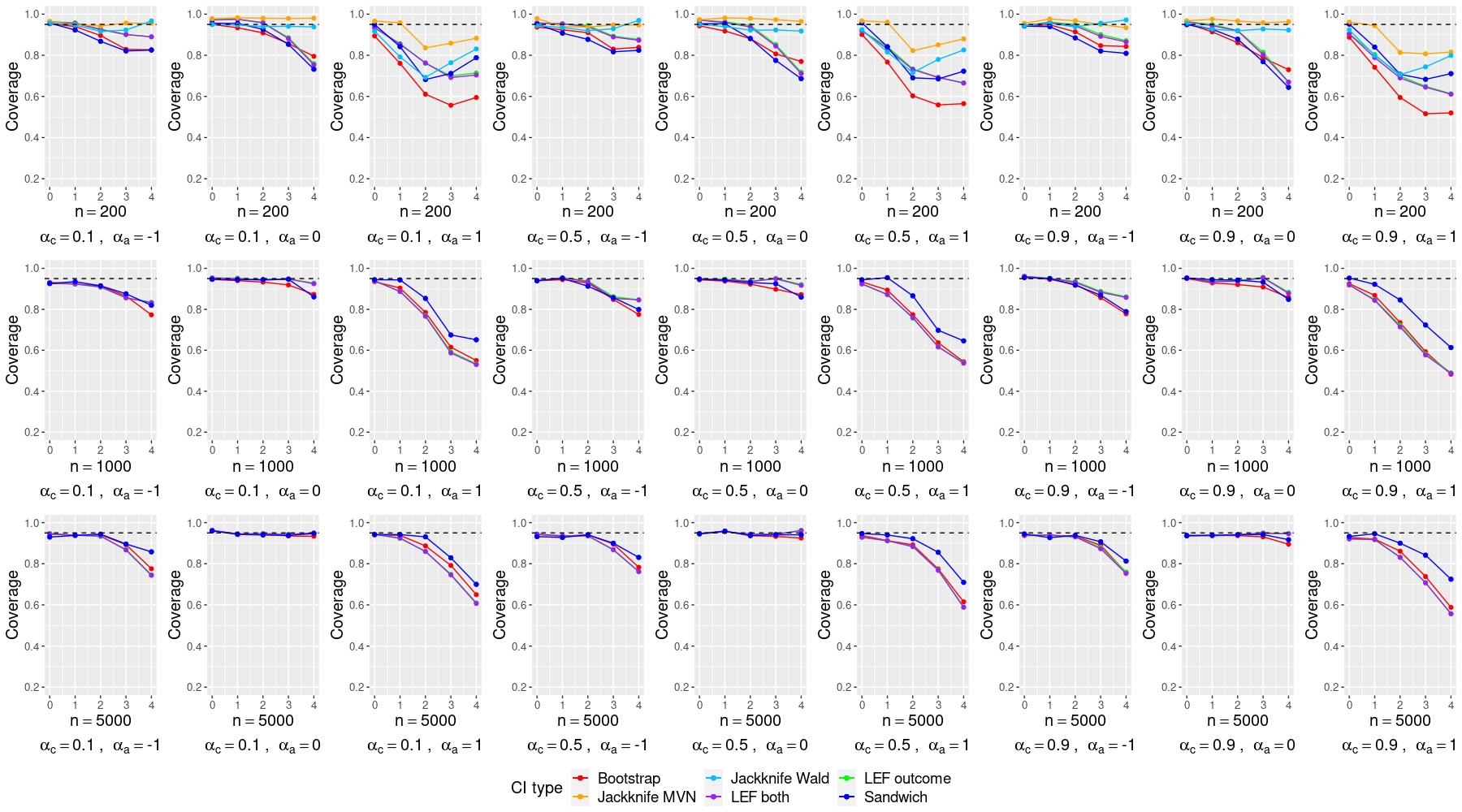}
 \caption{Coverage of the CIs under \textbf{high event rates}. Bootstrap: CIs constructed by nonparametric bootstrap; LEF both: CIs constructed by applying Approach 2 of LEF bootstrap; LEF outcome: CIs constructed by applying Approach 1 of LEF bootstrap; Jackknife Wald: CIs constructed by applying Approach 1 of Jackknife resampling; Jackknife MVN: CIs constructed by applying Approach 2 of Jackknife resampling; Sandwich: CIs based on the sandwich variance estimator}
 \label{fig:coverage_high}
\end{sidewaysfigure}

Figures~\ref{fig:coverage_med} and \ref{fig:coverage_high} present the empirical coverage rates of the CIs when the event rates were medium and high, respectively.

\clearpage
\subsection{Bias-eliminated CI coverage}
\begin{sidewaysfigure}[h]
 \centering
 \includegraphics[width = \textwidth]{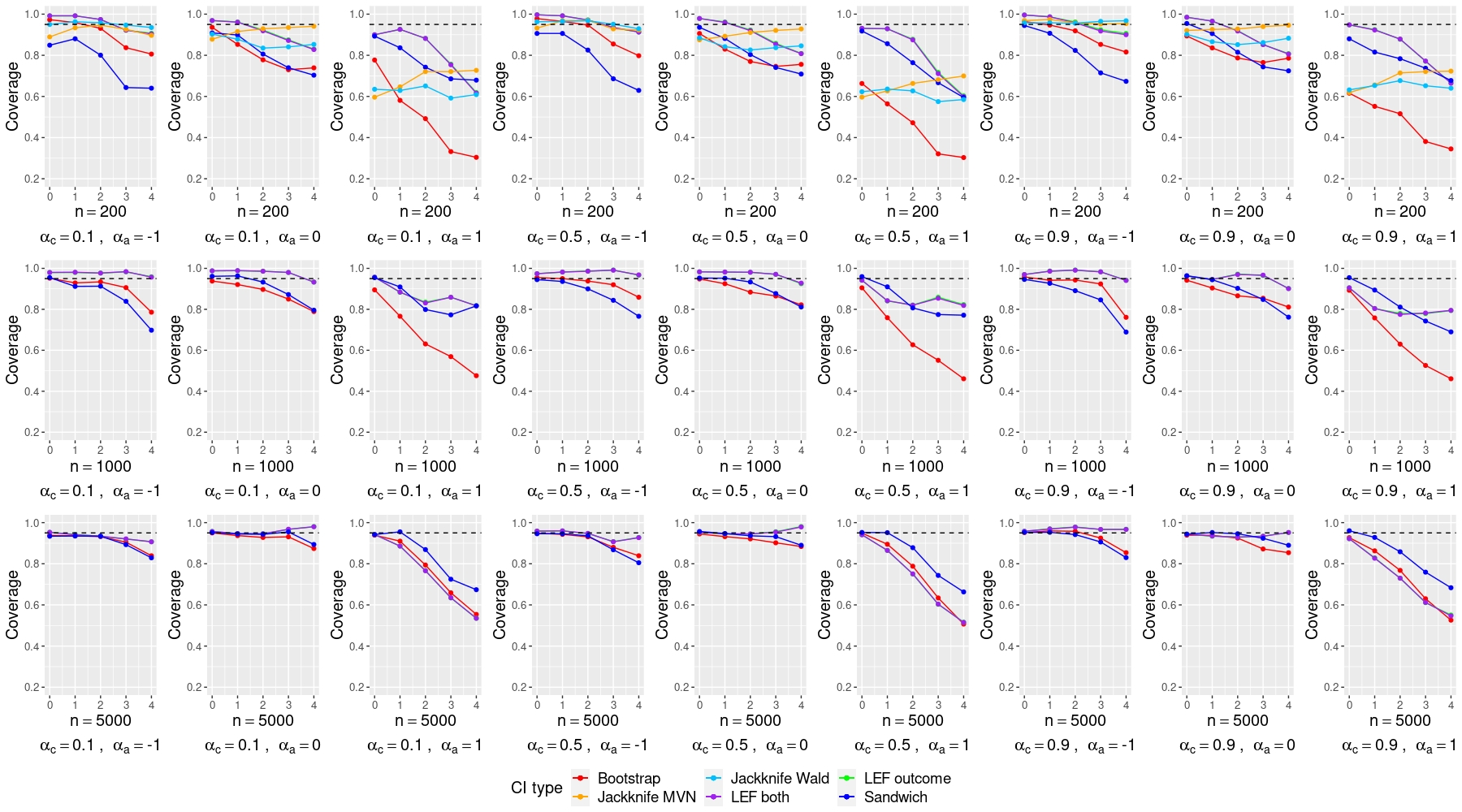}
 \caption{Bias-eliminated coverage of the CIs under \textbf{low event rates}. Bootstrap: CIs constructed by nonparametric bootstrap; LEF both: CIs constructed by applying Approach 2 of LEF bootstrap; LEF outcome: CIs constructed by applying Approach 1 of LEF bootstrap; Jackknife Wald: CIs constructed by applying Approach 1 of Jackknife resampling; Jackknife MVN: CIs constructed by applying Approach 2 of Jackknife resampling; Sandwich: CIs based on the sandwich variance estimator}
 \label{fig:bias_elim_coverage_low}
\end{sidewaysfigure}

\begin{sidewaysfigure}[h]
 \centering
 \includegraphics[width = \textwidth]{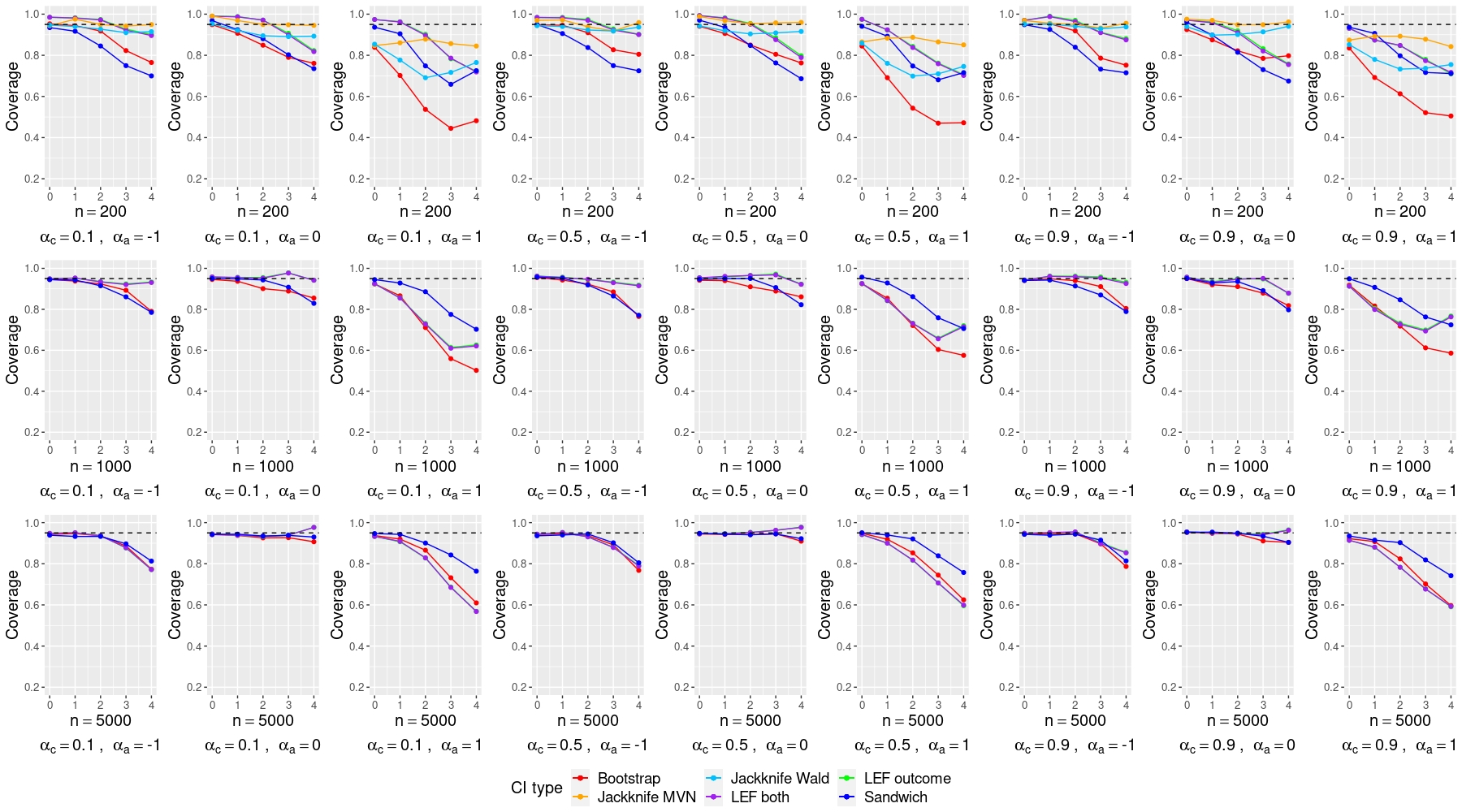}
 \caption{Bias-eliminated coverage of the CIs under \textbf{medium event rates}. Bootstrap: CIs constructed by nonparametric bootstrap; LEF both: CIs constructed by applying Approach 2 of LEF bootstrap; LEF outcome: CIs constructed by applying Approach 1 of LEF bootstrap; Jackknife Wald: CIs constructed by applying Approach 1 of Jackknife resampling; Jackknife MVN: CIs constructed by applying Approach 2 of Jackknife resampling; Sandwich: CIs based on the sandwich variance estimator}
 \label{fig:bias_elim_coverage_med}
\end{sidewaysfigure}

\begin{sidewaysfigure}[h]
 \centering
 \includegraphics[width = \textwidth]{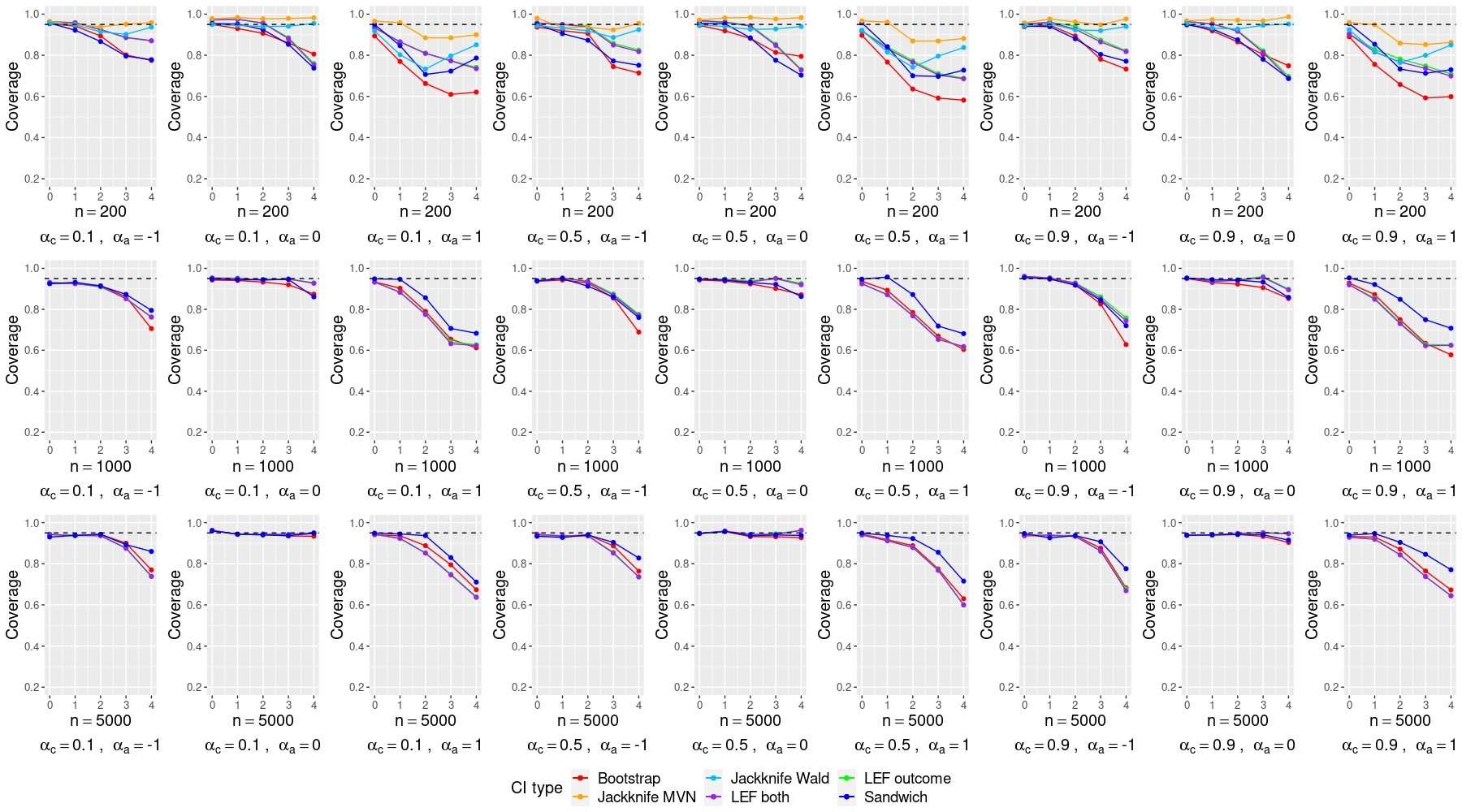}
 \caption{Bias-eliminated coverage of the CIs under \textbf{high event rates}. Bootstrap: CIs constructed by nonparametric bootstrap; LEF both: CIs constructed by applying Approach 2 of LEF bootstrap; LEF outcome: CIs constructed by applying Approach 1 of LEF bootstrap; Jackknife Wald: CIs constructed by applying Approach 1 of Jackknife resampling; Jackknife MVN: CIs constructed by applying Approach 2 of Jackknife resampling; Sandwich: CIs based on the sandwich variance estimator}
 \label{fig:bias_elim_coverage_high}
\end{sidewaysfigure}

Figures~\ref{fig:bias_elim_coverage_low}--\ref{fig:bias_elim_coverage_high} present bias-eliminated coverage rates under different outcome rate scenarios.

\clearpage
\subsection{Empirical standard deviation and MSE of the MRD estimates}

\begin{sidewaysfigure}[h!]
 \centering
 \includegraphics[width = \textwidth]{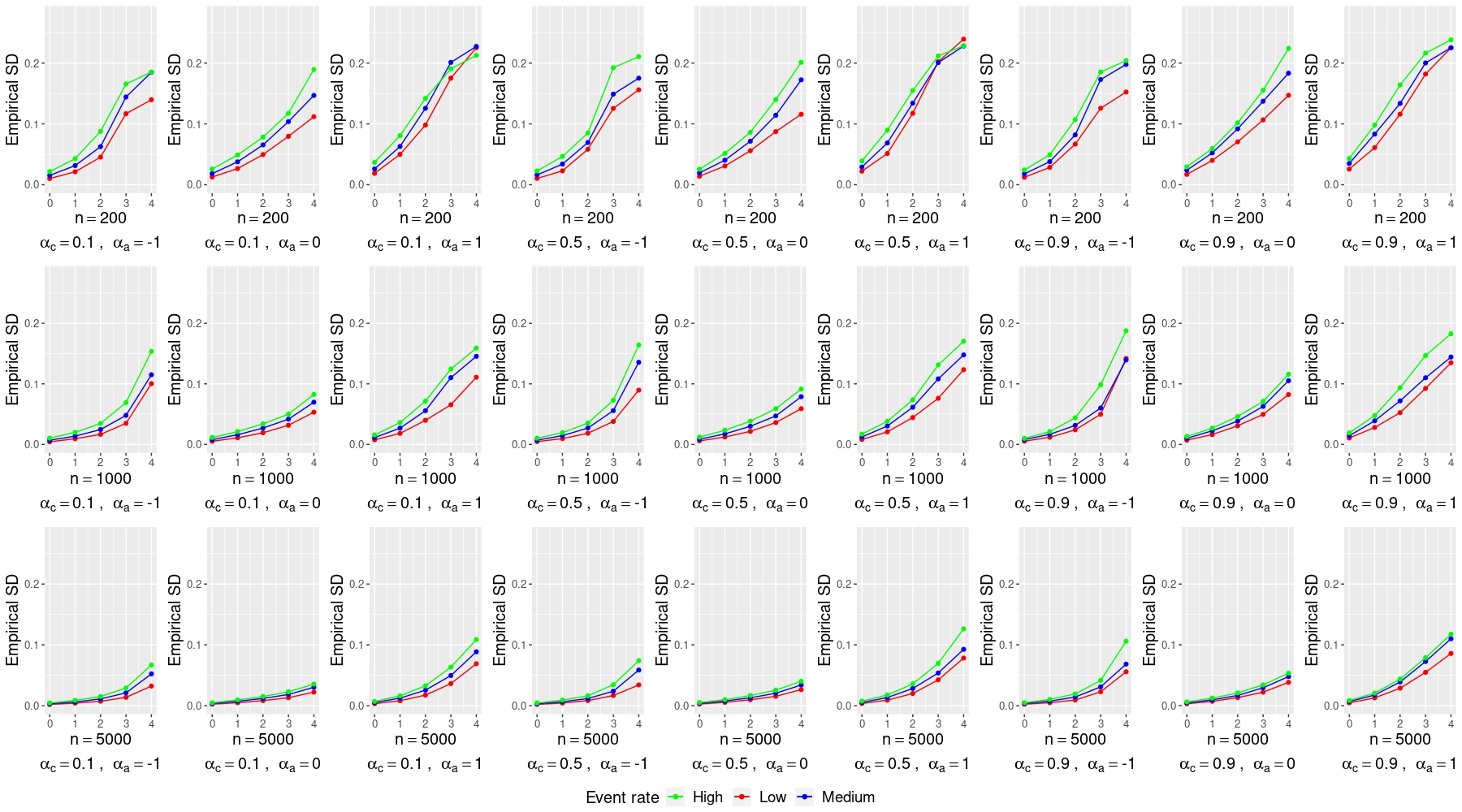}
 \caption{Empirical standard deviation of the MRD estimates in various simulation scenarios.}
 \label{fig:sd_all}
\end{sidewaysfigure}

For all scenarios examined, the empirical standard deviations (Figure~\ref{fig:sd_all}) were always higher at later visits than at earlier visits. 
There are a few possible reasons for this: 1) due to the non-collapsibility of logistic models, we were not able to specify an exactly correct MSM of the potential outcomes parameterised by the discrete-time hazard. Instead, we used a rich model stratified by trial visits, which led to  data sparsity and larger variability in MRD estimates at later visits. 2) Only earlier trials had data at longer follow-up trial visits, so the information was scarce by design.  3) As mentioned previously, we rely on IPW to estimate the MRD. 
The limited amount of observations at later visits paired with the increased imbalance between the treatment groups likely resulted in the estimated weights for later trials visits being larger and  more variable.
Consequently, the larger magnitude and variability of the weights at later visits could contribute to the unstable estimation of   the MRD. 

\begin{sidewaysfigure}[h!]
 \centering
 \includegraphics[width = \textwidth]{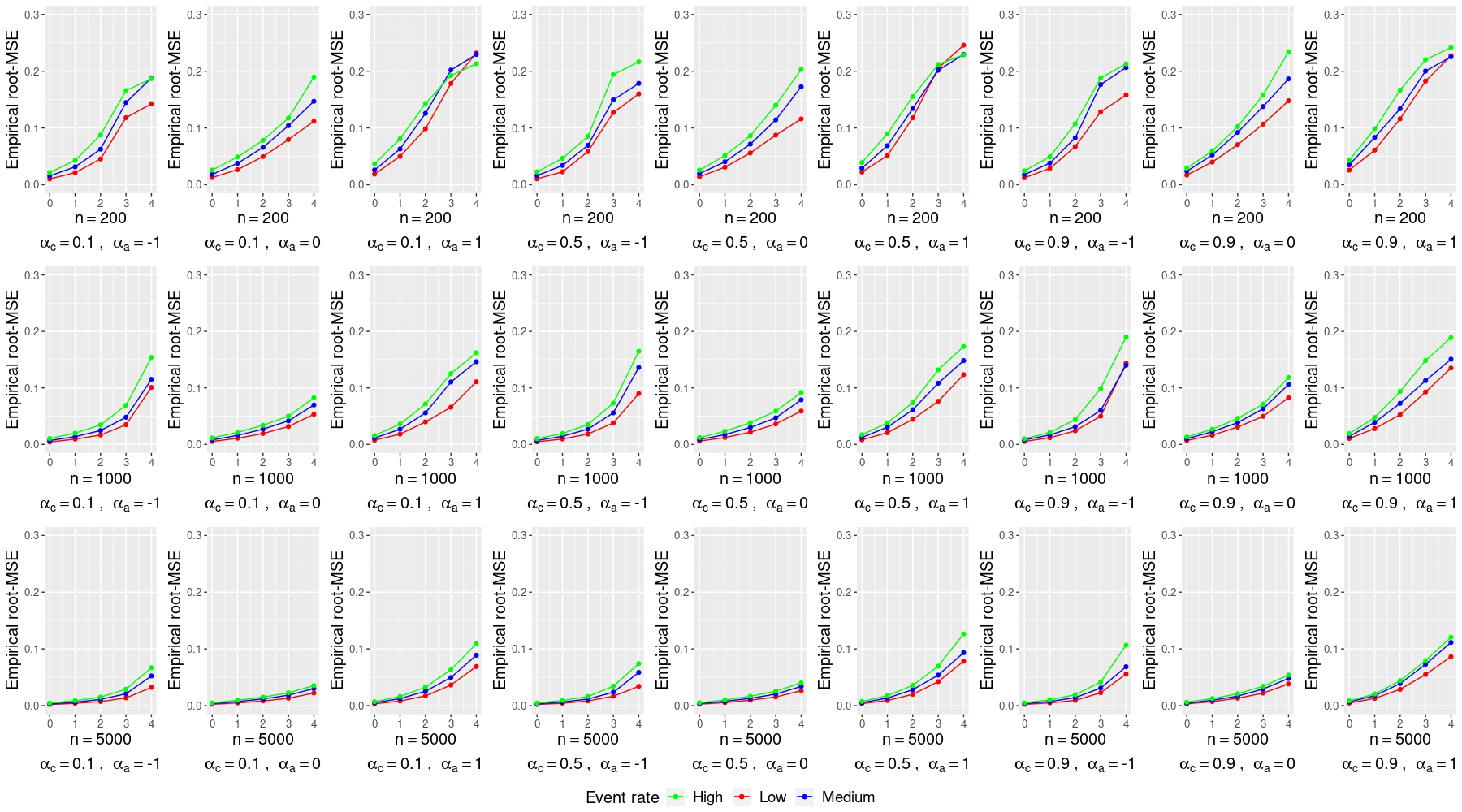}
 \caption{Root-mean-squared error (MSE) of the MRD estimates in various simulation scenarios.}
 \label{fig:mse_low}
\end{sidewaysfigure}

The patterns of empirical standard deviations largely followed those of empirical biases across treatment prevalence and confounding strength scenarios. That is, the scenarios with larger biases also exhibited larger standard deviations. Overall, the impact of treatment prevalence on empirical standard deviations was more prominent than that of confounding strength, possibly due to the small range we specified for the coefficients of the time-varying confounder in the treatment and outcome processes.  Interestingly, the empirical standard deviations increased when event rates were higher, when fixing the sample size, treatment prevalence and confounding strength. 
This phenomenon might be explained by that more patients from the untreated group were depleted from the at-risk sets due to more event occurrences, which resulted in larger treatment group imbalances at later visits. 
As expected, larger sample sizes mitigated this variability by providing more information.
Results for the root-MSE were consistent with the results for bias and empirical standard deviations and can be found in Figure~\ref{fig:mse_low}.



\clearpage

\clearpage

\subsection{Ratio of estimated standard errors to the empirical standard deviation of the MRD estimator }

\begin{sidewaysfigure}[h]
 \centering
 \includegraphics[width = \textwidth]{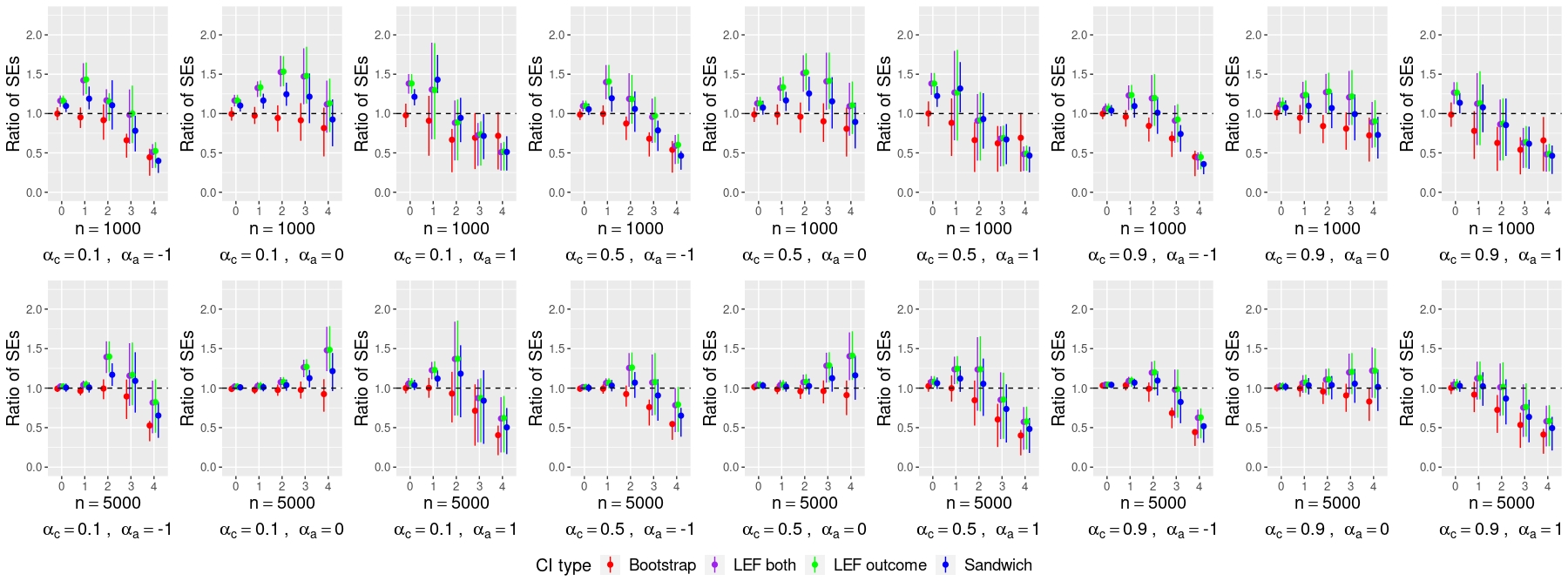}
 \caption{Ratio of the estimated standard error to empirical standard deviation of the MRD estimator (SE ratio) in  \textbf{low event rate and medium/large sample size scenarios} ($n = 1000, 5000$). The dots represent the averages of the ratio, with the bottom and top of the bar being the 1st and 3rd quartile of this ratio respectively. Bootstrap: CIs constructed by nonparametric bootstrap; LEF both: CIs constructed by applying Approach 2 of LEF bootstrap; LEF outcome: CIs constructed by applying Approach 1 of LEF bootstrap; Jackknife Wald: CIs constructed by applying Approach 1 of jackknife resampling; Sandwich: CIs based on the sandwich variance estimator.}
 \label{fig:se_ratio_low_1000_5000}
\end{sidewaysfigure}

\begin{figure}[h]
 \centering
 \includegraphics[width = \textwidth]{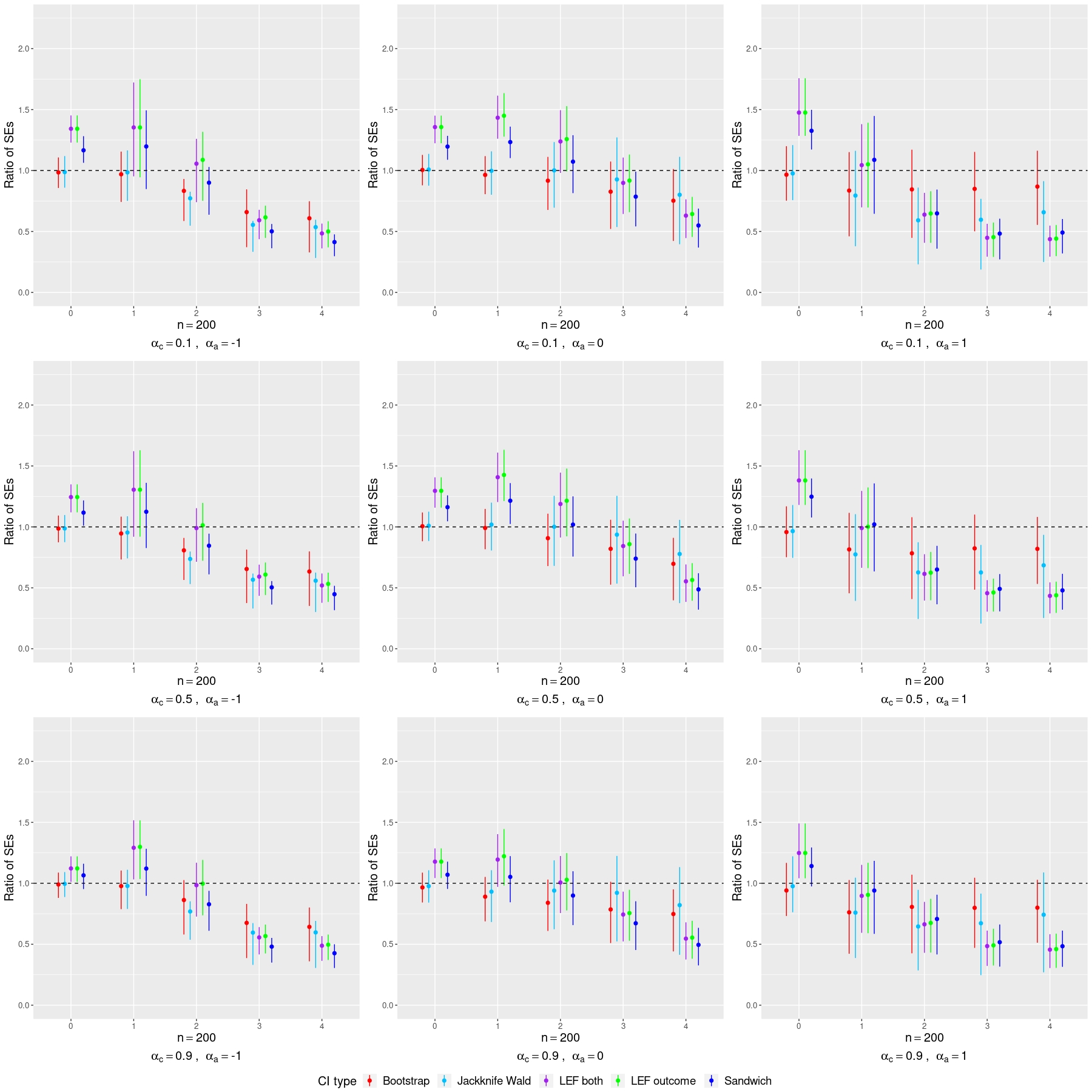}
 \caption{Ratio of the estimated standard error to empirical standard deviation of the MRD estimator (SE ratio) in \textbf{medium event rate and small sample size scenarios}. The dots represent the averages of the ratio, with the bottom and top of the bar being the 1st and 3rd quartile of this ratio respectively. Bootstrap: CIs constructed by nonparametric bootstrap; LEF both: CIs constructed by applying Approach 2 of LEF bootstrap; LEF outcome: CIs constructed by applying Approach 1 of LEF bootstrap; Jackknife Wald: CIs constructed by applying Approach 1 of jackknife resampling; Sandwich: CIs based on the sandwich variance estimator.}
 \label{fig:se_ratio_med_200}
\end{figure}

\begin{figure}
\centering
 \includegraphics[width = \textwidth]{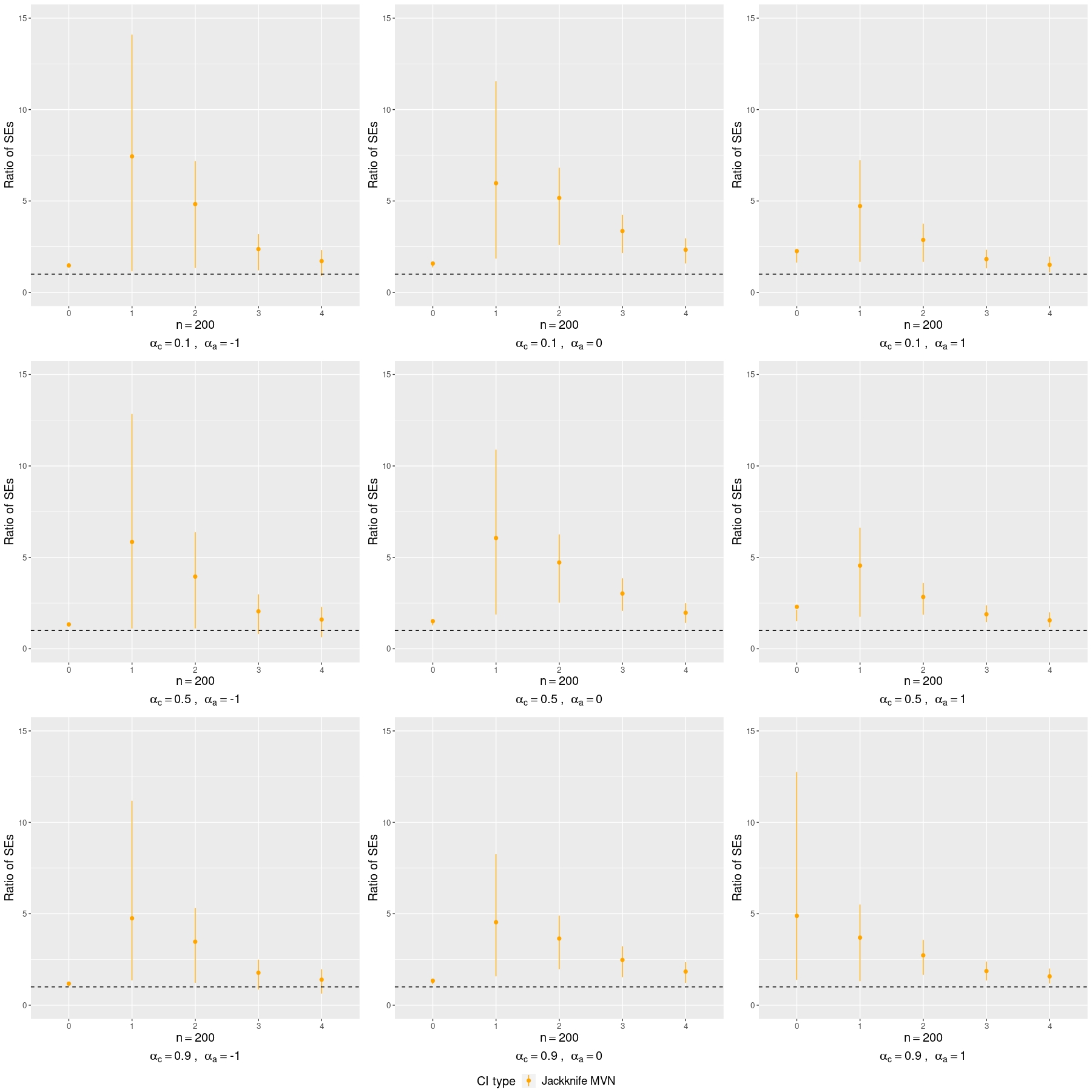}
 \caption{Ratio of the estimated standard error to empirical standard deviation of the MRD estimator (SE ratio) in \textbf{medium event rate and small sample size scenarios}. The dots represent the averages of the ratio, with the bottom and top of the bar being the 1st and 3rd quartile of this ratio respectively. Jackknife MVN: CIs constructed by applying Approach 2 of jackknife resampling. }
 \label{fig:se_ratio_med_jackknifeMVN}
\end{figure}

\begin{sidewaysfigure}[h]
 \centering
 \includegraphics[width = \textwidth]{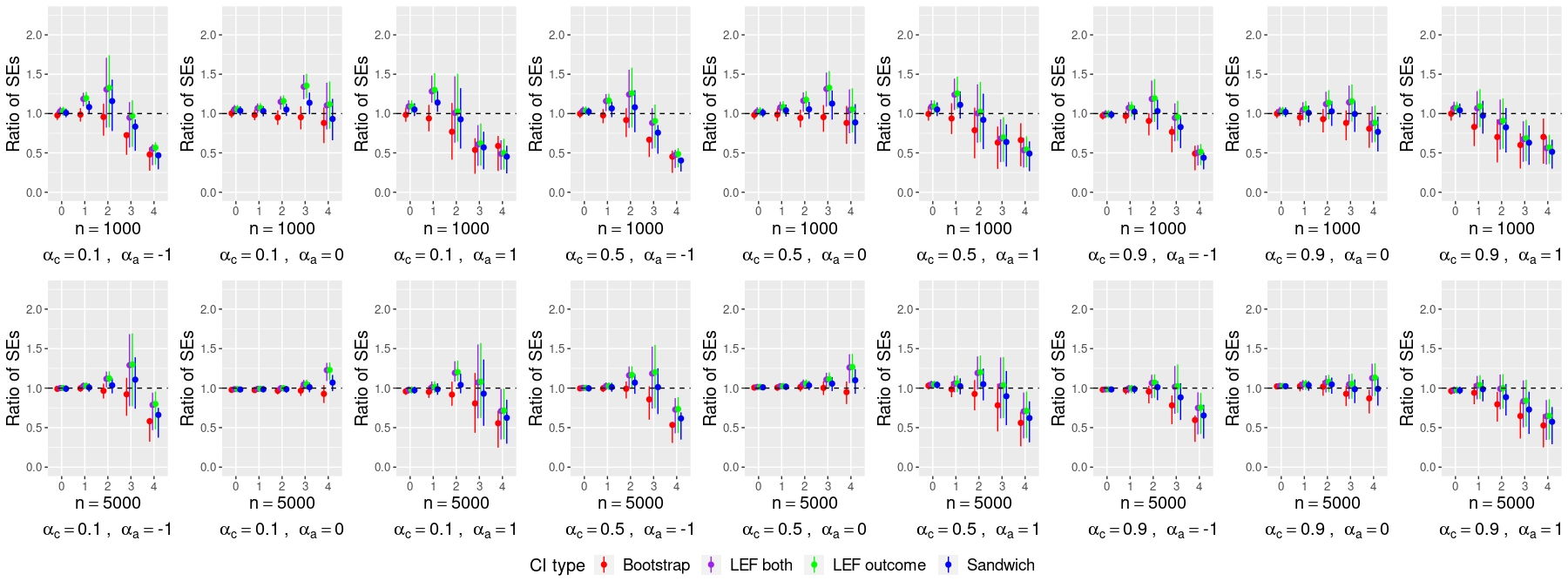}
 \caption{Ratio of the estimated standard error to empirical standard deviation of the MRD estimator (SE ratio) in  \textbf{medium event rate and medium/large sample size scenarios} ($n = 1000, 5000$). The dots represent the averages of the ratio, with the bottom and top of the bar being the 1st and 3rd quartile of this ratio respectively. Bootstrap: CIs constructed by nonparametric bootstrap; LEF both: CIs constructed by applying Approach 2 of LEF bootstrap; LEF outcome: CIs constructed by applying Approach 1 of LEF bootstrap; Jackknife Wald: CIs constructed by applying Approach 1 of jackknife resampling; Sandwich: CIs based on the sandwich variance estimator.}
 \label{fig:se_ratio_med_1000_5000}
\end{sidewaysfigure}

\begin{figure}[h]
 \centering
 \includegraphics[width = \textwidth]{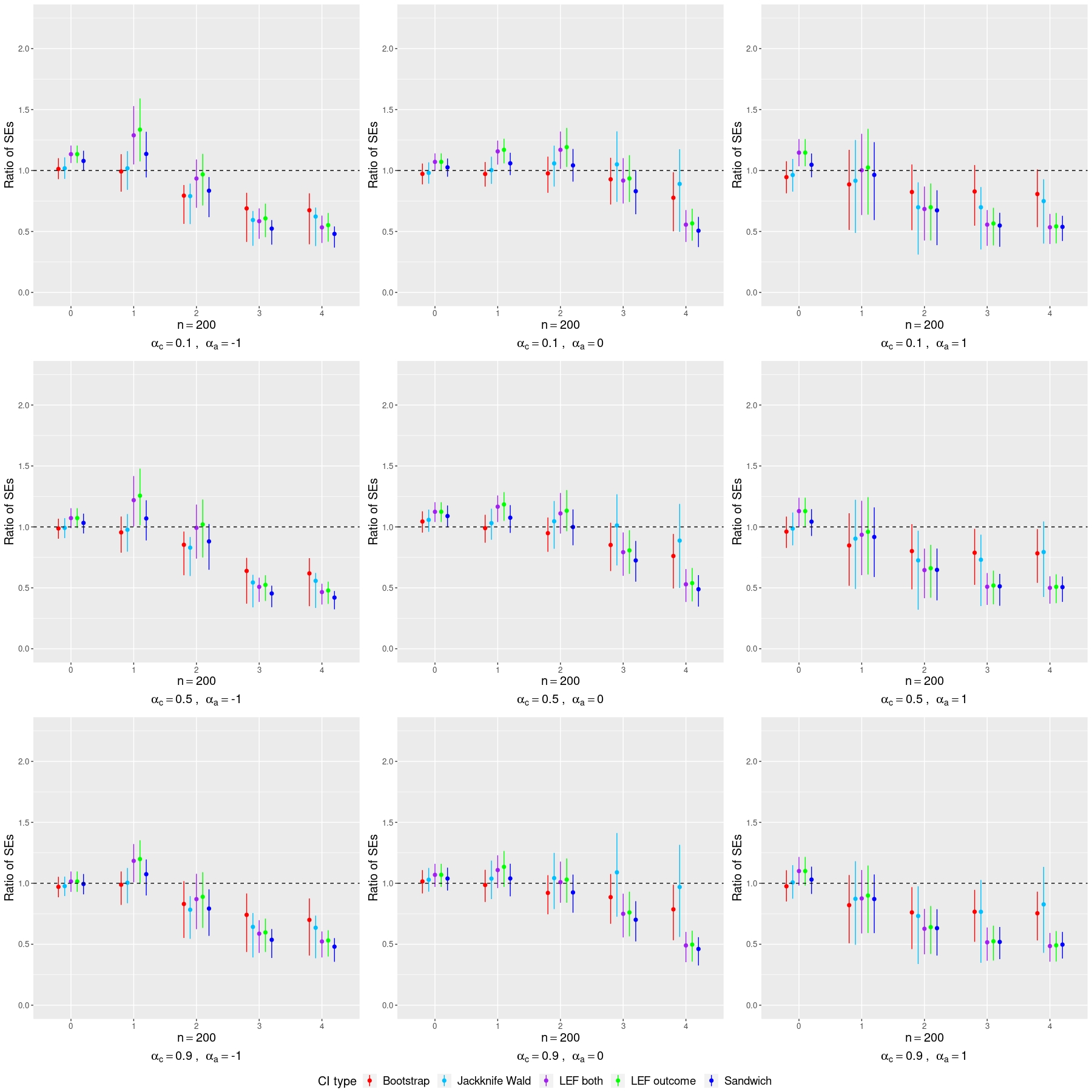}
 \caption{Ratio of the estimated standard error to empirical standard deviation of the MRD estimator (SE ratio) in \textbf{high event rate and small sample size scenarios}. The dots represent the averages of the ratio, with the bottom and top of the bar being the 1st and 3rd quartile of this ratio respectively. Bootstrap: CIs constructed by nonparametric bootstrap; LEF both: CIs constructed by applying Approach 2 of LEF bootstrap; LEF outcome: CIs constructed by applying Approach 1 of LEF bootstrap; Jackknife Wald: CIs constructed by applying Approach 1 of jackknife resampling; Sandwich: CIs based on the sandwich variance estimator.}
 \label{fig:se_ratio_high_200}
\end{figure}

\begin{figure}
 \includegraphics[width = \textwidth]{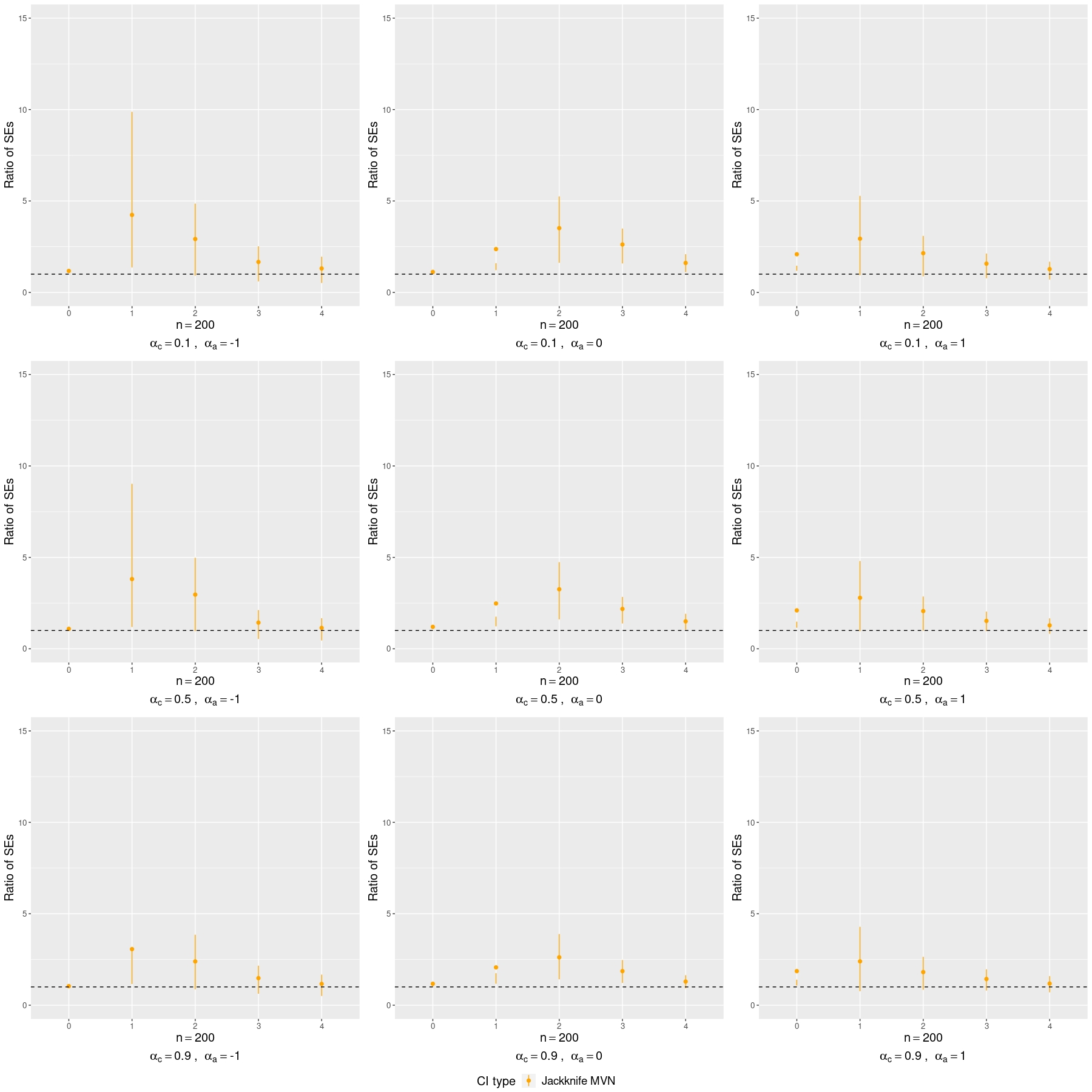}
 \caption{Ratio of the estimated standard error to empirical standard deviation of the MRD estimator (SE ratio) in \textbf{high event rate and small sample size scenarios}. The dots represent the averages of the ratio, with the bottom and top of the bar being the 1st and 3rd quartile of this ratio respectively. Jackknife MVN: CIs constructed by applying Approach 2 of jackknife resampling. }
 \label{fig:se_ratio_high_jackknifeMVN}
\end{figure}

\begin{sidewaysfigure}[h]
 \centering
 \includegraphics[width = \textwidth]{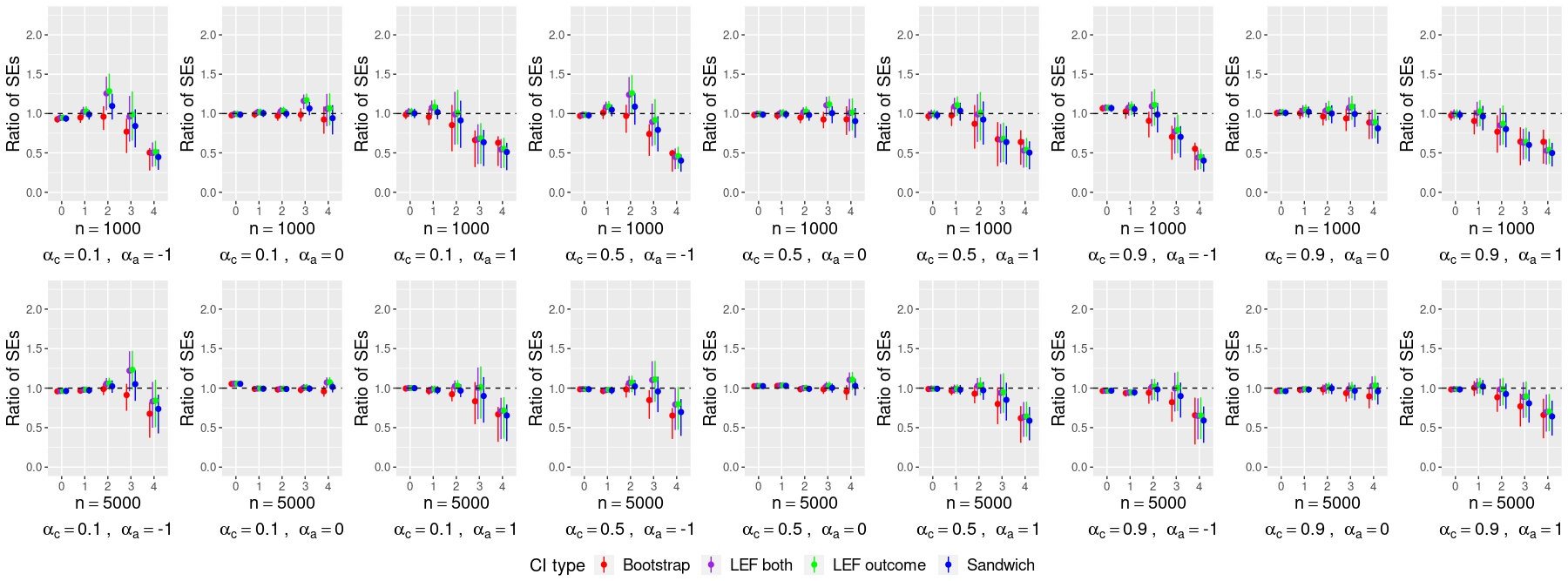}
 \caption{Ratio of the estimated standard error to empirical standard deviation of the MRD estimator (SE ratio) in  \textbf{high event rate and medium/large sample size scenarios} ($n = 1000, 5000$). The dots represent the averages of the ratio, with the bottom and top of the bar being the 1st and 3rd quartile of this ratio respectively. Bootstrap: CIs constructed by nonparametric bootstrap; LEF both: CIs constructed by applying Approach 2 of LEF bootstrap; LEF outcome: CIs constructed by applying Approach 1 of LEF bootstrap; Jackknife Wald: CIs constructed by applying Approach 1 of jackknife resampling; Sandwich: CIs based on the sandwich variance estimator.}
 \label{fig:se_ratio_high_1000_5000}
\end{sidewaysfigure}

Figure~\ref{fig:se_ratio_low_1000_5000} presents the SE ratio results in the scenarios with   low event rate and moderate/large sample sizes. 
Figure~\ref{fig:se_ratio_med_200}--\ref{fig:se_ratio_high_1000_5000} present the SE ratio results in the  medium and high event rate scenarios.

\clearpage
\subsection{Monte Carlo standard error of the CI coverage} \label{appendix: MCSE}
\begin{sidewaysfigure}[h]
 \centering
 \includegraphics[width = \textwidth]{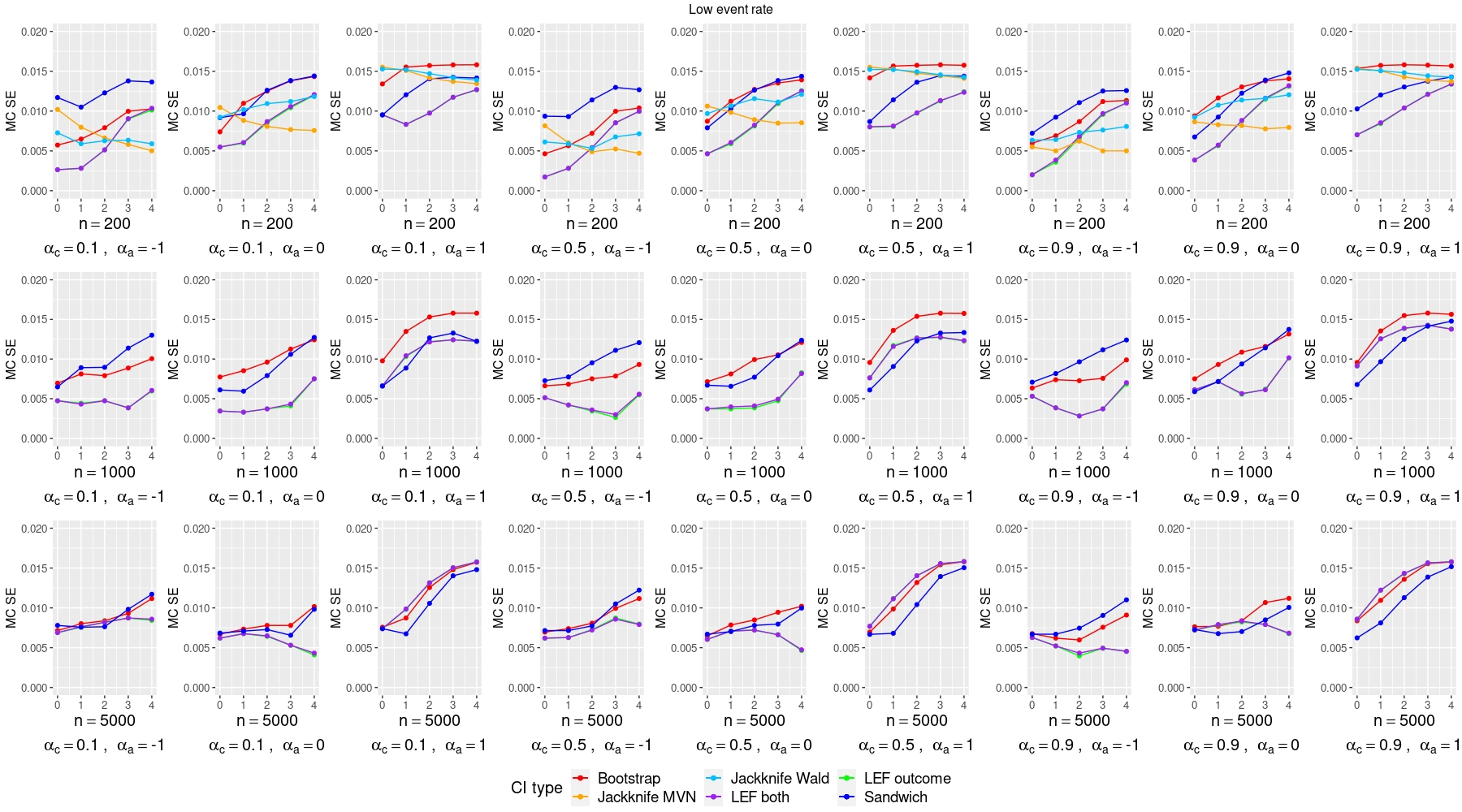}
 \caption{Monte Carlo standard error of the CI coverage under \textbf{low event rates}. Bootstrap: CIs constructed by nonparametric bootstrap; LEF both: CIs constructed by applying Approach 2 of LEF bootstrap; LEF outcome: CIs constructed by applying Approach 1 of LEF bootstrap; Jackknife Wald: CIs constructed by applying Approach 1 of Jackknife resampling; Jackknife MVN: CIs constructed by applying Approach 2 of Jackknife resampling; Sandwich: CIs based on the sandwich variance estimator}
 \label{fig:MCSE_low}
\end{sidewaysfigure}

\begin{sidewaysfigure}[h]
 \centering
 \includegraphics[width = \textwidth]{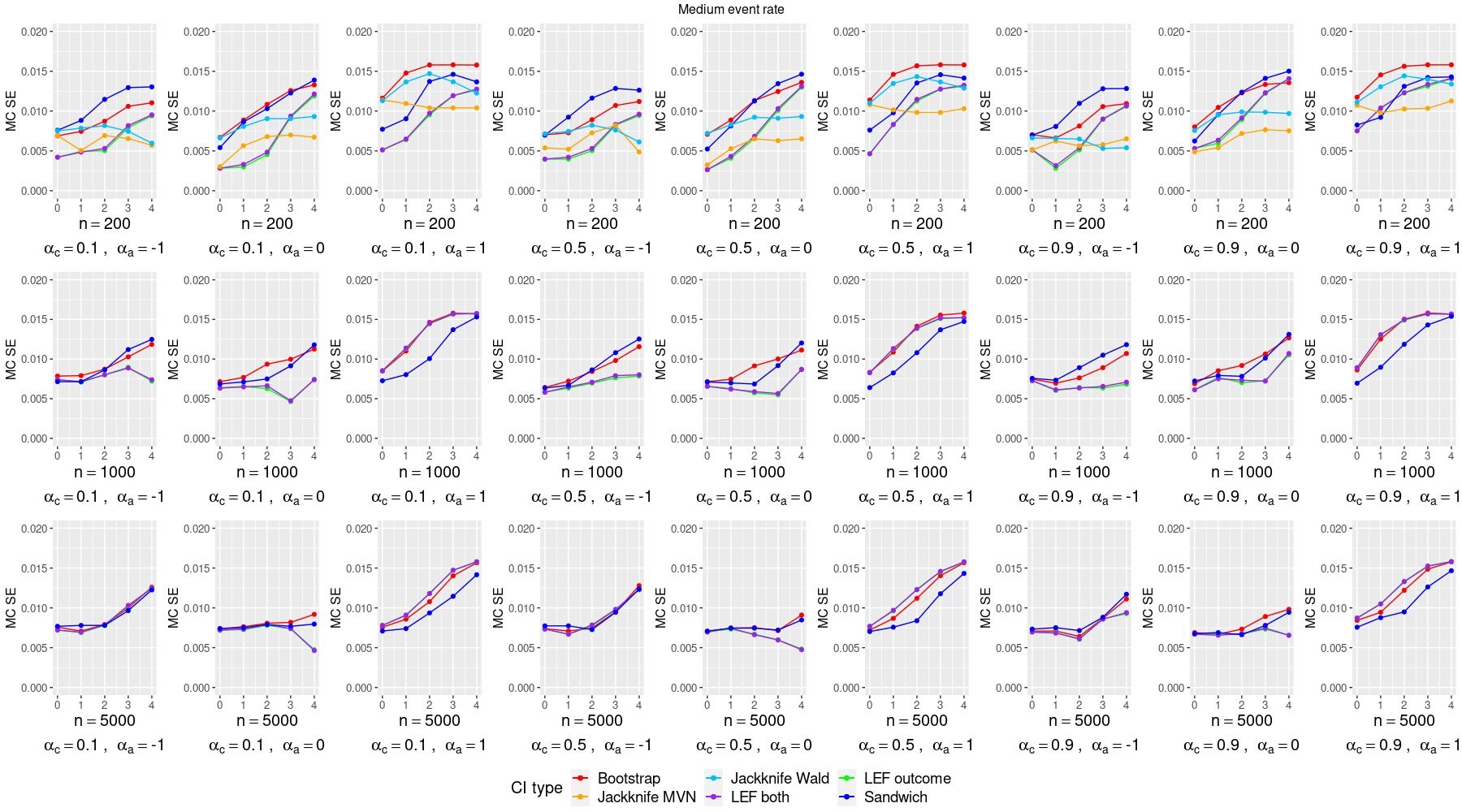}
 \caption{Monte Carlo standard error of  the CI coverage under \textbf{medium event rates}. Bootstrap: CIs constructed by nonparametric bootstrap; LEF both: CIs constructed by applying Approach 2 of LEF bootstrap; LEF outcome: CIs constructed by applying Approach 1 of LEF bootstrap; Jackknife Wald: CIs constructed by applying Approach 1 of Jackknife resampling; Jackknife MVN: CIs constructed by applying Approach 2 of Jackknife resampling; Sandwich: CIs based on the sandwich variance estimator}
 \label{fig:MCSE_med}
\end{sidewaysfigure}

\begin{sidewaysfigure}[h]
 \centering
 \includegraphics[width = \textwidth]{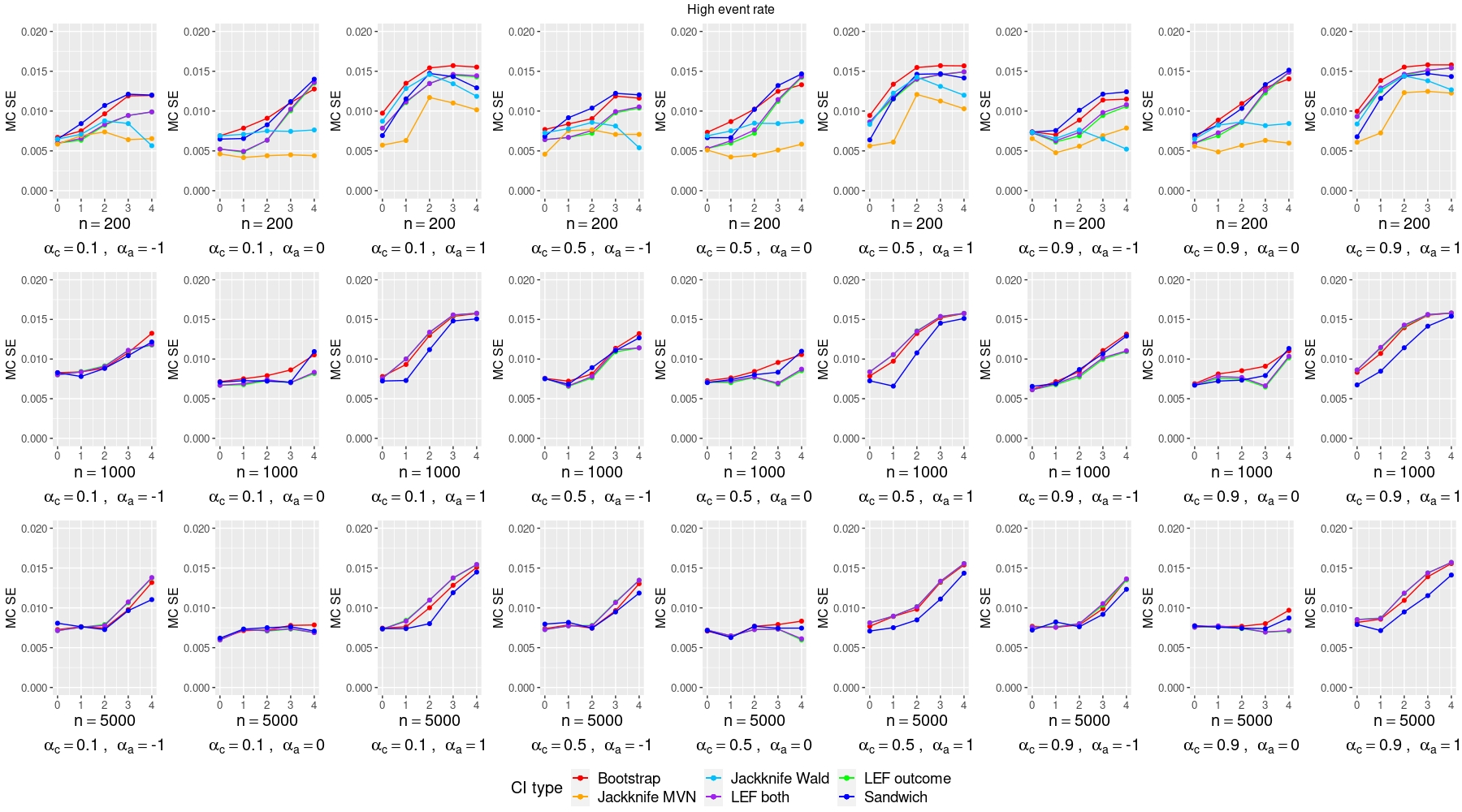}
 \caption{Monte Carlo standard error the CI coverage under \textbf{high event rates}. Bootstrap: CIs constructed by nonparametric bootstrap; LEF both: CIs constructed by applying Approach 2 of LEF bootstrap; LEF outcome: CIs constructed by applying Approach 1 of LEF bootstrap; Jackknife Wald: CIs constructed by applying Approach 1 of Jackknife resampling; Jackknife MVN: CIs constructed by applying Approach 2 of Jackknife resampling; Sandwich: CIs based on the sandwich variance estimator}
 \label{fig:MCSE_high}
\end{sidewaysfigure}

Finally, to assess whether lower-than-nominal coverage in some simulation scenarios may be due to uncertainty induced by Monte Carlo simulation error, we calculated and illustrated the standard errors of empirical coverage rates in Figures \ref{fig:MCSE_low}--\ref{fig:MCSE_high}. We observe minimal Monte Carlo standard errors across all simulation scenarios, with the error mostly remaining at 1.5\%, especially for small sample sizes where we did not observe much difference in Monte Carlo errors between the CI methods. According to the Monte Carlo standard error formula for CI coverage, if the expected coverage is 95\% and we want to acheive a  standard error of 1.5\% or less across the simulations, then $n_{sim} = \frac{95(100-95)}{1.5^2} = 211$ simulations are required \cite{morris_using_2019}. As a result, 1000 simulations were sufficient to theoretically achieve nominal coverage with a Monte Carlo error of at most 1.5\%  across the  simulations.
We also observed lower Monte Carlo errors at earlier visits in large  sample size ($n = 5000$) scenarios. As the outcome event rate increased, the Monte Carlo errors converged to similar values for all CI methods as well. 

\clearpage
\subsection{Convergence issues} \label{appendix: failure rates}
During the simulations, we encountered estimation errors when the \texttt{glm} function in \texttt{R} was used by the \texttt{trial\_msm} function of the \texttt{TrialEmulation} package to estimate the MSM parameters in small sample size scenarios ($n = 200$). The MSM parameters $\beta_{1,3}$ and $\beta_{1,4}$ in equation (11) of the main text 
were often inestimable in such scenarios due to data sparsity issues. When such errors occurred, the default in the \texttt{predict.glm} function used for estimating the MRD was to treat the inestimable parameters as equal to 0. Therefore, we are still  able to estimate the MRD  and construct nonparametric bootstrap CIs, LEF bootstrap CIs and Jackknife Wald CIs. However, the data sparsity issues also led to difficulty to construct Sandwich CIs because of non-positive definiteness  and non-symmetry of the variance matrices.

From Tables \ref{failures_low}, \ref{failures_med} and \ref{failures_high}, we see that, whilst we encountered little to no convergence issues when constructing nonparametric bootstrap CIs, LEF bootstrap CIs and Jackknife Wald CIs for all data scenarios, there were substantially more problems of non-positive definiteness  and non-symmetry when computing the sandwich variance matrices. This led to a high rate of failures, ranging from 3 to 81\%, in constructing  the Sandwich CIs as the multivariate normal distribution for sampling requires a symmetric, positive definite variance matrix. There were problems with these matrix properties for more than half of the simulations when the sample size was $n = 200$, whilst the failure rate for sample size $n = 1000$ was between 3\% and 37\%. Although we have not included the failure rates by confounding strength,  we note that the failure rate for the Sandwich CIs was further exacerbated in scenarios with $n = 200$ and high treatment prevalence, averaging at around 81\%. On occasion, we observed that this was due to \texttt{R} being too restrictive with the matrix properties. In some cases, the sandwich variance matrix was almost symmetric with differences between symmetric values at a negligible magnitude of $10^{-4}$, so discarding the  results based on  such a difference and forcing a symmetric matrix for Sandwich CIs may have been judicious, and may have modified the coverage rate results for these CIs slightly. For very large sample sizes $n = 5000$, the failure rate was negligible across all CI methods.

Jackknife MVN CIs also encountered many construction issues, where for low and high treatment prevalence scenarios, this method failed for roughly half of the simulations, making it unreliable for such simulation scenarios. These construction issues were often due to the treatment coefficients at the later visits ($k = 3,4$)  in the MSM being inestimable in some jackknife samples. Therefore the jackknife variance estimate in equation (11) of the main text could not be calculated.

\begin{table}[ht]
\centering
\caption{Proportions of simulations with  CI construction failures in \textbf{low event rate scenarios, aggregated by sample size and treatment prevalence}. Bootstrap: CIs constructed by nonparametric bootstrap; LEF both: CIs constructed by applying Approach 2 of LEF bootstrap; LEF outcome: CIs constructed by applying Approach 1 of LEF bootstrap; Jackknife Wald: CIs constructed by applying Approach 1 of Jackknife resampling; Jackknife MVN: CIs constructed by applying Approach 2 of Jackknife resampling; Sandwich: CIs based on the sandwich variance estimator}
\scalebox{0.8}{
\begin{tabular}{lrrrrrrrr}
  \hline
Outcome event & Sample size & Treatment & Bootstrap & LEF outcome & LEF both & Sandwich & Jackknife MVN & Jackknife Wald \\ 
rate &  & prevalence &  &  & & &  & \\ 
  \hline
  Low & 200 & -1 & 0.00 & 0.00 & 0.00 & 0.07 & 0.46 & 0.00 \\ 
   &  &  & 0.00 & 0.00 & 0.00 & 0.07 & 0.46 & 0.00 \\ 
   &  &  & 0.00 & 0.00 & 0.00 & 0.07 & 0.46 & 0.00 \\ 
   &  & 0 & 0.00 & 0.00 & 0.00 & 0.27 & 0.03 & 0.00 \\ 
   &  & & 0.00 & 0.00 & 0.00 & 0.27 & 0.03 & 0.00 \\ 
   &  &  & 0.00 & 0.00 & 0.00 & 0.27 & 0.03 & 0.00 \\ 
   &  & 1 & 0.00 & 0.00 & 0.00 & 0.81 & 0.51 & 0.00 \\ 
   &  &  & 0.00 & 0.00 & 0.00 & 0.81 & 0.51 & 0.00 \\ 
   &  &  & 0.00 & 0.00 & 0.00 & 0.81 & 0.51 & 0.00 \\ 
   & 1000 & -1 & 0.00 & 0.00 & 0.00 & 0.00 &  &  \\ 
   &  &  & 0.00 & 0.00 & 0.00 & 0.00 &  &  \\ 
   &  &  & 0.00 & 0.00 & 0.00 & 0.00 &  &  \\ 
   &  & 0 & 0.00 & 0.00 & 0.00 & 0.03 &  &  \\ 
   &  & & 0.00 & 0.00 & 0.00 & 0.03 &  &  \\ 
   &  &  & 0.00 & 0.00 & 0.00 & 0.03 &  &  \\ 
   &  & 1 & 0.00 & 0.00 & 0.00 & 0.37 &  &  \\ 
   &  &  & 0.00 & 0.00 & 0.00 & 0.37 &  &  \\ 
   &  &  & 0.00 & 0.00 & 0.00 & 0.37 &  &  \\ 
   & 5000 & -1 & 0.00 & 0.00 & 0.00 & 0.00 &  &  \\ 
   &  &  & 0.00 & 0.00 & 0.00 & 0.00 &  &  \\ 
   &  &  & 0.00 & 0.00 & 0.00 & 0.00 &  &  \\ 
   &  & 0 & 0.00 & 0.00 & 0.00 & 0.00 &  &  \\ 
   &  &  & 0.00 & 0.00 & 0.00 & 0.00 &  &  \\ 
   &  &  & 0.00 & 0.00 & 0.00 & 0.00 &  &  \\ 
   &  & 1 & 0.00 & 0.00 & 0.00 & 0.01 &  &  \\ 
   &  &  & 0.00 & 0.00 & 0.00 & 0.01 &  &  \\ 
   &  &  & 0.00 & 0.00 & 0.00 & 0.01 &  &  \\ 
\hline
\end{tabular}
}
\label{failures_low}
\end{table}

\begin{table}[ht]
\centering
\caption{Proportions of simulations with CI construction failures in \textbf{medium event rate scenarios, aggregated by sample size and treatment prevalence}. Bootstrap: CIs constructed by nonparametric bootstrap; LEF both: CIs constructed by applying Approach 2 of LEF bootstrap; LEF outcome: CIs constructed by applying Approach 1 of LEF bootstrap; Jackknife Wald: CIs constructed by applying Approach 1 of Jackknife resampling; Jackknife MVN: CIs constructed by applying Approach 2 of Jackknife resampling; Sandwich: CIs based on the sandwich variance estimator}
\scalebox{0.8}{
\begin{tabular}{lrrrrrrrr}
  \hline
Outcome event & Sample size & Treatment & Bootstrap & LEF outcome & LEF both & Sandwich & Jackknife MVN & Jackknife Wald \\ 
rate &  & prevalence &  &  & & &  & \\ 
  \hline
  Medium & 200 & -1 & 0.00 & 0.00 & 0.00 & 0.04 & 0.47 & 0.00 \\ 
   &  &  & 0.00 & 0.00 & 0.00 & 0.04 & 0.47 & 0.00 \\ 
   & &  & 0.00 & 0.00 & 0.00 & 0.04 & 0.47 & 0.00 \\ 
   &  & 0 & 0.00 & 0.00 & 0.00 & 0.12 & 0.06 & 0.00 \\ 
   &  &  & 0.00 & 0.00 & 0.00 & 0.12 & 0.06 & 0.00 \\ 
   &  &  & 0.00 & 0.00 & 0.00 & 0.12 & 0.06 & 0.00 \\ 
   &  & 1 & 0.00 & 0.00 & 0.00 & 0.63 & 0.51 & 0.00 \\ 
   &  &  & 0.00 & 0.00 & 0.00 & 0.63 & 0.51 & 0.00 \\ 
   &  &  & 0.00 & 0.00 & 0.00 & 0.63 & 0.51 & 0.00 \\ 
   & 1000 & -1 & 0.00 & 0.00 & 0.00 & 0.00 &  &  \\ 
   &  &  & 0.00 & 0.00 & 0.00 & 0.00 &  &  \\ 
   &  & & 0.00 & 0.00 & 0.00 & 0.00 &  &  \\ 
   &  & 0 & 0.00 & 0.00 & 0.00 & 0.01 &  &  \\ 
   &  &  & 0.00 & 0.00 & 0.00 & 0.01 &  &  \\ 
   &  &  & 0.00 & 0.00 & 0.00 & 0.01 &  &  \\ 
   &  & 1 & 0.00 & 0.00 & 0.00 & 0.10 &  &  \\ 
   &  &  & 0.00 & 0.00 & 0.00 & 0.10 &  &  \\ 
   &  &  & 0.00 & 0.00 & 0.00 & 0.10 &  &  \\ 
   & 5000 & -1 & 0.00 & 0.00 & 0.00 & 0.00 &  &  \\ 
   &  & & 0.00 & 0.00 & 0.00 & 0.00 &  &  \\ 
   &  &  & 0.00 & 0.00 & 0.00 & 0.00 &  &  \\ 
   &  & 0 & 0.00 & 0.00 & 0.00 & 0.00 &  &  \\ 
   &  &  & 0.00 & 0.00 & 0.00 & 0.00 &  &  \\ 
   &  &  & 0.00 & 0.00 & 0.00 & 0.00 &  &  \\ 
   &  & 1 & 0.00 & 0.00 & 0.00 & 0.00 &  &  \\ 
   &  &  & 0.00 & 0.00 & 0.00 & 0.00 &  &  \\ 
   &  &  & 0.00 & 0.00 & 0.00 & 0.00 &  &  \\ 
   \hline
\end{tabular}
}
\label{failures_med}
\end{table}

\begin{table}[ht]
\centering
\caption{Proportions of simulations with CI construction failures in \textbf{high event rate scenarios, aggregated by sample size and treatment prevalence}. Bootstrap: CIs constructed by nonparametric bootstrap; LEF both: CIs constructed by applying Approach 2 of LEF bootstrap; LEF outcome: CIs constructed by applying Approach 1 of LEF bootstrap; Jackknife Wald: CIs constructed by applying Approach 1 of Jackknife resampling; Jackknife MVN: CIs constructed by applying Approach 2 of Jackknife resampling; Sandwich: CIs based on the sandwich variance estimator}
\scalebox{0.8}{
\begin{tabular}{lrrrrrrrr}
  \hline
Outcome event & Sample size & Treatment & Bootstrap & LEF outcome & LEF both & Sandwich & Jackknife MVN & Jackknife Wald \\ 
rate &  & prevalence &  &  & & &  & \\ 
  \hline
High & 200 & -1 & 0.00 & 0.00 & 0.00 & 0.04 & 0.46 & 0.00 \\ 
   &  &  & 0.00 & 0.00 & 0.00 & 0.04 & 0.46 & 0.00 \\ 
   &  &  & 0.00 & 0.00 & 0.00 & 0.04 & 0.46 & 0.00 \\ 
   &  & 0 & 0.00 & 0.00 & 0.00 & 0.07 & 0.08 & 0.00 \\ 
   &  &  & 0.00 & 0.00 & 0.00 & 0.07 & 0.08 & 0.00 \\ 
   &  &  & 0.00 & 0.00 & 0.00 & 0.07 & 0.08 & 0.00 \\ 
   &  & 1 & 0.00 & 0.00 & 0.00 & 0.41 & 0.50 & 0.00 \\ 
   &  &  & 0.00 & 0.00 & 0.00 & 0.41 & 0.50 & 0.00 \\ 
   &  &  & 0.00 & 0.00 & 0.00 & 0.41 & 0.50 & 0.00 \\ 
   & 1000 & -1 & 0.00 & 0.00 & 0.00 & 0.00 &  &  \\ 
   &  &  & 0.00 & 0.00 & 0.00 & 0.00 &  &  \\ 
   &  &  & 0.00 & 0.00 & 0.00 & 0.00 &  &  \\ 
   &  & 0 & 0.00 & 0.00 & 0.00 & 0.00 &  &  \\ 
   &  &  & 0.00 & 0.00 & 0.00 & 0.00 &  &  \\ 
   &  &  & 0.00 & 0.00 & 0.00 & 0.00 &  &  \\ 
   &  & 1 & 0.00 & 0.00 & 0.00 & 0.03 &  &  \\ 
   &  &  & 0.00 & 0.00 & 0.00 & 0.03 &  &  \\ 
   &  &  & 0.00 & 0.00 & 0.00 & 0.03 &  &  \\ 
   & 5000  & -1 & 0.00 & 0.00 & 0.00 & 0.00 &  &  \\ 
   &  &  & 0.00 & 0.00 & 0.00 & 0.00 &  &  \\ 
   &  &  & 0.00 & 0.00 & 0.00 & 0.00 &  &  \\ 
   &  & 0 & 0.00 & 0.00 & 0.00 & 0.00 &  &  \\ 
   &  &  & 0.00 & 0.00 & 0.00 & 0.00 &  &  \\ 
   &  &  & 0.00 & 0.00 & 0.00 & 0.00 &  &  \\ 
   &  & 1 & 0.00 & 0.00 & 0.00 & 0.00 &  &  \\ 
   &  &  & 0.00 & 0.00 & 0.00 & 0.00 &  &  \\ 
   &  &  & 0.00 & 0.00 & 0.00 & 0.00 &  &  \\ 
  
   \hline
\end{tabular}
}
\label{failures_high}
\end{table}

\clearpage
\section{Target trial protocol to estimate the per-protocol
effect of HAART on all-cause mortality using the HERS data}

\begin{table}[ht]
\centering
\caption{Protocol of a  target trial to estimate the per-protocol
effect of HAART on all-cause mortality}
\small
\begin{tabular}{p{0.3\linewidth} p{0.6\linewidth}}
\toprule
\textit{Eligibility criteria} & HIV-infected women with no prior history of HAART. 
\\
\\
\textit{Treatment strategies} & Patients in the treatment arm will receive HAART throughout the trial.  Patients in the control arm will not receive HAART during the trial (but may take other antiviral treatments).\\
\\
\textit{Assignment procedures} & Patients will be randomly assigned to either treatment strategy at baseline and will be aware of the strategy to which they have been assigned. \\
\\
\textit{Follow-up period} & Patients are followed from treatment randomisation at baseline visit, for a course of 5 visits spaced 6 months apart, or until event or loss to follow-up, whichever occurs first.\\
\\
\textit{Outcome} & Death from any cause\\
\\
\textit{Causal contrast of interest} & Per-protocol effect of sustained HAART treatment on all-cause mortality\\
\\
\textit{Statistical Methods} & Patients' follow-up will be artificially censored at the time they deviate from their assigned treatment strategy at baseline. IPW using time-varying covariates will be used to address the selection bias caused by this artificial censoring and dependent censoring due to patient's loss to follow-up. \\
 \bottomrule
\end{tabular}

\label{HERS protocol}
\end{table}

\bibliographystyle{SageV}
\bibliography{references}